\documentclass[10pt,twocolumn,aps,prd,amsmath,amssymb,preprintnumbers,nofootinbib]{revtex4-1}

\usepackage{graphicx}
\usepackage{tensor}
\usepackage{xcolor}
\usepackage{braket}

\usepackage{multirow}

\begin{document}

\setcounter{footnote}{0}

\title{Hamiltonian gravity in tetrad-connection variables}
\author{Erick I. Duque}
\email{eqd5272@psu.edu}
\affiliation{Institute for Gravitation and the Cosmos,
The Pennsylvania State University, 104 Davey Lab, University Park,
PA 16802, USA}

\begin{abstract}
A systematic Hamiltonian formulation of the Einstein--Cartan system, based on the Hilbert--Palatini action with the Barbero--Immirzi and cosmological constants, is performed using the traditional ADM decomposition and without fixing the time gauge. This procedure results in a larger phase space compared to that of the Ashtekar--Barbero approach as well as a larger set of first-class constraints generating gauge transformations that are on-shell equivalent to spacetime diffeomorphisms and SO(1,3) transformations. The imbalance in the number of components between the tetrad and the connection is resolved by the identification of second-class constraints implied by the action, which can be implemented by use of Dirac brackets or by solving them directly. The Hamiltonian system remains well-defined off the second-class constraint surface in an extended phase space with additional degrees of freedom, implying a more general geometric theory.
Implications for canonical quantum gravity are discussed.
\end{abstract}

\maketitle

\section{Introduction}

Compared to Einstein's metric system, the Einstein--Cartan theory has a larger gravitational field content described by a tetrad $e^\mu_I$ and a connection 1-form $\tensor{\omega}{_\mu^I^J}$, which are kinematically independent from each other and related only by dynamics.
(Greek letters are used for spacetime indices and capital Latin letters $I,J,K,.\,.\,. \in \{0,1,2,3\}$ for internal Lorentz indices.)
In this system, the spacetime metric is no longer an elementary field, but a function of the tetrad given by $g^{\mu\nu}=\eta^{IJ}e^\mu_Ie^\nu_J$, where $\eta^{IJ}$ is the Minkowski metric of an internal space.
Two central advantages of the tetrad-connection variables over the metric ones are the ease with which fermions can be coupled and the possibility to describe dynamical spacetime geometries with torsion.
Therefore, a complete understanding of its Hamiltonian formulation, including its full gauge content, is paramount for canonical approaches to quantum or modified gravity with fermionic matter and torsion.

However, detailed analyses in canonical formulations of gravity are more complicated than in its Lagrangian cousin because it distinguishes the degrees of freedom between configuration and momentum variables, which are not necessarily spacetime tensors and hence transformation properties, such as covariance, are not manifest.
Typically, the ADM decomposition \cite{ADM,arnowitt2008republication} is adopted such that the spacetime region of interest is assumed hyperbolic, $M = \Sigma \times \mathbb{R}$, with a three-dimensional spatial manifold $\Sigma$.
The manifold $M$ is then foliated into spacelike hypersurfaces $\Sigma_t$ parametrized by $t\in{\mathbb R}$.
The spacetime metric $g_{\mu\nu}$ on $M$ then defines a unit vector field $n^{\mu}$ normal to $\Sigma_t$ and induces a spatial metric $q_{ab}(t)$ on $\Sigma_{t}$ given by the restriction of the spacetime tensor $q_{\mu\nu}=g_{\mu\nu}+n_{\mu}n_{\nu}$ to $\Sigma_{t}$.
(We use the Latin letters $a,b,c,\dots,h$ for spatial indices, and $i,j,k,.\,.\,.\in\{1,2,3\}$ will be reserved for internal Euclidean indices.)
Based on this foliation, the line element takes the form
\begin{equation}
  {\rm d} s^2 = - N^2 {\rm d} t^2 + q_{a b} ( {\rm d} x^a + N^a {\rm d} t )
  ( {\rm d} x^b + N^b {\rm d} t )
  \,,
  \label{eq:ADM line element}
\end{equation}
where $N$ and $N^a$ turn out to be Lagrange multipliers in the Einstein--Hilbert action when written in the ADM variables,
\begin{eqnarray}
    S_{\rm EH} 
    \!\!&=&\!\! \frac{1}{16\pi G}\int{\rm d}^4x\sqrt{-\det g} \left[R-2\Lambda\right]
    \nonumber\\
    \!\!&=&\!\! \int {\rm d}t {\rm d}^3x\left(p^{ab}\dot{q}_{ab}- H^{\rm EH}N - H^{\rm EH}_a N^a\right)\,,
\end{eqnarray}
which implies that $H^{\rm EH}$ and $H_a^{\rm EH}$ are constraints that must vanish on physical solutions and the Hamiltonian is given by ${\cal H}[N,N^a]=H^{\rm EH}[N]+H^{\rm EH}_a[N^a]$ (where $H^{\rm EH}[N]=\int {\rm d}^3 x H^{\rm EH}(x) N(x)$).
In this decomposition, the spatial metric $q_{ab}$ appears as the configuration variable with conjugate momentum $p^{ab}=\frac{\sqrt{\det q}}{16\pi G} \left(K^{ab}-K^c_c q^{ab}\right)$, where $K_{ab}$ is the extrinsic curvature of $\Sigma_t$.

At this stage, covariance is not manifest; however, because the Hamiltonian system has been derived from the Einstein--Hilbert action, it is indeed covariant, but a detailed analysis of the constraints is necessary to understand how covariance is realized canonically.
When smeared by the lapse and shift, the constraints act as the generator of time evolution via Poisson brackets, but when smeared by arbitrary functions $\epsilon^{\bar{0}}$ and $\epsilon^a$, they act as generators of gauge transformations, ${\cal H} [\epsilon^{\bar{0}},\epsilon^a]$:
Given a phase-space function $\mathcal{O}$, the constraints generate its gauge transformation via $\delta_\epsilon \mathcal{O} = \{ \mathcal{O} , H^{\rm EH}[\epsilon^{\bar{0}} , \epsilon^a]
\}$ and its time evolution via $\dot{\mathcal{O}} = \{\mathcal{O} , H^{\rm EH}[N , N^a]\}$.
(The index $\bar{0}$ is used as the normal component of spacetime indices and $t$ will be used later as the time component; $0$ will be reserved for the normal component of internal Lorentz indices.)
Furthermore, the Poisson brackets of the constraints with themselves form a first-class algebra,
\begin{eqnarray}
    \!\!\!\!\!\!\{H^{\rm EH}_a[N^a],H^{\rm EH}_b[\epsilon^b]\}
    \!\!&=&\!\! - H^{\rm EH}_a\left[\mathcal{L}_{\vec{\epsilon}}N^a\right]\,,
    \label{eq:Hypersurface deformation algebra - HaHa}\\
    \!\!\!\!\!\!\{H^{\rm EH}[N],H^{\rm EH}_a[\epsilon^a]\} \!\!&=&\!\! - H^{\rm EH}[\epsilon^a\partial_a N]\,,
    \\
    \!\!\!\!\!\!\{H^{\rm EH}[N],H^{\rm EH}[\epsilon^{\bar{0}}]\} \!\!&=&\!\! - H^{\rm EH}_a\!\left[q^{ab} \left(\epsilon^{\bar{0}}\partial_bN-N\partial_b\epsilon^{\bar{0}}\right)\right]
    \label{eq:Hypersurface deformation algebra - HH}
\end{eqnarray}
which implies that the constraints vanish consistently in all gauges and under time evolution.
Off shell, this algebra describes the gauge content, closely related to the underlying diffeomorphism covariance, of general relativity in terms of hypersurface deformations \cite{hojman1976geometrodynamics,kuchar1974geometrodynamics,EMGCov,HypDef}.
Notice that, unlike typical Lie algebras, (\ref{eq:Hypersurface deformation algebra - HaHa})-(\ref{eq:Hypersurface deformation algebra - HH}) possesses structure \emph{functions} given by the components of the inverse of the spatial metric $q^{ab}$ appearing in (\ref{eq:Hypersurface deformation algebra - HH}).

Earlier work in Ashtekar--Barbero variables \cite{Ashtekar1,Ashtekar2,JacobsonSmolin1,JacobsonSmolin2,Samuel,Barbero,Holst} has shown that the tetrad-connection system based on the Holst action---a generalization of the Hilbert--Palatini action by the addition of a topological term involving the Barbero--Immirzi parameter---has several similarities to an SO(3) Einstein--Yang--Mills system in canonical form.
It is therefore convenient to review the latter:
The canonical decomposition of the Yang--Mills action implies contributions $H^{\rm YM}$ and $H_a^{\rm YM}$ to the constraints $H=H^{\rm EH}+H^{\rm YM}$ and $H_a=H_a^{\rm EH}+H_a^{\rm YM}$ with the spatial components of the SU($n$)---or SO($n$)---connection $A^X_a$ as configuration variables (where $X,Y,Z\in\{1,2,\dots,n^2-1\}$) and conjugate momenta $P^a_X=\delta_{XY}\sqrt{\det q} \left( q^{a b} F_{\bar{0} b}^Y - \frac{\theta}{2} \epsilon^{a b c} F_{b c}^Y \right)=:-E^a_X-\theta B^a_X$ related to the components of the strength tensor field $F_{\mu\nu}^X=2\partial_{[\mu}A^X_{\nu]}+\tensor{f}{^X_Y_Z}A_\mu^YA^Z_\nu$, where $F^X_{\bar{0}a}=n^\mu F_{\mu a}^X$, and $\theta$ is the constant of the topological term.
The time components of the Yang--Mills connection $A^X_t$ appear as additional Lagrange multipliers that imply the Gauss constraint $G_X$,
\begin{eqnarray}
    \!\!\!S_{\rm YM} \!\!\!&=&\!\!\! - \frac{1}{4} \int{\rm} \!{\rm d}^4x \sqrt{- \det g}\left[F_X^{\mu\nu} F_{\mu\nu}^X - \frac{\theta}{2} \tensor{\epsilon}{^\mu^\nu_\alpha_\beta} F^X_{\mu \nu}F_X^{\alpha \beta}\right]
    \nonumber\\
    \!\!\!&=&\!\!\! \int{\rm} \!{\rm d}^4x \!\left[P^a_X\dot{A}^X_a - H^{\rm YM}N -H^{\rm YM}_aN^a-G_XA^X_t\right]
    \!.
\end{eqnarray}
The full Einstein--Yang--Mills system implies the constraint algebra
\begin{eqnarray}\label{eq:Constraint algebra - EYM - HaHa}
    \!\!\!\!\{H_a[N^a],H_b[\epsilon^b]\}
    \!\!&=&\!\! - H_a\left[\mathcal{L}_{\vec{\epsilon}}N^a\right]
    - G_X\left[\epsilon^aN^b F^X_{ab}\right],\,\,\,
    \\
    \!\!\!\!\{H[N],H_a[\epsilon^a]\} \!\!&=&\!\! - H[\epsilon^a\partial_a N]
    + G_X\left[N\epsilon^aF^X_{\bar{0}a}\right]\,,
    \\
    \!\!\!\!\{H[N],H[\epsilon^{\bar{0}}]\} \!\!&=&\!\! - H_a\left[q^{ab} \left(\epsilon^{\bar{0}}\partial_bN-N\partial_b\epsilon^{\bar{0}}\right)\right]\,,\label{eq:Constraint algebra - EYM - HH}
    \\
    \!\!\!\!\{G_X[A_t^X],G_Y[{\cal A}^Y]\}
    \!\!&=&\!\! - G_X \left[g \tensor{f}{^X_Y_Z} A_t^Y {\cal A}^Z\right]\,,\\
    \!\!\!\!\{H_a[N^a],G_X[{\cal A}^X]\} \!\!&=&\!\! 0
    \!\quad,\quad\!
    \{H[N],G_X[{\cal A}^X]\} = 0\,,\label{eq:Constraint algebra - EYM - HG}
\end{eqnarray}
which is still first class and contains additional structure functions given by the components of the strength tensor field.
While the Hamiltonian and vector constraints, $H$ and $H_a$, generate hypersurface deformations that are on-shell equivalent to spacetime diffeomorphisms in vacuum, the Gauss constraint $G_X$ generates SU($n$) or SO($n$) transformations in the presence of Yang--Mills fields---in the algebra, ${\cal A}^X$ is therefore an arbitrary smearing function acting as a corresponding gauge generator.

Indeed, an SO(3) Gauss constraint is obtained in Ashtekar--Barbero variables and the resulting first-class constraint algebra resembles (\ref{eq:Constraint algebra - EYM - HaHa})-(\ref{eq:Constraint algebra - EYM - HG}).
However, the Holst action is Lorentz invariant and the resulting constraint algebra should contain an $\mathfrak{so}(1,3)$ sub-algebra; that the Ashtekar--Barbero procedure results instead in an $\mathfrak{so}(3)$ sub-algebra, which can only generate rotations but not boosts, indicates that the system has been partially gauge-fixed.
Indeed, an important and ubiquitous ingredient in treatments of Ashtekar--Barbero variables is the imposition of the time gauge, which restricts certain components of the tetrad to match the Eulerian frame introduced by the ADM decomposition and results in a set of second-class constraints that are not all fundamental because the time-gauge restriction is itself not fundamental but a gauge-fixing choice.
Alternatives to the time gauge are sometimes used, where a deviation of the tetrad components with the Eulerian frame is chosen, but these constitute gauge-fixing procedures nonetheless.

The procedure of gauge fixing is sometimes necessary to obtain solutions to the equations of motion of the underlying theory.
However, fixing (or even partially fixing) a gauge before modifying the action or Hamiltonian that the theory is based on becomes problematic because it is not possible to show that the modified theory preserves the gauge symmetries of the original one---the reason being that some first-class constraints, which generate the gauge flow, trivialize after a gauge fixing.
Loop quantum gravity \cite{rovelli2004quantum,thiemann2008modern} falls into this category: The time gauge (or one of its variants) is fixed prior to the application of its quantization procedure and the modification of the Hamiltonian to write it in terms of holonomies.
A quantum or modified gravity theory is demonstrably covariant only if it preserves the full gauge content of the underlying classical theory, not just a reduced version of it.

One purpose of this study is to extend previous treatments to the full gauge content of the theory by dispensing with the time gauge and obtain an additional first-class constraint such that, together with the Gauss constraint, it forms an $\mathfrak{so}(1,3)$ sub-algebra.
This task has been considered in the literature, but the available discussions appear incomplete.
For instance, \cite{Alexandrov1} attempted to address this problem by starting with the gauge-fixed system of \cite{Alexandrov3} and only after the fixing does it extend the SO(3) subgroup to SO(4,$\mathbb{C}$); such a procedure is conceptually inequivalent to the actual decomposition of the Einstein--Cartan system without gauge fixing (even after the imposition of reality conditions) because the latter is based on a connection of the SO(1,3) group without being the subgroup of a larger one---the technical inequivalence between the systems is confirmed a posteriori by comparing the resulting Hamiltonians and the relation between the phase-space variables and the tetrad.
A more complete analysis was carried in the older study \cite{Ashtekar3} without the Barbero--Immirzi parameter: A canonical decomposition of the Hilbert--Palatini action was performed in the appendix of this reference, but then a partial gauge fixing was performed right after their Eq.~(A.13); the main text then works with the partially gauge fixed system.
Moreover, as many other references, the procedures of \cite{Alexandrov3,Ashtekar3} absorb some phase-space functions into the lapse and shift to turn the constraints polynomial; however, doing so changes the gauge content of the system---for instance, the original constraint algebra is modified by such rescalings---which generally has physical implications \cite{HypDef}. The results of these partially-gauged procedures explicitly lack the total number of (six) first-class constraints necessary to generate general SO(1,3) transformations.
On the other hand, \cite{BarberoEuclidean} explored this problem without fixing the time gauge but only in the Euclidean gravity system, obtaining the full SO(4) transformations rather than the reduced SO(3) of the traditional approach.
Meanwhile, the study in \cite{Lagraa} succeeded in obtaining a boost-generating constraint in the Lorentzian case and explicitly obtained an $\mathfrak{so}(1,3)$ sub-algebra.
However, such study does not use the ADM formalism and hence the variables are not easily related to the more familiar ADM or Ashtekar--Barbero variables.
Furthermore, the same study fixes the non-dynamical part of the connection to zero and hence the result can only be a special case of the full theory; it also uses Dirac brackets with the purpose to eliminate the structure function from the algebra, but doing so obscures its gauge content.
In contrast, \cite{Montesinos} solves the second-class constraints from the start; in doing this, the variables they obtain are not easily related to the more familiar variables either.
Furthermore, the brackets of the remaining constraints were not computed and hence their first-class status was not proved.
Finally, none of the previous studies showed that the first-class constraints generate spacetime diffeomorphisms and hence did not show whether their systems were diffeomorphism covariant in canonical form.

Here, we present a complete and consistent treatment by using a standard ADM decomposition, identifying the second-class constraints and their relation to the torsion tensor, and extending the phase space off the second-class constraint surface.
This results in variables that resemble two copies of the Ashtekar--Barbero connection with their respective conjugate momenta and we identify useful canonical transformations that relate them to boosted frames.
The introduction of the extended phase space facilitates several explicit computations of Poisson brackets and we show that ten constraints arise that are indeed first class and form an algebra that resembles the Einstein--Yang--Mills algebra (\ref{eq:Constraint algebra - EYM - HaHa})-(\ref{eq:Constraint algebra - EYM - HG}) with two Gauss-type constraints forming an $\mathfrak{so}(1,3)$ sub-algebra.
Furthermore, we study in detail the gauge transformations that the first-class constraints generate and identify them as linear combinations of spacetime diffeomorphisms and SO(1,3) transformations on shell.
The second-class constraints can be implemented by either solving them directly or by using Dirac brackets, and we show that they preserve the gauge content: Spacetime diffeomorphisms and SO(1,3) transformations are still generated when the second-class constraints are imposed.
Moreover, the theory remains well-defined off the second-class constraint surface---the spacetime metric, the tetrad, and the connection remain covariant---in which case it has six additional degrees of freedom compared to what the Hilbert--Palatini action implies; hence the canonical formulation in the extended phase space must be understood as a generalized geometric theory with a richer dynamical content for the torsion.
This analysis holds for arbitrary values of the Barbero--Immirzi parameter and the cosmological constant.
We expect that this work will serve as a firm basis for future canonical approaches to modified and quantum gravity in tetrad-connection variables preserving full diffeomorphism and Lorentz covariance.
For instance, here we discuss some implications for canonical approaches to quantum gravity, specifically for loop quantization.

This work is organized as follows.
We start with a brief review of the Lagrangian formulation of the Hilbert--Palatini action with the Barbero--Immirzi parameter in Sec.~\ref{sec:Lagrangian formulation}.
In Sec.~\ref{sec:Foliation}, we foliate the spacetime to parametrize the components of the tetrad into expressions suitable for the ADM decomposition.
The canonical analysis is performed in Sec.~\ref{sec:Canonical formulation}, where we identify the symplectic structure, the non-dynamical variables, and introduce a useful extension of the phase space.
In Sec.~\ref{sec:First-class constraints}, we compute the first-class constraints and their algebra under Poisson brackets of the extended phase space; we study in detail the gauge transformations they generate and explicitly relate them to spacetime diffeomorphisms and SO(1,3) transformations; and we identify a set of (nonlocal) Dirac observables that imply conserved currents.
In Sec.~\ref{sec:Second-class constraint surface}, we obtain the second-class constraints that reduce the extended phase space to that implied by the action; the second-class constraints are explicitly solved, but we find that it is computationally more convenient to use Dirac brackets.
In Sec.~\ref{sec:Cosmological}, we incorporate the cosmological constant contributions.
In Sec.~\ref{sec:Geometry}, we derive geometric objects related to the area and volume and discuss implications for loop quantum gravity.
We end with concluding remarks in Sec.~\ref{sec:Conclusions}.
Technical details necessary to follow the procedure are included in the main text and additional explicit computations can be found in the appendices for intermediate steps.

In the entirety of this work, we  assume four dimensions with Lorentzian signature and set $c=1$.

\section{Lagrangian formulation}
\label{sec:Lagrangian formulation}

The Einstein--Cartan theory postulates two kinematically independent fields to describe geometry: The tetrad $e^\mu_I$ and the connection 1-form $\tensor{\omega}{_\mu^I^J}=-\tensor{\omega}{_\mu^J^I}$.
The tetrads define the geometry through the postulate that the spacetime metric is given by the inverse of
\begin{equation}\label{eq:Inverse metric in tetrads}
    g^{\mu\nu}=\eta^{IJ}e_I^\mu e_J^\nu\,,
\end{equation}
where $\eta_{IJ}$ is the Minkowski metric; this, in turn, implies
\begin{equation}
    g_{\mu\nu} e^\mu_I e^\nu_J = \eta_{IJ}\,.
\end{equation}
Spacetime indices are raised and lowered with $g_{\mu\nu}$ and internal Lorentz indices with $\eta_{IJ}$.

The spacetime metric defines the derivative operator $\nabla_\mu$ by the compatibility condition
\begin{eqnarray}
    \nabla_\alpha g_{\mu\nu}=0\,,
\end{eqnarray}
which is used for the parallel transport of spacetime tensors.
On the other hand, the connection 1-form is used for the parallel transport in the internal Minkowski space:
The covariant derivative of SO(1,3)-valued tensor fields $f^{I_1,I_2,\dots,I_n}$ is given by
\begin{equation}
    D_\mu f^{I_1,\dots,I_n} = \nabla_\mu f^{I_1,\dots,I_n} + \sum_{k=1}^n \tensor{\omega}{_\mu^{I_k}_J} f^{I_1,\dots,J,\dots,I_n}\,.
\end{equation}
The antisymmetry of the connection implies that the covariant derivative is compatible with the internal Minkowski metric,
\begin{equation}
    D_\mu \eta_{IJ}=0\,.
\end{equation}
The torsion of the spacetime is defined by
\begin{eqnarray}\label{eq:Torsion}
    \tensor{T}{^I_\mu_\nu} \!\!&\equiv&\!\! D_{[\mu} e_{\nu]}^I
    \nonumber\\
    \!\!&=&\!\! \partial_{[\mu} e_{\nu]}^I
    + \tensor{\omega}{_{[\mu}^I_{|K|}} e^K_{\nu]}\,.
\end{eqnarray}

The dynamics is generated by the action \cite{Holst}
\begin{eqnarray}\label{eq:Holst action}
    S[e,\omega] \!\!&=&\!\! \int {\rm d}^4x\; \frac{|\det e|}{16\pi G} e^\mu_I e^\nu_J \tensor{P}{^I^J_K_L} F^{KL}_{\mu\nu}
    \\
    \!\!&=&\!\! \int \frac{{\rm d}^4x}{32\pi G} \epsilon^{\mu\nu\alpha\beta}\epsilon_{IJMN}e^M_\alpha e^N_\beta \tensor{P}{^I^J_K_L} F^{KL}_{\mu\nu}
    \nonumber
\end{eqnarray}
where $\det e$ is the determinant of the co-tetrad $e_\mu^I$---such that $\epsilon_{\mu\nu\alpha\beta}=|\det e|^{-1}\epsilon_{IJKL}e^I_\mu e^J_\nu e^K_\alpha e^L_\beta$ is the spacetime-volume form and $\epsilon_{IJKL}$ is totally antisymmetric with $\epsilon_{0123}=-1$, and we used the identity $2ee^{[\mu}_Ie^{\nu]}_J=\epsilon^{\mu\nu\alpha\beta}\epsilon_{IJMN}e^M_\alpha e^N_\beta$ to relate the two lines---
\begin{equation}
    F^{IJ}_{\mu\nu} = 2 \partial_{[\mu} \omega_{\nu]}^{IJ}
    + 2 \omega_{[\mu}^{IK} \omega_{\nu]}^{LJ} \eta_{KL}
\end{equation}
(or $F^{IJ}={\rm d} \omega^{IJ}+ \eta_{IJ}\omega^{IK}\wedge \omega^{LJ}$ in differential form notation) is the strength tensor field associated with the connection, and
\begin{equation}
    \tensor{P}{^I^J_K_L} = \delta^{[I}_K \delta^{J]}_L
    - \frac{\zeta}{2} \tensor{\epsilon}{^I^J_K_L}\,,
\end{equation}
where $\zeta\in\mathbb{R}$ is the inverse of the Barbero--Immirzi parameter---notice that choosing the notation $\epsilon_{0123} = 1$ instead results in the same expressions with $\zeta$ replaced by $-\zeta$.
Here, $\tensor{P}{^I^J_K_L}$ may be interpreted as a mapping from the tensor product of two Minkowski spaces into itself, with inverse
\begin{equation}
    \tensor{(P^{-1})}{_I_J^K^L} = \frac{1}{1+\zeta^2} \left(\delta^{[K}_I \delta^{L]}_J
    + \frac{\zeta}{2} \tensor{\epsilon}{_I_J^K^L}\right)
    \,,
\end{equation}
such that $\tensor{P}{^I^J_M_N}\tensor{(P^{-1})}{^M^N_K_L}=\delta^{[I}_K\delta^{J]}_L$.

Using $\delta F^{IJ}_{\mu\nu}=2 D_{[\mu} \delta \omega^{KL}_{\nu]}$, the variation of the action with respect to the connection 1-form, upon some simplification, is given by
\begin{equation}\label{eq:Torsion-constraints}
    \tensor{\epsilon}{^I^J^K^L} \tensor{(P^{-1})}{_K_L^M^N} \frac{\delta S[e,\omega]}{\delta \omega_\mu^{MN}}
    = - \frac{\epsilon^{\mu\nu\alpha\beta}}{2\pi G} e^{[I}_\alpha \tensor{T}{^{J]}_\nu_\beta}\,,
\end{equation}
where we neglected boundary terms.
The equation of motion $\delta S[e,\omega]/\delta \omega_\mu^{IJ}=0$ is, therefore, equivalent to zero torsion, which uniquely determines the connection as the one compatible with the tetrad:
\begin{equation}\label{eq:Tetrad compatible}
    \tensor{\omega}{_\mu^J^K}=e^{\nu J}\nabla_\mu e_\nu^K
    \,,
\end{equation}
such that $D_\mu e_\nu^I =0$---in the presence of fermionic matter, the torsion is not necessarily vanishing.
This implies 24 independent equations that uniquely determine the 24 components of the connection one-form in terms of the tetrad.

As the equivalent to the equations of motion, only the 12 independent timelike components of the torsion $\tensor{T}{^I_t_a}=0$ contain time derivatives, while the 12 purely spatial independent components $\tensor{T}{^I_a_b}=0$ do not.
Therefore, the latter are not evolution equations but rather constraints, 6 of which must be related to the variation with respect to the non-dynamical components $\omega_t^{IJ}$---their non-dynamical status can be readily deduced from the fact that the action does not contain time derivatives of them---while the remaining 6 constraints point to the existence of 6 additional non-dynamical components of the connection 1-form---this will be clarified by the subsequent canonical analysis---leaving a total of only 12 dynamical components of the connection.

On the other hand, the tetrad has 16 components, 12 of which must be identified as the conjugate momenta of the dynamical components of the connection 1-form, and the remaining 4 constitute the familiar, non-dynamical lapse and shift of the ADM variables, giving rise to constraints of their own.
With this counting, we conclude that the Legendre transformation of the action must imply a symplectic structure of 12 canonical pairs, and 16 constraints---as will be made clear after the canonical decomposition is performed, 10 of these constraints are first class and the remaining 6 are second class.

\section{Foliation}
\label{sec:Foliation}

The canonical formulation of the Einstein--Hilbert action theory is based on the ADM decomposition by foliating the spacetime manifold, a procedure that we adopt in the following and extend it in order to treat the internal Minkowski space.

Given a globally hyperbolic spacetime, $M = \Sigma \times \mathbb{R}$, the line element in ADM form is given by (\ref{eq:ADM line element}), where $N$ is the lapse, $N^a$ the shift, and $q_{ab}$ is the spatial metric induced on the three-dimensional hypersurface $\Sigma$.

The coordinate frame refers to an observer with time-evolution vector field 
\begin{equation}
    t^\mu = N n^\mu + N^a s_a^\mu
    \,,
    \label{eq:Time-evolution vector field}
\end{equation}
where $n^\mu$ is a unit vector normal to the hypersurface $\Sigma$, and $s^\mu_a$ are three basis vectors tangential to $\Sigma$.
Therefore, the lapse is understood as the normal component of the time-evolution vector field, and the shift as the tangential components.
The inverse spacetime metric can then be written as
\begin{equation}
    g^{\mu \nu} = - n^\mu n^\nu
    + q^{a b} s^\mu_a s^\nu_a
    \,.\label{Inverse ADM metric}
\end{equation}

In the same spirit of the ADM decomposition, we define an internal normal vector $\hat{n}^I$ and internal spatial basis vectors $\hat{s}^I_i$, such that $\eta_{IJ}\hat{n}^I\hat{n}^J=-1$, $\eta_{IJ}\hat{s}^I_i\hat{s}^J_j=\delta_{ij}$, and $\eta_{IJ}\hat{n}^I\hat{s}^J_j=0$.
Notice that, in general, $\hat{n}^I\neq n^I$, where $n^I\equiv e^I_\mu n^\mu$.
We proceed to decompose the internal tensors in this basis:
The Minkowski metric can be written as
\begin{equation}\label{eq:Internal foliation}
    \eta_{IJ} = - \hat{n}_I \hat{n}_J + \delta_{ij} \hat{s}^i_I \hat{s}^j_J
\end{equation}
and the tetrad as
\begin{eqnarray}\label{eq:Tetrad decomposition 1}
    e_I^\mu = - \phi \hat{n}_I n^\mu + \varepsilon^a_i \hat{s}^i_I s^\mu_a
    - \Theta^a \hat{n}_I s^\mu_a
    + \Phi_i \hat{s}^i_I n^\mu
    \,.
\end{eqnarray}
In what follows, the internal index $0$ denotes the internal normal component---notice the sign that has to be taken into account in the contractions $T_0^I=\hat{n}^JT_J^I$ or $T^0_J = - \eta_{KI}\hat{n}^K T^I_J$---similarly, use of lowercase Latin indices $i,j,k,\dots$ in Lorentzian tensors denote contraction with an internal spatial basis vector---such that $T_j^I=\hat{s}^J_j T_J^I$ or $T_J^i=\hat{s}^i_I T_J^I$.
Internal Euclidean indices are raised and lowered with $\delta_{ij}$.

Substituting (\ref{eq:Tetrad decomposition 1}) into (\ref{eq:Inverse metric in tetrads}) we obtain
\begin{eqnarray}
    g^{\mu\nu} \!\!&=&\!\! - \left(\phi^2 - \Phi_i\Phi_i\right) n^\mu n^\nu
    + \left(\varepsilon^a_i \varepsilon^b_i-\Theta^a\Theta^b\right) s^\mu_a s^\nu_b
    \nonumber\\
    \!\!&&\!\! + \left(\varepsilon^a_i \Phi^i-\phi \Theta^a\right) \left(n^\mu s^\nu_a + n^\nu s^\mu_a\right)
    \,.
\end{eqnarray}
A comparison with (\ref{Inverse ADM metric}) implies the relations
\begin{eqnarray}
    \phi^2 - \Phi^i\Phi_i \!\!&=&\!\! 1
    \,,\\
    \varepsilon^a_i \Phi^i-\phi \Theta^a \!\!&=&\!\! 0
    \,,\\
    \varepsilon^a_i \varepsilon^b_i-\Theta^a\Theta^b \!\!&=&\!\! q^{ab}
    \,.
\end{eqnarray}
The solution to the first two equations can be parametrized by an internal velocity $v^i$ such that
\begin{eqnarray}
    \phi \!\!&=&\!\! \gamma\,,\\
    \Phi^i \!\!&=&\!\! - \gamma v^i\,,\\
    \Theta^a \!\!&=&\!\! - \varepsilon^a_i v^i
    \,,
\end{eqnarray}
where $\gamma = 1/\sqrt{1-v^2}$ and $v^2= v^iv_i$.

We can therefore write the tetrad and the spatial metric in terms of the triad $\varepsilon^a_i$ and the internal velocity $v^i$ as follows,
\begin{eqnarray}\label{eq:Tetrad expression}
    e_I^\mu \!\!&=&\!\! - \gamma \left(\hat{n}_I + v_i \hat{s}^i_I\right) n^\mu
    + \varepsilon^a_i \left(v^i \hat{n}_I+\hat{s}^i_I\right) s^\mu_a
    \nonumber\\
    \!\!&=:&\!\! - n_I n^\mu + \varepsilon^a_i s^i_I s^\mu_a
    \,,\\
    q^{ab} \!\!&=&\!\! \varepsilon^a_i \varepsilon^b_j \left( \delta^{ij} - v^i v^j\right)
    \,,\label{eq:Inverse spatial metric in triad/velocity}
\end{eqnarray}
where we defined $s^i_I = v^i \hat{n}_I + \hat{s}^i_I$.
The tetrad is therefore described by the 16 quantities $N$, $N^a$, $\varepsilon^a_i$, and $v^i$.
Standard treatments of Ashtekar--Barbero variables implicitly set the velocity parameters to zero from the start---implied by the common choice of $\hat{n}^I=n^I$, which is usually denominated as the \textit{time gauge}---to facilitate the canonical analysis.
In doing so, however, the canonical pairs related to $v^i$ are eliminated; such a procedure would imply a reduced phase space that cannot have access to the full gauge content of the theory: Fixing the internal Lorentz frame breaks Lorentz covariance, which explains the lack of first-class constraints that generate internal boosts in the standard treatments.
Alternatively, one can fix $v^i=\chi^i(t,x)$ to some non-vanishing, non-dynamical function $\chi^i(t,x)$, but this still implies a phase-space reduction and the trivialization of some first-class constraints as in \cite{Ashtekar3,Alexandrov3}.
As we show in the following sections, such Lorentz constraints indeed arise in the canonical formulation if $v^i$ is not fixed a priori, resulting in a larger phase space compared to the Ashtekar--Barbero variables.

Inspection of the frame basis, defined by
\begin{equation}\label{eq:Tetrad frame}
    n_I = \gamma \hat{n}_I + \gamma v_i \hat{s}^i_I\quad,\quad
    s^i_I = v^i \hat{n}_I + \hat{s}^i_I\,,
\end{equation}
reveals that it is boosted with respect to the internal one, defined by $(\hat{n}_I,\hat{s}^i_I)$, by a relative velocity $v^i$.

It is useful for the following to have the relations
\begin{eqnarray}\label{eq:Lorentz contracion spatial basis}
    \eta^{IJ} s^i_I s^j_J \!\!&=&\!\! \delta^{ij} - v^i v^j
    \,,\\
    \varepsilon_a^i\varepsilon^a_j
    \!\!&=&\!\! \delta^i_j+\gamma^2v^iv_j
    \,,\\
    \delta^i_j \!\!&=&\!\! \left(\delta^i_k+\gamma^2v^iv_k\right) \left(\delta^k_j-v^kv_j\right)
    \,,\label{eq:Inverse Lorentz contractor}
\end{eqnarray}
where the spatial indices are raised and lowered with $q_{ab}$.
(Notice that while the internal basis $(\hat{n}_I,\hat{s}_I^i)$ is orthonormal, this is not the case for the frame basis $(n_I,s_I^i)$ because the latter's spatial basis vectors suffer from a Lorentz contraction according to (\ref{eq:Lorentz contracion spatial basis}).)
Also, it is useful to write the tetrad in the coordinate and internal bases,
\begin{eqnarray}\label{eq:Tetrad coordinate basis}
    e^\mu_I
    \!\!&=&\!\! - \gamma N^{-1} \left(\hat{n}_I + v_i \hat{s}^i_I \right) t^\mu
    + \left(\gamma N^{-1} N^a + \varepsilon^a_k v^k\right) \hat{n}_I s^\mu_a
    \nonumber\\
    \!\!&&\!\!
    + \left(\gamma N^{-1} N^a v_i + \varepsilon^a_i \right) \hat{s}^i_I s^\mu_a
    \,.
\end{eqnarray}
Notice that $e^t_0 = \hat{n}^I e^t_I = \gamma N^{-1}$ and $e^a_0 = \hat{n}^I e^a_I = - \left(\gamma N^{-1} N^a + \varepsilon^a_k v^k\right)$.
From this we obtain
\begin{eqnarray}\label{eq:Determinant e}
    (\det e)^{-1}
    = - \frac{\det \varepsilon}{\gamma N}
    \quad,\quad
    \sqrt{\det q} = \gamma/\det\varepsilon\,,
\end{eqnarray}
where $\det \varepsilon$ denotes the determinant of the triad $\varepsilon^a_i$.

Using $e_\mu^Ie_I^\nu=\delta^\nu_\mu$, we obtain the co-tetrad components,
\begin{eqnarray}\label{eq:co-tetrad components 1}
    e_t^0 \!\!&=&\!\! \gamma N
    + v_j N^b \varepsilon_b^j
    \,,\\
    e^i_t
    \!\!&=&\!\! \gamma N v^i
    + N^b \varepsilon_b^i
    \,,\\
    e_a^i
    \!\!&=&\!\! \varepsilon_a^i
    \,,\\
    e_a^0
    \!\!&=&\!\! v_i \varepsilon_a^i\,.\label{eq:co-tetrad components 4}
\end{eqnarray}

\section{Canonical formulation}
\label{sec:Canonical formulation}

We are now ready to perform the canonical decomposition of Einstein--Cartan gravity.
In doing this, we will use the tetrad in the coordinate and internal bases (\ref{eq:Tetrad coordinate basis}).

\subsection{Symplectic structure}

Using $F^{KL}_{ta}\supset\dot{\omega}_a^{KL}$, the symplectic contribution to the action is given by
\begin{equation}
    \int {\rm d}^4x \left[\tilde{\cal P}^a_i \dot{K}_a^i + \tilde{\cal K}^a_i \dot{\Gamma}_a^i \right]
    \,,
\end{equation}
where the configuration variables are the connection components
\begin{equation}\label{eq:Canonical connections}
    K^i_a =\omega^{0i}_a
    \quad,\quad
    \Gamma^i_a=\frac{1}{2} \tensor{\epsilon}{^i_k_l} \omega_a^{kl}\,,
\end{equation}
and their respective conjugate momenta are given by
\begin{eqnarray}\label{eq:Momenta Gauss}
    \tilde{\cal P}^a_i \!\!&=&\!\! {\cal P}^a_i+ \zeta {\cal K}^a_i
    \,,\\
    \tilde{\cal K}^a_i \!\!&=&\!\! {\cal K}^a_i - \zeta {\cal P}^a_i
    \,,\label{eq:Momenta Lorentz}
\end{eqnarray}
where
\begin{eqnarray}\label{eq:Momenta Ashtekar-Barbero 1}
    {\cal P}^a_i \!\!&=&\!\! \frac{|\det e|}{8\pi G} \left(e^t_0 e^a_i - e^t_i e^a_0\right)
    = \frac{|\det e|}{8\pi G} 2e^{[t}_0 e^{a]}_i
    \nonumber\\
    \!\!&=&\!\! \frac{\gamma^2/(8\pi G)}{\det \varepsilon} \varepsilon^a_j \left(\delta^j_i - v^j v_i \right)
    \,,\\
    {\cal K}^a_i \!\!&=&\!\! \frac{|\det e|}{8\pi G} e^t_k e^a_l \tensor{\epsilon}{^k^l_i}
    \nonumber\\
    \!\!&=&\!\! \frac{\gamma^2/(8\pi G)}{\det \varepsilon}  \varepsilon^a_j \tensor{\epsilon}{_i^j^k} v_k
    \,.\label{eq:Momenta Ashtekar-Barbero 2}
\end{eqnarray}

It is useful to invert the triad relation
\begin{equation}
    \varepsilon^a_i = \frac{\sqrt{(\det {\cal P})^{-1}}}{\sqrt{8\pi G}} \left(\delta_i^j+\gamma^2v_iv^j\right) {\cal P}^a_j
\end{equation}
with which we can obtain the spatial density relation
\begin{equation}
    \det {\cal P} = \frac{\gamma^4 (\det \varepsilon)^{-2}}{(8\pi G)^3}
    = \frac{\gamma^2 \det q}{(8\pi G)^3}\,,
\end{equation}
as well the inverse relations
\begin{eqnarray}
    \left({\cal P}^{-1}\right)_a^j \!\!&=&\!\! \frac{\det \varepsilon}{\gamma^2 /(8\pi G)} \varepsilon_a^j\,,\\
    \varepsilon_a^j \!\!&=&\!\! \sqrt{8\pi G \det {\cal P}} \left({\cal P}^{-1}\right)_a^j\,,
\end{eqnarray}
satisfying $\left({\cal P}^{-1}\right)_a^i {\cal P}^a_j=\delta^i_j$ and $\left({\cal P}^{-1}\right)_b^i {\cal P}^a_i=\delta^a_b$.

There exist two important canonical transformations.
The first one leads to the new configuration variables
\begin{eqnarray}\label{eq:Connection Ashtekar-Barbero 1}
    A_a^i \!\!&=&\!\! K^i_a
    - \zeta \Gamma^i_a\,,\\
    B_a^i \!\!&=&\!\! \Gamma^i_a+\zeta K^i_a
    \,,\label{eq:Connection Ashtekar-Barbero 2}
\end{eqnarray}
whose conjugate momenta are given by (\ref{eq:Momenta Ashtekar-Barbero 1}) and (\ref{eq:Momenta Ashtekar-Barbero 2}), respectively:
\begin{equation}
    \int {\rm d}^4x \left[\tilde{\cal P}^a_i \dot{K}_a^i + \tilde{\cal K}^a_i \dot{\Gamma}_a^i \right]
    = \int {\rm d}^4x \left[{\cal P}^a_i \dot{A}_a^i + {\cal K}^a_i \dot{B}_a^i \right]\,.
\end{equation}
The variable (\ref{eq:Connection Ashtekar-Barbero 1}) resembles the Ashtekar--Barbero connection and hence these configuration variables are useful to connect the results to traditional approaches; however, the new symplectic structure implies a larger phase space that includes the new variable (\ref{eq:Connection Ashtekar-Barbero 2}).

Moreover, the connection components (\ref{eq:Canonical connections}) sum up to a total of 18 independent components---as well as the new variables (\ref{eq:Connection Ashtekar-Barbero 1}) and (\ref{eq:Connection Ashtekar-Barbero 2})---while the momenta (\ref{eq:Momenta Gauss}) and (\ref{eq:Momenta Lorentz})---or (\ref{eq:Momenta Ashtekar-Barbero 1}) and (\ref{eq:Momenta Ashtekar-Barbero 2})---sum up to a total of 12 independent components (9 from the triad $\varepsilon^a_i$ and 3 from the internal velocity $v^i$).
Therefore, we expect to get 6 second-class constraints to eliminate the extra 6 kinematical components of the connections.
In particular, the relation
\begin{equation}\label{eq:KP relation}
    {\cal K}^a_i = \tensor{\epsilon}{_i^j^k} {\cal P}^a_j v_k\,,
\end{equation}
derived from comparing (\ref{eq:Momenta Ashtekar-Barbero 1}) and (\ref{eq:Momenta Ashtekar-Barbero 2}), implies that the momentum ${\cal K}^a_i$ has only 3 independent components.

To resolve the excess connection components, with respect to those of the tetrad, as well as to restrict the number of canonical variables to the actual kinematical degrees of freedom, it is useful to perform a second canonical transformation with the new configuration variables
\begin{eqnarray}\label{eq:New connection D}
    {\cal D}_a^i \!\!&=&\!\! A_a^i - \tensor{\epsilon}{_j^k^i} B_a^j v_k
    \,,\\
    {\cal E}^i \!\!&=&\!\! \tensor{\epsilon}{_j^k^i} B_a^j {\cal P}^a_k\,,
    \label{eq:New connection E}
\end{eqnarray}
and their respective conjugate momenta, given by ${\cal P}^a_i$ and the co-velocity $v_i$:
\begin{equation}\label{eq:Symplectic structure D,E}
    \int {\rm d}^4x \left[\tilde{\cal P}^a_i \dot{K}_a^i + \tilde{\cal K}^a_i \dot{\Gamma}_a^i \right]
    = \int {\rm d}^4x \left[{\cal P}^a_i \dot{\cal D}_a^i + v_i \dot{\cal E}^i \right]\,.
\end{equation}

It is useful to decompose the $B$ connection as
\begin{eqnarray}\label{eq:B dynamical decomposition}
    B_a^i
    &=& \left({\cal P}^{-1}\right)^j_a \left({\cal B}^i_j + \frac{1}{2} \tensor{\epsilon}{^i_j_k}{\cal E}^k\right)\,,
\end{eqnarray}
where the symmetric internal tensor ${\cal B}_{ij}={\cal B}_{ji}$, defined by
\begin{equation}\label{eq:B non-dynamical}
    {\cal B}_{ik} = B_b^j \delta_{j(i} {\cal P}^b_{k)} \,,
\end{equation}
implies 6 non-dynamical components of the connection 1-form because they do not appear in the symplectic structure (\ref{eq:Symplectic structure D,E}).
Similarly, we can decompose
\begin{eqnarray}\label{eq:K dec in v,K}
    {\cal K}^a_i &=& {\cal P}^a_j \left(\tensor{\epsilon}{_i^j^k} v_k
    + \mathfrak{K}^j_i\right)\,,
\end{eqnarray}
where the symmetric tensor $\mathfrak{K}_{ij}=\mathfrak{K}_{ji}$ is vanishing according to the relation (\ref{eq:KP relation}), hence its absence from the symplectic structure (\ref{eq:Symplectic structure D,E}).

For completeness and future use, we decompose the spatial metric and the torsion in terms of the phase-space variables.
First, the inverse spatial metric (\ref{eq:Inverse spatial metric in triad/velocity}) can be written as
\begin{equation}\label{eq:Inverse spatial metric - canonical}
    q^{ab}
    = \frac{(\det {\cal P})^{-1}}{8\pi G} \left(\delta^{ij}+\gamma^2 v^iv^j \right) {\cal P}^a_i {\cal P}^b_j
    \,,
\end{equation}
which is always positive, or in the more symmetric form
\begin{equation}\label{eq:Inverse spatial metric - canonical extended}
    q^{ab}=\frac{\gamma^2 (\det {\cal P})^{-1}}{8\pi G} \delta^{ij} \left({\cal P}^a_i {\cal P}^b_j-{\cal K}^a_i {\cal K}^b_j\right) \bigg|_{\mathfrak{K}=0}\,,
\end{equation}
where we used
\begin{equation}
    \delta^{ij} {\cal K}^a_i {\cal K}^b_j\big|_{\mathfrak{K}=0} = \left(\delta^{ij} v^2-v^iv^j\right) {\cal P}^a_i {\cal P}^b_j\,.
\end{equation}
The latter expression holds only if $\mathfrak{K}_{ij}=0$ and hence it can be used only in the corresponding context.

Relevant torsion components for the forthcoming canonical analysis can be expressed as the symmetric internal tensor
\begin{eqnarray}\label{eq:Torsion-spatial-sym}
    {\cal T}^{ij}
    \!\!&=&\!\! - \frac{\delta_k^{(i} \tensor{\epsilon}{^{j)}^p^q}{\cal P}^a_p {\cal P}^b_q \tensor{T}{^k_{ab}}}{\sqrt{8\pi G \det {\cal P}}}
    \nonumber\\
    \!\!&=&\!\!
    (\delta^{ij}\delta^p_q-\delta^{p(i}\delta_q^{j)}) {\cal P}^d_p \Gamma_d^q - \delta_q^{(i}\delta^{j)p} {\cal K}^d_p K_d^q
    \nonumber\\
    \!\!&&\!\!
    + \delta_k^{(i} \tensor{\epsilon}{^{j)}^p^q} {\cal P}^c_p {\cal P}^d_q \partial_d \left({\cal P}^{-1}\right)_c^k
    \,.
\end{eqnarray}

\subsection{Non-dynamical variables}

In addition to (\ref{eq:B non-dynamical}), the 6 components $\omega_t^{IJ}$ are non-dynamical because their time derivatives do not appear in the action.
In the following, we will use
\begin{equation}\label{eq:Non-dynamical time components of connection}
    K_t^i = \omega_t^{0i} \quad,\quad \Gamma_t^i = \frac{1}{2} \tensor{\epsilon}{^i_k_l} \omega^{kl}_t\,.
\end{equation}
Together with ${\cal B}_{ij}$, this sums up to a total of 12 non-dynamical components of the connection, as expected from our counting in the Lagrangian formulation, which implies the existence of 12 constraints.

The non-dynamical components (\ref{eq:B non-dynamical}) imply 6 second-class constraints given by
\begin{equation}\label{eq:Second-class constraint S}
    {\cal C}_{ij} = \frac{\delta S[\omega,e]}{\delta {\cal B}^{ij}} = - \frac{\delta {\cal H}[\omega,e]}{\delta {\cal B}^{ij}} = 0\,,
\end{equation}
where ${\cal H}[\omega,e]$ denotes the Hamiltonian, while the components (\ref{eq:Non-dynamical time components of connection}) imply 6 first-class constraints that we expect to be related to Gauss-type constraints generating Lorentz transformations.
This leaves only 12 dynamical components of the connection given by the components (\ref{eq:New connection D}) and (\ref{eq:New connection E}).

On the other hand, from the 16 components of the tetrad, 4 are given by the non-dynamical lapse and shift, which give rise to the first-class Hamiltonian and vector constraints, leaving only 12 canonical variables of the tetrad, such that we have a total of 12 canonical pairs given by $({\cal D}_b^j,{\cal P}^a_i)$ and $({\cal E}^i,v^j)$.

Since we have a total of 10 first-class constraints, the 12 canonical pairs reduce to the expected 2 degrees of freedom of standard general relativity in dynamical solutions.
In summary, the expected conclusion of the full canonical analysis is to obtain 6 second-class constraints and 10 first-class constraints---including the Hamiltonian constraint $H$ and 3 vector constraints $H_a$ jointly generating spacetime diffeomorphisms, 3 Gauss constraints $G_i$ generating internal rotations, and 3 Lorentz constraints $L_i$ generating internal boosts.

\subsection{Extended phase space}

In terms of the canonical variables, the action takes the form
\begin{eqnarray}
    S \!\!&=&\!\! \int {\rm d}^4x \left({\cal P}^a_i \dot{\cal D}_a^i + v_i \dot{\cal E}^i \right)
    \\
    \!\!&&\!\!\qquad
    - {\cal H} [{\cal P},v,{\cal D},{\cal E};N,\vec{N},K_t,\Gamma_t,{\cal B}]\,,\nonumber
\end{eqnarray}
such that the Poisson bracket of any two phase-space functionals ${\cal O}$ and ${\cal U}$ is given by
\begin{eqnarray}
    \{{\cal O},{\cal U}\} \!\!&=&\!\! \int{\rm d}^3z \Bigg[\frac{\delta {\cal O}}{\delta {\cal D}^i_c(z)} \frac{\delta {\cal U}}{\delta {\cal P}_i^c(z)}
    - \frac{\delta {\cal O}}{\delta {\cal P}_i^c(z)} \frac{\delta {\cal U}}{\delta {\cal D}^i_c(z)}
    \nonumber\\
    \!\!&&\!\!\qquad\quad
    + \frac{\delta {\cal O}}{\delta {\cal E}^i(z)} \frac{\delta {\cal U}}{\delta v_i(z)}
    - \frac{\delta {\cal O}}{\delta v_i(z)} \frac{\delta {\cal U}}{\delta {\cal E}^i(z)}
    \Bigg]\,,\quad
\end{eqnarray}
and the variables $N,N^a,K_t^i,\Gamma_t^i$, and ${\cal B}_{ij}$ are all Lagrange multipliers giving rise to constraints.

However, as will be clear in the following sections, the resulting constraints have a very complicated dependence on these canonical variables.
They are much simpler in the $(\tilde{\cal P},\tilde{\cal K},K,\Gamma)$ or $({\cal P},{\cal K},A,B)$ variables.
The latter can be used in the Poisson brackets by extending the phase space to include the non-dynamical components $\mathfrak{K}_{ij}$ and ${\cal B}^{ij}$ in the symplectic structure and reduce the phase-space at the end with the additional second-class constraint $\mathfrak{K}_{ij}=0$.

Therefore, we start with an extended phase space with coordinates $({\cal P},{\cal K},A,B)$ and \emph{define}
\begin{eqnarray}\label{eq:K frak}
    v_i \!\!&\equiv&\!\! - \frac{1}{2} \tensor{\epsilon}{_i_m^n} \left({\cal P}^{-1}\right)^m_b {\cal K}^b_n
    \,,\\
    \mathfrak{K}_{ij} \!\!&\equiv&\!\! \left({\cal P}^{-1}\right)^m_b \delta_{m(i} {\cal K}^b_{j)}
    \,,
\end{eqnarray}
such that (\ref{eq:K dec in v,K}) holds for $\mathfrak{K}_{ij}\neq0$.
Using this, the extended symplectic term takes the form
\begin{eqnarray}\label{eq:Symplectic structure - extended}
    \!\!&&\!\!
    \int {\rm d}^4x \left[\tilde{\cal P}^a_i \dot{K}_a^i + \tilde{\cal K}^a_i \dot{\Gamma}_a^i \right]
    \\
    \!\!&&\!\!\qquad\qquad\qquad
    = \int {\rm d}^4x \left[{\cal P}^a_i \dot{A}_a^i + {\cal K}^a_i \dot{B}_a^i \right]
    \nonumber\\
    \!\!&&\!\!\qquad\qquad\qquad
    = \int {\rm d}^4x \bigg[
    {\cal P}^a_i \left(A_a^i+ \tensor{\epsilon}{_j^i^k} v_k B_a^j
    + \mathfrak{K}^i_j B_a^j\right)^\bullet
    \nonumber\\
    \!\!&&\!\!\qquad\qquad\qquad\qquad\qquad\quad
    + \left(\tensor{\epsilon}{_i^j^k} v_k
    + \mathfrak{K}^j_i\right) \left({\cal P}^a_j B_a^i\right)^\bullet
    \bigg]
    \nonumber\\
    \!\!&&\!\!\qquad\qquad\qquad
    = \int {\rm d}^4x \left[{\cal P}^a_i \dot{\cal D}_a^i + v_i \dot{\cal E}^i
    + \mathfrak{K}_{ij} \dot{\cal B}^{ij} \right]
    \,,\nonumber
\end{eqnarray}
where we used integrations of parts in the time coordinate and neglected boundary terms to obtain the second equality, and defined
\begin{eqnarray}\label{eq:New connection D - extended}
    {\cal D}_a^i \!\!&=&\!\! A_a^i+ \tensor{\epsilon}{_j^i^k} v_k B_a^j
    + \mathfrak{K}^i_j B_a^j
    \,,\\
    {\cal E}^i \!\!&=&\!\! \tensor{\epsilon}{_j^k^i} B_a^j {\cal P}^a_k
    \,,\\
    {\cal B}_{ik} \!\!&=&\!\! B_b^j \delta_{j(i} {\cal P}^b_{k)}
    \,.
\end{eqnarray}
While the expression of ${\cal D}_a^i$ receives a contribution of $\mathfrak{K}_{ij}$ compared to (\ref{eq:New connection D}), the expressions of ${\cal E}^i$ and ${\cal B}_{ij}$ remain unchanged, hence the decomposition (\ref{eq:B dynamical decomposition}) still holds.

The symplectic term (\ref{eq:Symplectic structure - extended}) implies that the sets of canonical pairs $\{(K,\tilde{\cal P}),(\Gamma,\tilde{\cal K})\}$, $\{(A,{\cal P}),(B,{\cal K})\}$, and $\{({\cal D},{\cal P}),({\cal E},v),({\cal B},\mathfrak{K})\}$ are related to one another by canonical transformations and hence we may use any of them in the canonical analysis with equivalent results.
To recover the original phase space, it suffices to implement the second-class constraint
\begin{equation}\label{eq:Second-class constraint - primary}
    \mathfrak{K}_{ij} \big|_{\rm OS} =0\,,
\end{equation}
where we use the subscript "OS", standing for "on shell", to denote that the expression is evaluated on physical solutions, where all the constraints are imposed and the equations of motion hold.
The consistency of this constraint under time evolution requires $\{\mathfrak{K}_{ij},{\cal H}\}|_{\rm OS}=0$, which implies the secondary second-class constraint
\begin{equation}
    {\cal C}_{ij} = - \frac{\delta {\cal H}}{\delta {\cal B}^{ij}}\,,
\end{equation}
which must vanish on shell and matches (\ref{eq:Second-class constraint S}).
As will be shown in Section~\ref{sec:Second-class constraint surface}, this secondary second-class constraint can be solved for the non-dynamical variable ${\cal B}_{ij}$ and, therefore, does not require a tertiary second-class constraint imposing $\{{\cal C}_{ij},{\cal H}\}|_{\rm OS}=0$.

In the extended phase space, we have 18 canonical pairs with 10 first-class and 12 second-class constraints, implying still the standard 2 degrees of freedom---each first-class constraint determines a canonical pair, while second-class constraints fix individual phase-space variables and hence determine a total of 6 canonical pairs corresponding to $(\mathfrak{K},{\cal B})$.

Unlike the first-class constraints, the second-class constraints may be solved before computing the dynamics and do not generate a gauge flow in the reduced phase space.
Therefore, the evaluation of the correct dynamics while working with the extended phase space requires the use of corresponding Dirac brackets---to eliminate the spurious gauge flow generated by the contributions of the second-class constraints to the first-class constraints---which are computed in Section~\ref{sec:Second-class constraint surface}.
Before doing so, we focus on the first-class constraints in the next section.

\section{First-class constraints}
\label{sec:First-class constraints}

\subsection{Lorentz--Gauss constraints}
\label{sec:Gauss and Lorentz constraints}

The action (\ref{eq:Holst action}) can be written as
\begin{widetext}
\begin{eqnarray}\label{eq:Holst action canonical 1}
    S[e,\omega]
    \!\!\!&=&\!\!\! \int {\rm d}^4x \left[\tilde{\cal P}^a_i F_{ta}^{0i}
    + \tilde{\cal K}^a_i {\cal F}_{ta}^i
    + \frac{|\det e|}{16\pi G} e^a_I e^b_J \tensor{P}{^I^J_K_L} F^{KL}_{ab}\right]
    \nonumber\\
    \!\!\!&=&\!\!\! \int {\rm d}^4x \left[
    \tilde{\cal P}^a_i \dot{\tilde{K}}_a^i + {\cal K}^a_i \dot{\Gamma}_a^i \right]
    - H[N]
    - H_a[N^a]
    - L_i[K_t^i]
    - G_i[\Gamma_t^i]
    \,,
\end{eqnarray}
\end{widetext}
which is linear in the Lagrange multipliers $N$, $N^a$, $K_t^i$, and $\Gamma_t^i$ and, therefore, all the local expressions of $H$, $H_a$, $L_i$, and $G_i$ are constraints.
Only the first two terms in the first line contribute to the symplectic terms and to $L_i$ and $G_i$, while only the last term in the first line contributes to $H$ and $H_a$.
Here, we defined
\begin{equation}
    {\cal F}^i_{ta}
    = \frac{1}{2} \tensor{\epsilon}{^i_k_l} F^{kl}_{ta}\,.
\end{equation}

We first focus on the $L_i$ and $G_i$ constraints, given by
\begin{eqnarray}\label{eq:Lorentz constraint}
    L_i \!\!&=&\!\! - \partial_a \tilde{\cal P}^a_i
    - \left( \tilde{\cal P}^a_k \Gamma^l_a
    - \tilde{\cal K}^a_k K^l_a \right)\tensor{\epsilon}{^k_l_i}
    \\
    \!\!&=&\!\! - \partial_a {\cal P}^a_i
    - \zeta \partial_a {\cal K}^a_i
    - \left({\cal P}^a_k B_a^l
    - {\cal K}^a_k A_a^l\right)\tensor{\epsilon}{^k_l_i}
    \,,
    \nonumber\\
    \!\!&=&\!\! - \left[\delta^j_i + \zeta \left(\tensor{\epsilon}{_i^j^k} v_k+\mathfrak{K}^j_i\right)\right] \partial_a {\cal P}^a_j
    \nonumber\\
    \!\!&&\!\!
    - \zeta {\cal P}^a_j \left(\tensor{\epsilon}{_i^j^k} \partial_a v_k+\partial_a \mathfrak{K}^j_i\right)
    \nonumber\\
    \!\!&&\!\!
    + {\cal P}^a_j \mathfrak{K}^j_k {\cal D}_a^l \tensor{\epsilon}{^k_l_i}
    + 2 {\cal D}_a^l {\cal P}^a_{[l} v_{i]}
    + \left( \delta_{ij} - v_i v_j \right) {\cal E}^j
    \,,\nonumber
\end{eqnarray}
and
\begin{eqnarray}\label{eq:Gauss constraint}
    G_i \!\!&=&\!\! - \partial_a \tilde{\cal K}^a_i
    - \left(\tilde{\cal P}^a_k K^l_a
    + \tilde{\cal K}^a_k \Gamma^l_a \right) \tensor{\epsilon}{^k_l_i}
    \\
    \!\!&=&\!\! - \partial_a {\cal K}^a_i
    + \zeta \partial_a {\cal P}^a_i
    - \left({\cal P}^a_k A_a^l
    + {\cal K}^a_k B_a^l \right) \tensor{\epsilon}{^k_l_i}
    \nonumber\\
    \!\!&=&\!\! - \left[\tensor{\epsilon}{_i^j^k} v_k
    + \mathfrak{K}^j_i
    - \zeta \delta^j_i\right] \partial_a {\cal P}^a_j
    - {\cal P}^a_j \left[\tensor{\epsilon}{_i^j^k} \partial_a v_k
    + \partial_a \mathfrak{K}^j_i\right]
    \nonumber\\
    \!\!&&\!\!
    + \tensor{\epsilon}{_i_j^k} \left({\cal D}_a^j {\cal P}^a_k
    + {\cal E}^j v_k\right)
    - {\cal P}^a_j \mathfrak{K}^j_k B_a^l \tensor{\epsilon}{^k_l_i}\,.
\end{eqnarray}

Notice that the transformations they generate on the velocity variable
\begin{eqnarray}
    \{v_i , L_j[\beta^j]\} \!\!&=&\!\! - \left( \delta_{ij} - v_i v_j \right) \beta^j
    \,,\\
    \{v_i , G_j[\theta^j]\} \!\!&=&\!\! \tensor{\epsilon}{_i_j^k} \theta^j v_k
\end{eqnarray}
are consistent with Einstein's velocity addition law and vector rotation.
Completing a four-vector $u^I=\gamma (1,v^i)$, we obtain
\begin{equation}
    \{u^I,L_j[\beta^j]\} =-\gamma (v_j \beta^j,\beta^i)
    = \tensor{\Omega}{^I_J} u^J\,,
\end{equation}
where
\begin{equation}
    \tensor{\Omega}{^I_J} = - \delta^I_0 \delta^j_J \beta_j
    - \delta^I_i \delta^0_J \beta^i
\end{equation}
corresponds to the first-order contribution to a Lorentz boost:
\begin{eqnarray}
    \tensor{\Lambda}{^I_J} \!\!&=&\!\!
    \tensor{\delta}{^I_J}
    + \tensor{\Omega}{^I_J}
    + O\left(\beta^2\right)
    \nonumber\\
    \!\!&=&\!\! \left(\begin{matrix}
        1 && - \beta_j\\
        - \beta^i &&\tensor{\delta}{^i_j}
    \end{matrix}\right)
\end{eqnarray}
with rapidity $\beta^i$: If $\beta^i=\eta \delta^i_1$, then
\begin{equation}
    \tensor{\left(\exp \Omega\right)}{^I_J}=\left(\begin{matrix}
        \cosh\eta && -\sinh\eta && 0 && 0\\
        -\sinh\eta &&\cosh\eta && 0 && 0\\
        0 && 0 && 1 && 0\\
        0 && 0 && 0 && 1\\
    \end{matrix}\right)\,.
\end{equation}
Furthermore, writing the tetrad components in terms of the canonical variables,
\begin{eqnarray}
    e^a_0 \!\!&=&\!\! - \left(\gamma N^{-1} N^a + \frac{\gamma\sqrt{(\det {\cal P})^{-1}}}{\sqrt{8\pi G}} \gamma v^j {\cal P}^a_j \right)
    \,,\qquad\\
    e^a_i \!\!&=&\!\! \frac{N^a}{N} \gamma v_i + \frac{\sqrt{(\det {\cal P})^{-1}}}{\sqrt{8\pi G}} \left(\delta_i^j+\gamma^2v_iv^j\right) {\cal P}^a_j
    \,,\qquad
\end{eqnarray}
we can compute the transformations
\begin{eqnarray}
    \{e^a_0,L_k[\beta^k]\} \!\!&=&\!\! - v_j \beta^j e^a_0 + \frac{\sqrt{(\det {\cal P})^{-1}}}{\sqrt{8\pi G}} {\cal P}_j^a \beta^j
    \nonumber\\
    \!\!&=&\!\!
    \beta^j e_i^a
    = -\tensor{\Omega}{^I_0}e^a_I
    \,,\\
    \{e^a_i,L_k[\beta^k]\}
    \!\!&=&\!\!
    - \left(\gamma N^{-1} N^a + \frac{\gamma\sqrt{(\det {\cal P})^{-1}}}{\sqrt{8\pi G}} \gamma v^j {\cal P}^a_j\right) \beta_i
    \nonumber\\
    \!\!&=&\!\!
    \beta_i e^a_0 = -\tensor{\Omega}{^I_i}e^a_I
    \,,
\end{eqnarray}
which imply that the transformation of the triad $e^a_J$ is an infinitesimal Lorentz boost of an internal co-vector:
\begin{equation}
    \{e^a_J,L_k[\beta^k]\}=-\tensor{\Omega}{^I_J}e^a_I\,.
\end{equation}
Lastly, computing
\begin{eqnarray}
    \{F^{0i}_{ab},L_k[\beta^k]\} \!\!&=&\!\! - {\cal F}^m_{ab} \tensor{\epsilon}{^i_m_n} \beta^n
    = \tensor{\Omega}{^0_k} F^{ki}_{ab}
    \,,\qquad\\
    \{{\cal F}^{i}_{ab},L_k[\beta^k]\} \!\!&=&\!\! F^{0m}_{ab} \tensor{\epsilon}{^i_m_n} \beta^n
    = \tensor{\Omega}{^j_0} F^{0k}_{ab}
    + \tensor{\Omega}{^k_0} F^{j0}_{ab}
    \,,\qquad
\end{eqnarray}
where we defined
\begin{equation}
    {\cal F}^i_{ab}
    = \frac{1}{2} \tensor{\epsilon}{^i_k_l} F^{kl}_{ab}\,,
\end{equation}
we conclude that
\begin{equation}
    \{F^{IJ}_{ab},L_k[\beta^k]\} = \tensor{\Omega}{^I_K} F^{KJ}_{ab}
    + \tensor{\Omega}{^J_K} F^{IK}_{ab}
\end{equation}
corresponds to an infinitesimal Lorentz boost too.

Similarly, the constraint $G_i$ generates SO(3) rotations in the internal indices.
Furthermore, as will be shown in Subsection~\ref{sec:Constraint algebra}, the brackets between the constraints $L_i$ and $G_i$ satisfy the Lorentz algebra.
For these reasons, we refer to them as the Lorentz--Gauss constraints in what follows.

\subsection{Hamiltonian and vector constraints}

The vector and Hamiltonian constraints are respectively given by
\begin{eqnarray}\label{eq:Vector constraint}
    H_a
    \!\!&=&\!\! \tilde{\cal P}^b_i F^{0i}_{ab}
    + \tilde{\cal K}^b_i {\cal F}^i_{ab}
    \\
    \!\!&=&\!\! {\cal P}^b_i \left(F^{0i}_{ab} - \zeta {\cal F}^i_{ab}\right)
    + {\cal K}^b_i \left({\cal F}^i_{ab} + \zeta F^{0i}_{ab}\right)
    \,,\nonumber
\end{eqnarray}
and
\begin{equation}\label{eq:Hamiltonian constraint}
    H = H^{(1)} + H^{(2)}\,,
\end{equation}
where
\begin{equation}
    \label{eq:H1}
    H^{(1)} = H_a \frac{\gamma \sqrt{(\det {\cal P})^{-1}}}{\sqrt{8\pi G}} v^q {\cal P}^a_q
    \,,
\end{equation}
\begin{equation}
    \label{eq:H2}
    H^{(2)} = - \frac{\gamma\sqrt{(\det {\cal P})^{-1}}}{\sqrt{8\pi G}} \frac{1}{2\gamma^2}  {\cal P}^a_p {\cal P}^b_q \tensor{\epsilon}{^p^q_i} \left({\cal F}^i_{ab} + \zeta F^{0i}_{ab}\right)
    \,,
\end{equation}
and
\begin{eqnarray}
    F_{ab}^{0i}
    \!\!&=&\!\! 2 \partial_{[a} K_{b]}^i - 2 \tensor{\epsilon}{^i_j_k} K_{[a}^j \Gamma_{b]}^{k}
    \,,\\
    {\cal F}^i_{ab}
    \!\!&=&\!\! 2 \partial_{[a} \Gamma_{b]}^i + \tensor{\epsilon}{^i_j_k} K_a^j K_b^k - \tensor{\epsilon}{^i_j_k} \Gamma_a^j \Gamma_b^k\,.
\end{eqnarray}
The linear combinations with the Barbero--Immirzi parameter can be written more compactly in the $A,B$ variables:
\begin{eqnarray}\label{eq:F lin comb - 1}
    \!\!\!\!&&\!\!\!\!
    F^{0i}_{ab} - \zeta {\cal F}^i_{ab}
    =
    2 \partial_{[a} A_{b]}^i
    \\
    \!\!\!\!&&\!\!\!\!\qquad
    + \frac{2 \tensor{\epsilon}{^i_k_l}}{1+\zeta^2} \left( \frac{\zeta}{2} \left(A_{[a}^k A_{b]}^l - B_{[a}^k B_{b]}^l\right)
    - A_{[a}^k B_{b]}^l \right)
    \,,\nonumber\\
    \!\!\!\!&&\!\!\!\!
    \label{eq:F lin comb - 2}
    {\cal F}^i_{ab} + \zeta F^{0i}_{ab}
    = 2 \partial_{[a} B_{b]}^i
    \\
    \!\!\!\!&&\!\!\!\!\qquad
    + \frac{2 \tensor{\epsilon}{^i_k_l}}{1+\zeta^2} \left( \frac{1}{2} \left(A_{[a}^k A_{b]}^l - B_{[a}^k B_{b]}^l\right)
    + \zeta A_{[a}^k B_{b]}^l \right)
    \,.\nonumber
\end{eqnarray}

Since $H_a$ vanishes independently of $H$, we find that the constraint $H=0$ implies $H^{(2)}=0$ on shell.

\subsection{Constraint algebra}
\label{sec:Constraint algebra}

We now show that the set of constraints given by $L_i$, $G_i$, $H_a$, and $H$ is first class in the extended phase space.
Detailed intermediate steps can be found in App.~\ref{app:Constraint brackets}.

First, the Lorentz--Gauss constraints satisfy the Lorentz algebra,
\begin{eqnarray}\label{eq:JJ}
    \{G_i[\theta_1^i],G_j[\theta_2^j]\}
    \!\!&=&\!\! G_k \left[\tensor{\epsilon}{^k_i_j} \theta_1^i \theta_2^j\right]\,,
    \\
    \label{eq:JL}
    \{G_i[\theta^i],L_j[\beta^j]\}
    \!\!&=&\!\! L_k \left[\tensor{\epsilon}{^k_i_j} \theta^i \beta^j\right]\,,
    \\
    \label{eq:LL}
    \{L_i[\beta_1^i],L_j[\beta_2^j]\}
    \!\!&=&\!\! - G_k \left[\tensor{\epsilon}{^k_i_j} \beta_1^i \beta_2^j\right]\,.
\end{eqnarray}

The vector constraint commutes with the Lorentz--Gauss constraints,
\begin{eqnarray}
    \{H_c[N^c],G_k[\theta^k]\}
    \!\!&=&\!\! 0
    \,,\label{eq:HaJ}\\
    \{H_c[N^c],L_k[\beta^k]\}
    \!\!&=&\!\! 0\,,\label{eq:HaL}
\end{eqnarray}
while its bracket with itself yields
\begin{eqnarray}\label{eq:HaHa}
    \{H_a[N^a],H_c[\epsilon^c]\} 
    \!\!&=&\!\! - H_a \left[\mathcal{L}_{\vec \epsilon} N^a\right]
    - L_i\left[\epsilon^a N^b F^{0i}_{ab}\right]
    \nonumber\\
    \!\!&&\!\!
    - G_i\left[\epsilon^a N^b {\cal F}^i_{ab}\right]
    \,.
\end{eqnarray}
This last bracket suggests that the linear combination $\mathfrak{D}_a[N^a]=H_a[N^a]+G_i[\Gamma_a^iN^a]+L_i[K_a^iN^a]$ generates spatial diffeomorphisms.
Indeed, such generator can be written as
\begin{eqnarray}
\label{eq:Spatial diffeomorphism generator}
    \mathfrak{D}_a
    \!\!&=&\!\! {\cal P}^b_i \partial_a A_b^i
    + {\cal K}^b_i \partial_a B_b^i
    - \partial_b \left({\cal P}^b_i A_a^i+{\cal K}^b_i B_a^i\right)
    \\
    \!\!&=&\!\!
    {\cal P}^b_i \partial_a {\cal D}_b^i
    - {\cal E}_i \partial_a v^i
    - {\cal B}^{ij} \partial_a \mathfrak{K}_{ij}
    - \partial_b \left({\cal P}^b_i {\cal D}_a^i\right)
    \,,\nonumber
\end{eqnarray}
and, therefore, it generates spatial diffeomorphisms along the shift,
\begin{eqnarray}
    \{A_a^i,\mathfrak{D}_b[N^b]\}
    \!\!&=&\!\! \mathcal{L}_{\vec{N}} A_a^i
    \,,\\
    \{B_a^i,\mathfrak{D}_b[N^b]\} \!\!&=&\!\! \mathcal{L}_{\vec{N}} B_a^i
    \,,\\
    \{{\cal P}^a_i,\mathfrak{D}_b[N^b]\} \!\!&=&\!\! \mathcal{L}_{\vec{N}} {\cal P}^a_i
    \,,\\
    \{{\cal K}^a_i,\mathfrak{D}_b[N^b]\} \!\!&=&\!\! \mathcal{L}_{\vec{N}} {\cal K}^a_i
    \,,
\end{eqnarray}
on and off the second-class constraint surface for all phase-space variables.
It follows that
\begin{eqnarray}\label{eq:Vector constraint brakcet}
    \{{\cal O},H_b[N^b]\} 
    \!\!&=&\!\! \mathcal{L}_{\vec N} {\cal O}
    + \{G_i\left[\Gamma_b^iN^b\right],{\cal O}\}
    \nonumber\\
    \!\!&&\!\!
    + \{L_i\left[K_b^iN^b\right],{\cal O}\}\,,
\end{eqnarray}
for an arbitrary phase-space function ${\cal O}$.

Because the Lorentz--Gauss constraints generate proper Lorentz transformations of $e^a_I$ and $F^{IJ}_{ab}$ even off shell, it follows from the Lorentz invariance of $\frac{|e|}{16\pi G} e^a_I e^b_J \tensor{P}{^I^J_K_L} F^{KL}_{ab}=-H[N]-H_a[N^a]$---using (\ref{eq:HaJ}) and (\ref{eq:HaL})---that they commute with the Hamiltonian constraint too:
\begin{eqnarray}\label{eq:HJ}
    \{H [N],G_k[\theta^k]\} \!\!&=&\!\! 0\,,\\
    \{H [N],L_k[\beta^k]\} \!\!&=&\!\! 0\,.\label{eq:HL}
\end{eqnarray}
Use of (\ref{eq:Vector constraint brakcet}) then implies the bracket
\begin{eqnarray}
    \{H[N],H_c[\epsilon^c]\}
    \!\!&=&\!\! - H[\mathcal{L}_{\vec \epsilon} N]
    + G_i\left[N\epsilon^a {\cal F}^i_{\bar{0}a}\right]
    \nonumber\\
    \!\!&&\!\!
    + L_i\left[N\epsilon^a F^{0i}_{\bar{0}a}\right]\,,
\end{eqnarray}
where we defined the expressions
\begin{equation}
    F_{\bar{0}a}^{0i} = \frac{1}{N} \{K_a^i,H[N]\}
    \quad,\quad
    {\cal F}_{\bar{0}a}^i = \frac{1}{N} \{\Gamma_a^i,H[N]\}\,,
\end{equation}
which are on-shell compatible with the identification of the normal components of the strength tensor field
\begin{eqnarray}
    n^\mu F^{0i}_{\mu a} \!\!&=&\!\! \frac{1}{N} F_{ta}^{0i} - \frac{N^b}{N} F_{ba}^{0i}
    =
    \frac{1}{N} \{K_a^i,H[N]\}
    \nonumber\,,\\
    n^\mu {\cal F}_{\mu a}^i \!\!&=&\!\! \frac{1}{N} {\cal F}_{ta}^i - \frac{N^b}{N} {\cal F}_{ba}^i
    = \frac{1}{N} \{\Gamma_a^i,H[N]\}\,,
\end{eqnarray}
where we substituted $\dot{K}_a^i=\{K_a^i,H[N]+H_b[N^b]+L_j[K_t^j]+G_j[\Gamma_t^j]\}$ and $\dot{\Gamma}_a^i=\{K_a^i,H[N]+H_b[N^b]+L_j[K_t^j]+G_j[\Gamma_t^j]\}$ in the time components of the strength tensor.

Using all the above, we compute the brackets of the $H^{(1)}$ and $H^{(2)}$ components of the Hamiltonian constraint:
\begin{eqnarray}
    \{H^{(1)}[N],H^{(1)}[\epsilon^{\bar{0}}]\}
    = 0
    \,,
\end{eqnarray}
\begin{eqnarray}
    \!\!&&\!\!
    \{H^{(2)}[N],H^{(1)}[\epsilon^{\bar{0}}]\}+\{H^{(1)}[N],H^{(2)}[\epsilon^{\bar{0}}]\}
    \nonumber\\
    \!\!&&\!\!\qquad
    = - H^{(2)} \left[\frac{\sqrt{(\det {\cal P})^{-1}}}{\sqrt{8\pi G}} \gamma v^q {\cal P}^a_q (\epsilon^{\bar{0}}\partial_a N-N\partial_a \epsilon^{\bar{0}})\right]
    \nonumber\\
    \!\!&&\!\!\qquad\quad
    - H_a\left[q^{ab} \left(\epsilon^{\bar{0}}\partial_bN-N\partial_b\epsilon^{\bar{0}}\right)\right]
    \,,
\end{eqnarray}
where $q^{ab}$ is the inverse spatial metric (\ref{eq:Inverse spatial metric - canonical}), and
\begin{eqnarray}
    \!\!&&\!\!\{H^{(2)}[N],H^{(2)}[\epsilon^{\bar{0}}]\}
    \\
    \!\!&&\!\!
    \qquad=
    H^{(2)}\left[\frac{\sqrt{(\det {\cal P})^{-1}}}{\sqrt{8\pi G}} \gamma v^q {\cal P}^b_q \left(\epsilon^{\bar{0}}\partial_bN-N\partial_b\epsilon^{\bar{0}}\right)\right]\,.\nonumber
\end{eqnarray}
Therefore, the full bracket of the Hamiltonian constraint with itself is given by
\begin{equation}\label{eq:HH}
    \{H[N],H[\epsilon^{\bar{0}}]\}=- H_a\left[q^{ab} \left(\epsilon^{\bar{0}}\partial_bN-N\partial_b\epsilon^{\bar{0}}\right)\right]\,.
\end{equation}

In summary, the constraint algebra, using Poisson brackets in the extended phase space, is given by
\begin{eqnarray}
    \label{eq:HaHa-Poisson}
    \{H_a[N^a],H_c[\epsilon^c]\}
    \!\!&=&\!\! - H_a \left[\mathcal{L}_{\vec \epsilon} N^a\right]
    - G_i\left[\epsilon^a N^b {\cal F}^i_{ab}\right]
    \nonumber\\
    \!\!&&\!\!
    - L_i\left[\epsilon^a N^b F^{0i}_{ab}\right]
    \,,\\
    \label{eq:HHa-Poisson}
    \{H[N],H_c[\epsilon^c]\}
    \!\!&=&\!\! - H[\mathcal{L}_{\vec \epsilon} N]
    + G_i\left[N\epsilon^a {\cal F}^i_{\bar{0}a}\right]
    \nonumber\\
    \!\!&&\!\!
    + L_i\left[N\epsilon^a F^{0i}_{\bar{0}a}\right]
    \,,\\
    \label{eq:HH-Poisson}
    \{H[N],H[\epsilon^{\bar{0}}]\}
    \!\!&=&\!\! - H_a\left[q^{ab} \left(\epsilon^{\bar{0}}\partial_bN-N\partial_b \epsilon^{\bar{0}}\right)\right]
    \,,\qquad\\
    \label{eq:HaJ-Poisson}
    \{H_c[N^c],G_k[\theta^k]\}
    \!\!&=&\!\! 0
    \,,\\
    \label{eq:HaL-Poisson}
    \{H_c[N^c],L_k[\beta^k]\}
    \!\!&=&\!\! 0
    \,,\\
    \label{eq:HJ-Poisson}
    \{H[N],G_k[\theta^k]\}
    \!\!&=&\!\! 0
    \,,\\
    \label{eq:HL-Poisson}
    \{H[N],L_k[\beta^k]\}
    \!\!&=&\!\! 0
    \,,\\
    \label{eq:JJ-Poisson}
    \{G_i[\Gamma_t^i],G_j[\theta^j]\} \!\!&=&\!\! G_k \left[\tensor{\epsilon}{^k_i_j} \Gamma_t^i \theta^j\right]
    \,,\\
    \label{eq:JL-Poisson}
    \{G_i[\Gamma_t^i],L_j[\beta^j]\}
    \!\!&=&\!\! L_k \left[\tensor{\epsilon}{^k_i_j} \Gamma_t^i \beta^j\right]
    \,,\\
    \label{eq:LL-Poisson}
    \{L_i[K_t^i],L_j[\beta^j]\}
    \!\!&=&\!\! - G_k \left[\tensor{\epsilon}{^k_i_j} K_t^i \beta^j\right]
    \,.
\end{eqnarray}
This algebra is first class and is precisely what one would expect from an extrapolation of the Einstein--Yang--Mills system's algebra (\ref{eq:Constraint algebra - EYM - HaHa})-(\ref{eq:Constraint algebra - EYM - HG}).

\subsection{Gauge transformations and covariance}

We say that the spacetime is covariant if the canonical gauge transformation is equivalent to a diffeomorphism of the spacetime metric,
\begin{equation}
    \delta_{\epsilon,\theta,\beta} g_{\mu \nu} \big|_{\rm OS} =
    \mathcal{L}_{\xi} g_{\mu \nu} \big|_{\rm OS}
    \,,
    \label{eq:Spacetime covariance condition}
\end{equation}
where "OS" indicates an evaluation on shell, involving not only the vanishing of the constraints but also use of Hamilton's equations of motion on the right-hand side for the time-derivatives: $\dot{q}_{ab}=\{q_{ab},H[N,\vec{N}]\}$.
The gauge parameters $(\epsilon^{\bar{0}}, \epsilon^a)$ in (\ref{eq:Spacetime covariance condition}) are related to the four-vector field $\xi^\mu$ that acts as the diffeomorphism generator by a change of basis from the observer's frame to the Eulerian frame associated to the foliation:
\begin{eqnarray}
    \xi^\mu \!\!&=&\!\! \xi^t t^\mu + \xi^a s^\mu_a
    = \epsilon^{\bar{0}} n^\mu + \epsilon^a s^\mu_a
    \,,\nonumber
    \\
    \xi^t \!\!&=&\!\! \frac{\epsilon^{\bar{0}}}{N}
    \quad,\quad
    \xi^a = \epsilon^a - \frac{\epsilon^{\bar{0}}}{N} N^a
    \,.
\label{eq:Diffeomorphism generator projection}
\end{eqnarray}

The components of the spacetime metric in (\ref{eq:Spacetime covariance condition}) are given by those of the ADM line element (\ref{eq:ADM line element}), which are expressed in terms of $N$, $N^a$, and $q_{ab}$.
Therefore, the left-hand side of (\ref{eq:Spacetime covariance condition}) requires gauge transformations of $N$ and $N^a$ which are not directly provided by the Poisson brackets because they do not have momenta.
Instead, their gauge transformations, as well as those of $K_t^i$ and $\Gamma_t^i$, are defined by the preservation of the form of Hamilton's equations of motion, hence they are determined by the constraint algebra (\ref{eq:HaHa-Poisson})-(\ref{eq:LL-Poisson}) and given by \cite{pons1997gauge,salisbury1983realization,HypDef}
\begin{eqnarray}\label{eq:Gauge transf - N}
    \delta_{\epsilon,\theta,\beta} N \!\!&=&\!\! \dot{\epsilon}^0 + \epsilon^b\partial_b N - N^b \partial_b \epsilon^{\bar{0}}
    \,,\\
    \delta_{\epsilon,\theta,\beta} N^a \!\!&=&\!\! \dot{\epsilon}^a + \epsilon^b\partial_b N^a - N^b \partial_b \epsilon^a
    \nonumber\\
    \!\!&&\!\!
    + q^{ab} \left(\epsilon^{\bar{0}}\partial_b N - N \partial_b \epsilon^{\bar{0}}\right)
    \,,\label{eq:Gauge transf - Na}\\
    \delta_{\epsilon,\theta,\beta} K_t^i \!\!&=&\!\! \dot{\beta}^i
    + \tensor{\epsilon}{^i_j_k} \left(\beta^j \Gamma_t^k+\theta^j K_t^k\right)
    + \epsilon^a N^b F^{0i}_{ab}
    \nonumber\\
    \!\!&&\!\!
    + \left(\epsilon^{\bar{0}}N^a-N\epsilon^a\right) F^{0i}_{\bar{0}a}
    \,,\label{eq:Gauge transf - Kt}\\
    \delta_{\epsilon,\theta,\beta} \Gamma_t^i \!\!&=&\!\! \dot{\theta}^i
    + \tensor{\epsilon}{^i_j_k} \left(\theta^j \Gamma_t^k
    - \beta^j K_t^k\right)
    + \epsilon^a N^b {\cal F}^i_{ab}
    \nonumber\\
    \!\!&&\!\!
    + \left(\epsilon^{\bar{0}}N^a-N\epsilon^a\right) {\cal F}^i_{\bar{0}a}
    \,.\label{eq:Gauge transf - Gammat}
\end{eqnarray}

The transformations (\ref{eq:Gauge transf - N}) and (\ref{eq:Gauge transf - Na}) correspond precisely to spacetime diffeomorphisms of the lapse and shift, $\mathcal{L}_\xi N$ and $\mathcal{L}_\xi N^a$, respectively.
On the other hand, the canonical decomposition of spacetime diffeomorphisms and proper Lorentz transformations---the latter denoted by $\delta^{\rm SO(1,3)}_{\theta,\beta}$ in what follows---of the connection components are respectively given by
\begin{eqnarray}
    \mathcal{L}_\xi K_t^i
    \!\!&=&\!\! \delta^{\rm SO(1,3)}_{\xi^\mu \Gamma_\mu,\xi^\mu K_\mu} K_t^i
    + \epsilon^a N^b F_{ab}^{0i}
    \nonumber\\
    \!\!&&\!\!
    + \left(\epsilon^{\bar{0}}N^a-N\epsilon^a\right) F_{0 a}^{0i}
    \,,\\
    \mathcal{L}_\xi \Gamma_t^i \!\!&=&\!\! \delta^{\rm SO(1,3)}_{\xi^\mu \Gamma_\mu,\xi^\mu K_\mu} \Gamma_t^i
    + \epsilon^a N^b {\cal F}_{ab}^i
    \nonumber\\
    \!\!&&\!\!
    + \left(\epsilon^{\bar{0}}N^a-N\epsilon^a\right) {\cal F}_{0 a}^i
    \,,
\end{eqnarray}
and
\begin{eqnarray}
    \delta^{\rm SO(1,3)}_{\theta,\beta} K_t^i \!\!&=&\!\! \dot{\beta}^i
    + \tensor{\epsilon}{^i_j_k} \left( \beta^j \Gamma_t^k
    + \theta^j K_t^k\right)
    \,,\\
    \delta^{\rm SO(1,3)}_{\theta,\beta} \Gamma_t^i \!\!&=&\!\!
    \dot{\theta}^i
    + \tensor{\epsilon}{^i_j_k} \left(\theta^j \Gamma_t^k - \beta^j K_t^k\right)\,.
\end{eqnarray}
(See App.~\ref{app:Lorentz transformations} and \ref{app:Lie derivatives} for detailed derivations.)
A comparison with the canonical gauge transformations (\ref{eq:Gauge transf - N})-(\ref{eq:Gauge transf - Gammat}) reveals that the latter correspond to linear combinations of diffeomorphisms and Lorentz transformations for the Lagrange multipliers; specifically,
\begin{equation}
    \delta_{\epsilon,\theta,\beta} = \mathcal{L}_{\xi} + \delta^{\rm SO(1,3)}_{\theta-\xi^\mu \Gamma_\mu,\beta-\xi^\mu K_\mu}\,.
\end{equation}
For the full system to be covariant, this transformation must hold for the full expressions of all physical tensor fields.

As expected, the lapse, the shift, and the structure function, and hence the line element, are invariant to Lorentz transformations,
\begin{eqnarray}
    \{q^{ab} , G_i[\theta^i]\} \!\!&=&\!\! 0
    \,,\\
    \{q^{ab} , L_i[\beta^i]\} \!\!&=&\!\! 0
    \,.
\end{eqnarray}
This, in turn, implies that the vector constraint does generate spatial diffeomorphisms of the spatial metric on shell:
\begin{eqnarray}
    \{q^{ab}(x),H_c[\epsilon^c]\} \!\!&=&\!\! \mathcal{L}_{\vec \epsilon}\, q^{ab}
    - L_i\left[\{q^{ab}(x),K_b^iN^b\}\right]
    \nonumber\\
    \!\!&&\!\!
    - G_i\left[\{q^{ab}(x),\Gamma_b^iN^b\}\right]
    \,.
\end{eqnarray}
Furthermore, its canonical gauge transformation does not contain derivatives of the normal gauge function---see Appendix~\ref{sec:Covariance condition} for the detailed computation---
\begin{equation}\label{eq:Spacetime covariance condition - reduced}
    \frac{\partial \{q^{ab},H[\epsilon^{\bar{0}}]\}}{\partial (\partial_{c_1} \epsilon^{\bar{0}})} \bigg|_{\rm OS} = \frac{\partial \{q^{ab},H[\epsilon^{\bar{0}}]\}}{\partial (\partial_{c_1} \partial_{c_2} \epsilon^{\bar{0}})} \bigg|_{\rm OS} = \dotsi = 0\,,
\end{equation}
a necessary condition for (\ref{eq:Spacetime covariance condition}) to hold that is beyond the requirement that the algebra be first class \cite{EMGCov}.

Similarly, use of (\ref{eq:Gauge transf - N})-(\ref{eq:Gauge transf - Gammat}) and our previous observations that the Lorentz--Gauss constraints generate proper Lorentz transformations of the spatial components, as well as that the vector constraint generates the appropriate linear combination of spatial diffeomorphisms and SO(1,3) transformations, imply that the covariance condition of the tetrad
\begin{equation}
    \delta_{\epsilon,\theta,\beta} e_\mu^I \big|_{\rm OS} =
    \mathcal{L}_{\xi} e_\mu^I + \delta^{\rm SO(1,3)}_{\theta-\xi^\mu \Gamma_\mu,\beta-\xi^\mu K_\mu} e_\mu^I \big|_{\rm OS}\,,
    \label{eq:Tetrad covariance condition}
\end{equation}
reduces to the conditions
\begin{eqnarray}
    \frac{\partial \{v_i,H[\epsilon^{\bar{0}}]\}}{\partial(\partial_c\epsilon^{\bar{0}})} \bigg|_{\rm OS} 
    \!\!&=&\!\! \frac{\sqrt{(\det {\cal P})^{-1}}}{\gamma\sqrt{8\pi G}} {\cal P}^c_i \bigg|_{\rm OS} 
    \,,\label{eq:Tetrad covariance condition - reduced - 1}\\
    \frac{\partial \{v_i,H[\epsilon^{\bar{0}}]\}}{\partial((\partial_{c_1}\partial_{c_2}\epsilon^{\bar{0}})} \bigg|_{\rm OS} 
    \!\!&=&\!\! 0
    \,,\label{eq:Tetrad covariance condition - reduced - 2}
\end{eqnarray}
and
\begin{eqnarray}
    \!\!&&\!\!
    \left(\delta^a_c\delta^k_i
    - \frac{1}{2} {\cal P}^a_i ({\cal P}^{-1})^k_c \right) \frac{\partial\{{\cal P}^c_k,H[\epsilon^{\bar{0}}]\}}{\partial(\partial_d \epsilon^{\bar{0}})} \bigg|_{\rm OS}
    \nonumber\\
    \!\!&&\!\!\qquad=
    - \frac{\gamma \sqrt{(\det {\cal P})^{-1}}}{\sqrt{8\pi G}} v^j {\cal P}^a_j {\cal P}^d_i \bigg|_{\rm OS}
    \,,\label{eq:Tetrad covariance condition - reduced - 3}\\
    \!\!&&\!\!
    \left(\delta^a_c\delta^k_i
    - \frac{1}{2} {\cal P}^a_i ({\cal P}^{-1})^k_c \right) \frac{\partial \{{\cal P}^c_k,H[\epsilon^{\bar{0}}]\}}{\partial(\partial_{d_1} \partial_{d_2} \epsilon^{\bar{0}})} \bigg|_{\rm OS} = 0
    \,,\qquad
    \label{eq:Tetrad covariance condition - reduced - 4}
\end{eqnarray}
while the covariance condition of the connection
\begin{equation}
    \delta_{\epsilon,\theta,\beta} \tensor{\omega}{_\mu^I^J} \big|_{\rm OS} =
    \mathcal{L}_{\xi} \tensor{\omega}{_\mu^I^J} + \delta^{\rm SO(1,3)}_{\theta-\xi^\mu \Gamma_\mu,\beta-\xi^\mu K_\mu} \tensor{\omega}{_\mu^I^J} \big|_{\rm OS}
    \,,
    \label{eq:Connection covariance condition}
\end{equation}
reduces to
\begin{equation}\label{eq:Connection covariance condition - reduced}
    \frac{\partial \{\tensor{\omega}{_a^I^J},H[\epsilon^{\bar{0}}]\}}{\partial (\partial_{c_1} \epsilon^{\bar{0}})} \bigg|_{\rm OS} = \frac{\partial \{\tensor{\omega}{_a^I^J},H[\epsilon^{\bar{0}}]\}}{\partial (\partial_{c_1} \partial_{c_2} \epsilon^{\bar{0}})} \bigg|_{\rm OS} = \dotsi = 0\,.
\end{equation}

It can be readily verified that the covariance condition of the connection (\ref{eq:Connection covariance condition - reduced}) holds because the Hamiltonian constraint has no derivatives of the momenta.
On the other hand, it requires more work to evaluate the covariance conditions of the tetrad (\ref{eq:Tetrad covariance condition - reduced - 1})-(\ref{eq:Tetrad covariance condition - reduced - 4}), but they indeed hold.
See Appendix~\ref{sec:Covariance condition} for a detailed proof.

Therefore, the full canonical gauge transformations of the spacetime metric, the tetrad, and the connection indeed correspond to linear combinations of spacetime diffeomorphisms and SO(1,3) transformations; all geometric objects constructed from these, such as curvature or torsion tensors, then transform in a corresponding manner; hence the theory is indeed fully covariant in its canonical formulation.
This is true on and off the second-class constraint surface: As we show in the next section, use of Dirac brackets to impose the second-class constraints preserves the above outcome.

\subsection{Dirac observables}

The Lorentz--Gauss constraints, being first class, imply that the functionals
\begin{eqnarray}\label{eq:Lorentz observable}
    \mathfrak{L}_i[\alpha^i] \!\!&=&\!\!
    \int{\rm d}^3x\; \alpha^i \tensor{\epsilon}{_i_k^l} \left( \Gamma^k_a \tilde{\cal P}^a_l
    - K^k_a \tilde{\cal K}^a_l \right)
    \\
    \!\!&=&\!\! \int{\rm d}^3x\; \alpha^i \tensor{\epsilon}{_i_k^l} \left(B_a^l {\cal P}^a_k
    - A_a^l {\cal K}^a_k\right)
    \,,\nonumber
\end{eqnarray}
and
\begin{eqnarray}\label{eq:Gauss observable}
    \mathfrak{G}_i [\sigma^i] \!\!&=&\!\!
    \int{\rm d}^3x\; \sigma^i \tensor{\epsilon}{_i_k^l} \left( K^k_a \tilde{\cal P}^a_l
    + \Gamma^k_a \tilde{\cal K}^a_l \right)
    \\
    \!\!&=&\!\!
    \int{\rm d}^3x\; \sigma^i \tensor{\epsilon}{_i_k^l} \left(A_a^k {\cal P}^a_l
    + B_a^k {\cal K}^a_l\right)
    \,,\nonumber
\end{eqnarray}
with constants $\alpha^i$ and $\sigma^i$, are (nonlocal) Dirac observables because they are identical to the Lorentz--Gauss constraints up to boundary terms when smeared by constants and hence commute with all the constraints up to boundary terms on shell: The brackets of the local versions of the observables with the constraints yield
\begin{eqnarray}
    \{\mathfrak{L}_i,L_j[K_t^j]\} \!\!&=&\!\!
    - \tensor{\epsilon}{_i_j^k} K_t^j G_k
    - \partial_a \left(\tensor{\epsilon}{_i_k^l} K_t^k\tilde{\cal K}^a_l\right)
    \,,\\
    \{\mathfrak{L}_i,G_j[\Gamma_t^j]\} \!\!&=&\!\!
    \tensor{\epsilon}{_i_j^k} \Gamma_t^j L_k
    + \partial_a \left( \tensor{\epsilon}{_i_k^l} \Gamma_t^k \tilde{\cal P}^a_l \right)
    \,,
\end{eqnarray}
\begin{eqnarray}\label{eq:LHa obs}
    \{\mathfrak{L}_i,H_a[N^a]\} \!\!&=&\!\! \partial_a\left( \mathfrak{L}_i N^a\right)
    \\
    \!\!&&\!\!
    - \partial_a\left(\tensor{\epsilon}{_i_j^k} \left(\Gamma^j_b \tilde{\cal P}^a_k-K^j_b \tilde{\cal K}^a_k\right) N^b\right)
    \nonumber
    \,,
\end{eqnarray}
and
\begin{eqnarray}
    \{\mathfrak{G}_i,L_j[K_t^j]\} \!\!&=&\!\! \tensor{\epsilon}{_i_j^k} K_t^j L_k
    + \partial_a \left( \tensor{\epsilon}{_i_k^l} K_t^k \tilde{\cal P}^a_l\right)
    \,,\\
    \{\mathfrak{G}_i,G_j[\Gamma_t^j]\} \!\!&=&\!\! \tensor{\epsilon}{_i_j^k} \Gamma_t^j G_k
    + \partial_a \left(\tensor{\epsilon}{_i_k^l} \Gamma_t^k \tilde{\cal K}^a_l \right)
    \,,
\end{eqnarray}
\begin{eqnarray}
    \{\mathfrak{G}_i,H_a[N^a]\} \!\!&=&\!\! - \partial_a \left( \mathfrak{G}_i N^a\right)
    \\
    \!\!&&\!\!
    + \partial_a\left(\tensor{\epsilon}{_i_k^l} \left(K^k_b \tilde{\cal P}^a_l + \Gamma^k_b \tilde{\cal K}^a_l \right) N^b\right)
    \,.\nonumber
\end{eqnarray}
as well as
\begin{widetext}
\begin{equation}
    \{\mathfrak{L}_i,H[N]\}
    = \partial_a \left( N \frac{\gamma \sqrt{(\det {\cal P})^{-1}}}{\sqrt{8\pi G}} \left[\mathfrak{L}_i v^q {\cal P}^a_q-\tensor{\epsilon}{_i_j^k} \left(K^j_b \tilde{\cal K}^a_k 
    - \Gamma^j_b \tilde{\cal P}^a_k\right) v^q {\cal P}^b_q
    - \frac{2}{\gamma^2} {\cal P}^{[a}_i {\cal P}^{b]}_q \left(K^q_b - \zeta \Gamma^q_b\right)\right]\right)
    \,,
\end{equation}
\begin{equation}
    \{\mathfrak{G}_i,H[N]\} = - \partial_a \left( N \frac{\gamma \sqrt{(\det {\cal P})^{-1}}}{\sqrt{8\pi G}} \left[\mathfrak{G}_i v^q {\cal P}^a_q -  \tensor{\epsilon}{_i_k^l} \left(K^k_b \tilde{\cal P}^a_l + \Gamma^k_b \tilde{\cal K}^a_l \right) v^q {\cal P}^b_q
    - \frac{2}{\gamma^2} {\cal P}^{[a}_i {\cal P}^{b]}_q \left(\Gamma^q_b + \zeta K^q_b\right)\right]\right)
    \,.\label{eq:JH obs}
\end{equation}

The boundary terms imply conserved densitized currents: Defining
\begin{eqnarray}
    \mathcal{L}_i^t = \mathfrak{L}_i
    \quad,\quad
    \mathcal{L}_i^a \!\!&=&\!\! - N \frac{\gamma \sqrt{(\det {\cal P})^{-1}}}{\sqrt{8\pi G}} \left[\mathfrak{L}_i v^q {\cal P}^a_q
    - \tensor{\epsilon}{_i_j^k} \left(K^j_b \tilde{\cal K}^a_k 
    - \Gamma^j_b \tilde{\cal P}^a_k\right) v^q {\cal P}^b_q
    - \frac{2}{\gamma^2} {\cal P}^{[a}_i {\cal P}^{b]}_q \left(K^q_b - \zeta \Gamma^q_b\right)\right]
    \nonumber\\
    \!\!&&\!\!
    - \mathfrak{L}_i N^a
    + \tensor{\epsilon}{_i_j^k} \left(\Gamma^j_b \tilde{\cal P}^a_k-K^j_b \tilde{\cal K}^a_k\right) N^b
    + \tensor{\epsilon}{_i_k^l} \left(K_t^k\tilde{\cal K}^a_l
    - \Gamma_t^k \tilde{\cal P}^a_l\right)
    \,,
\end{eqnarray}
and
\begin{eqnarray}
    \mathcal{G}_i^t = \mathfrak{G}_i
    \quad,\quad
    \mathcal{G}_i^a \!\!&=&\!\! N \frac{\gamma \sqrt{(\det {\cal P})^{-1}}}{\sqrt{8\pi G}} \left[\mathfrak{G}_i v^q {\cal P}^a_q
    - \tensor{\epsilon}{_i_k^l} \left(K^k_b \tilde{\cal P}^a_l + \Gamma^k_b \tilde{\cal K}^a_l \right) v^q {\cal P}^b_q
    - \frac{2}{\gamma^2} {\cal P}^{[a}_i {\cal P}^{b]}_q \left(\Gamma^q_b + \zeta K^q_b\right)\right]
    \nonumber\\
    \!\!&&\!\!
    + \mathfrak{G}_i N^a -  \tensor{\epsilon}{_i_k^l} \left(K^k_b \tilde{\cal P}^a_l + \Gamma^k_b \tilde{\cal K}^a_l \right) N^b
    - \tensor{\epsilon}{_i_k^l} \left(K_t^k \tilde{\cal P}^a_l
    - \Gamma_t^k \tilde{\cal K}^a_l\right)\,,
\end{eqnarray}
\end{widetext}
the densitized four-currents ${\cal L}_i^\mu=({\cal L}_i^t,{\cal L}_i^a)$ and ${\cal G}_i^\mu=({\cal G}_i^t,{\cal G}_i^a)$ are conserved because the brackets (\ref{eq:LHa obs})-(\ref{eq:JH obs}) imply $\dot{\cal L}_i^t=-\partial_a{\cal L}_i^a$ and $\dot{\cal G}_i^t=-\partial_a{\cal G}_i^a$ on shell, and hence $\partial_\mu{\cal L}_i^\mu=\nabla_\mu {\cal L}_i^\mu=0$ and $\partial_\mu{\cal G}_i^\mu=\nabla_\mu {\cal G}_i^\mu=0$.

Moreover, these observables form a local Lorentz algebra,
\begin{eqnarray}\label{eq:JJ - obs}
    \{\mathfrak{G}_i(x),\mathfrak{G}_j(y)\}
    \!\!&=&\!\! \tensor{\epsilon}{_i_j^k} \mathfrak{G}_k \delta^3(x-y)\,,
    \\
    \label{eq:JL - obs}
    \{\mathfrak{G}_i(x),\mathfrak{L}_j(y)\}
    \!\!&=&\!\! \tensor{\epsilon}{_i_j^k} \mathfrak{L}_k \delta^3(x-y)\,,
    \\
    \label{eq:LL - obs}
    \{\mathfrak{L}_i(x),\mathfrak{L}_j(y)\}
    \!\!&=&\!\! - \tensor{\epsilon}{_i_j^k} \mathfrak{G}_k \delta^3(x-y)\,.
\end{eqnarray}

\section{Second-class constraints}
\label{sec:Second-class constraint surface}

\subsection{Primary and secondary second-class constraints}

We now proceed to obtain the contribution of each first-class constraint to the second-class constraint ${\cal C}_{ij}$.
First, the Lorentz--Gauss constraints do not contribute on the second-class constraint surface:
\begin{eqnarray}
    {\cal C}_{ij}^{(G)} \!\!&=&\!\! - \frac{\delta G_k[\Gamma_t^k]}{\delta {\cal B}^{ij}}
    \nonumber\\
    \!\!&=&\!\!
    - 2 \tensor{\epsilon}{^p_m_{(i}} \delta^q_{j)} 
    \Gamma_t^m \mathfrak{K}_{pq}
    \,,\\
    {\cal C}_{ij}^{(L)}
    \!\!&=&\!\! - \frac{\delta L_k[K_t^k]}{\delta {\cal B}^{ij}}
    \nonumber\\
    \!\!&=&\!\!
    - 4 \delta^p_{(i} \delta_{j)[m} v_{q]} K_t^m \mathfrak{K}_p^q\,.
\end{eqnarray}

Using (\ref{eq:Spatial diffeomorphism generator}), we obtain the vector constraint contribution
\begin{eqnarray}
    {\cal C}_{ij}^{(V)} \!\!&=&\!\! - \frac{\delta H_a[N^a]}{\delta {\cal B}^{ij}}
    \\
    \!\!&=&\!\! - \frac{\delta \mathfrak{D}_a[N^a]}{\delta {\cal B}^{ij}}
    - \frac{\delta G_k[\Gamma_a^kN^a]}{\delta {\cal B}^{ij}}
    - \frac{\delta L_k[K_a^kN^a]}{\delta {\cal B}^{ij}}
    \nonumber\\
    \!\!&=&\!\! \mathcal{L}_{\vec{N}} \mathfrak{K}_{ij}
    - 2 \tensor{\epsilon}{^p_m_{(i}} \delta^q_{j)} 
    N^a\Gamma_a^m \mathfrak{K}_{pq}
    \nonumber\\
    \!\!&&\!\!
    - \frac{G_k N^a}{1+\zeta^2} \left({\cal P}^{-1}\right)^r_a \delta_{r(i} \left(\delta^k_{j)}+\zeta \left(\tensor{\epsilon}{_{j)}^k^n} v_n
    + \mathfrak{K}^k_{j)}\right)\right)
    \nonumber\\
    \!\!&&\!\!
    - 4 \delta^p_{(i} \delta_{j)[m} v_{q]} N^aK_a^m \mathfrak{K}_p^q
    \nonumber\\
    \!\!&&\!\!
    + \frac{L_kN^a}{1+\zeta^2}  \left({\cal P}^{-1}\right)^r_a \delta_{r(i} \left[\tensor{\epsilon}{_{j)}^k^n} v_n + \mathfrak{K}^k_{j)}-\zeta \delta^k_{j)}\right]
    \,.\nonumber
\end{eqnarray}
When all the first-class constraints vanish and $\mathfrak{K}_{ij}=0$, this vanishes identically and hence neither do $H_a$ or $H^{(1)}$ contribute to ${\cal C}_{ij}$ on shell.

Therefore, the only nontrivial contribution comes from the second term of the Hamiltonian constraint,
\begin{eqnarray}
    {\cal C}_{ij}^{(2)} \!\!&=&\!\! - \frac{\delta H^{(2)}[N]}{\delta {\cal B}^{ij}}
    \nonumber\\
    \!\!&=&\!\! - N \frac{\sqrt{(\det {\cal P})^{-1}}}{\gamma\sqrt{8\pi G}} {\cal T}_{ij}\,,
\end{eqnarray}
where ${\cal T}^{ij}$ is given precisely by the torsion components (\ref{eq:Torsion-spatial-sym}).
This can be written as
\begin{equation}\label{eq:Second-class constraint}
    {\cal T}_{ij} = \frac{1}{1+\zeta^2} \tensor{V}{_i_j^k^l} \left({\cal B}_{kl}-\bar{\cal B}_{kl}\right)\,,
\end{equation}
such that ${\cal B}^{kl}=\bar{\cal B}^{kl}$ solves the second-class constraint.
In the expression above,
\begin{equation}
    \tensor{V}{^k^l_i_j} = \left( \delta_{pq} - v_p v_q \right) \tensor{\epsilon}{^k^p_{(i}} \tensor{\epsilon}{^l^q_{j)}}
    =: V_{pq} \tensor{\epsilon}{^k^p_{(i}} \tensor{\epsilon}{^l^q_{j)}}\,,
\end{equation}
using the abbreviation $V_{mn} = \delta_{mn} - v_nv_m$, and
\begin{equation}
    \bar{\cal B}_{ij} = \tensor{\left(V^{-1}\right)}{_i_j^k^l} \mathfrak{b}_{kl}
    \,,
\end{equation}
where
\begin{eqnarray}
    \mathfrak{b}_{kl} \!\!&=&\!\! \left(\delta_{p(k} {\cal K}^d_{l)} 
    + \zeta \left(\delta^q_p\delta_{kl}-\delta_{p(k}\delta^q_{l)}\right) {\cal P}^d_q \right) {\cal D}_d^p
    - \zeta v_{(k} {\cal E}_{l)}
    \nonumber\\
    \!\!&&\!\!
    - (1+\zeta^2) \delta_{m(k} \tensor{\epsilon}{_{l)}^p^q} \left({\cal P}^{-1}\right)_c^m {\cal P}_p^d \partial_d {\cal P}^c_q\,,
\end{eqnarray}
and
\begin{equation}
    \tensor{\left(V^{-1}\right)}{^r^s_k_l}
    = \frac{1}{2\gamma^2} \left(V^{rs} V_{kl} - 2 V^r_{(k}V^s_{l)}\right)
\end{equation}
is the inverse of $\tensor{V}{^k^l_i_j}$ in the sense that
\begin{equation}
    \tensor{\left(V^{-1}\right)}{^r^s_k_l} \tensor{V}{^k^l_i_j} = \delta^{(r}_k\delta^{s)}_l\,.
\end{equation}

Because we are able to solve the second-class constraints, we do not need Dirac brackets, but only to substitute this solution into the Hamiltonian constraint, obtaining the reduced expression
\begin{widetext}
\begin{eqnarray}
    \bar{H}^{(2)}
    \!\!&=&\!\!
    - \frac{\sqrt{(\det {\cal P})^{-1}}}{\gamma\sqrt{8\pi G}} \Bigg[
    - {\cal P}^a_j \partial_a {\cal E}^j
    + \frac{1}{2} {\cal P}^a_j {\cal E}^j \left({\cal P}^{-1}\right)^k_c \partial_a {\cal P}^c_k
    - \frac{1}{2} {\cal E}^k \partial_a {\cal P}^a_k
    \\
    \!\!&&\!\!\qquad\qquad\qquad\quad
    + \frac{1}{1+\zeta^2} \Bigg( {\cal P}^a_{[j} {\cal P}^b_{k]} {\cal D}_a^j {\cal D}_b^k
    - \left( {\cal P}^a_{(j} v_{k)}
    + \frac{\zeta}{2} {\cal P}^a_l \tensor{\epsilon}{^l_j_k} \right) {\cal D}_a^j {\cal E}^k
    - \frac{1}{4} V_{jk} {\cal E}^j {\cal E}^k
    + \frac{1}{2} \tensor{\epsilon}{_p_q^r} \tensor{\epsilon}{^m^n^s} V_{rs} \bar{\cal B}^p_m \bar{\cal B}^q_n \Bigg)
    \Bigg]\,,\nonumber
\end{eqnarray}
\end{widetext}
and similarly for the other constraints.
We can then work entirely in the reduced phase space spanned by the canonical pairs $({\cal P},{\cal D})$ and $(v,{\cal E})$.
However, the reduced constraints are much less symmetric than in its extended phase-space version, complicating the explicit computation of the brackets.
It is more convenient to work in the extended phase space and instead use Dirac brackets to preserve the second-class constraints under time evolution or gauge transformations as we show below.

\subsection{Dirac brackets}
\label{sec:Dirac brackets}

To formulate the Dirac brackets we first compute the Poisson brackets between the second-class constraints:
\begin{equation}
    \{\mathfrak{K}_{ij}(x),\mathfrak{K}^{kl}(y)\} = 0
    \,,
\end{equation}
\begin{equation}
    \{\mathfrak{K}_{ij}(x),{\cal T}^{kl}(y)\} = - \frac{1}{1+\zeta^2} \tensor{V}{_i_j^k^l} \delta^3(x-y)
    \,,
\end{equation}
and
\begin{eqnarray}
    \{{\cal T}_{ij}(x),{\cal T}^{kl}(y)\}|_{\rm SCS}
    \!\!&=&\!\! \tensor{X}{_i_j^k^l} \delta^3(x-y)
    \label{eq:TT bracket}\\
    \!\!&&\!\!
    + \tensor{Y}{^d_i_j^k^l} \frac{\partial \delta^3(x-y)}{\partial x^d} \,,\nonumber
\end{eqnarray}
where the subscript "SCS" denotes an evaluation on the second-class constraint surface given by $\mathfrak{K}_{ij}=0={\cal T}_{ij}$, and
\begin{equation}
    \tensor{Y}{^d_i_j^k^l} = \frac{2}{1+\zeta^2} v^s \tensor{\epsilon}{_{s}_{(i}^{(k}} \tensor{\epsilon}{^{l)}_{j)}^p} {\cal P}^d_p\,,
\end{equation}
and
\begin{widetext}
\begin{eqnarray}
    \tensor{X}{_i_j^k^l} \!\!&=&\!\! - \frac{1}{1+\zeta^2} \Bigg[ \left(\delta_{ij} \delta^{p(k} \delta_q^{l)}
    - \delta^{kl} \delta^p_{(i} \delta_{j)q}
    \right) \left(\tilde{\cal P}^a_p K_a^q - \tilde{\cal K}^a_p \Gamma_a^q\right)
    + \delta^{(k}_{(i} \tensor{\epsilon}{_{j)}^{l)}^r} \left(G_r+\partial_a\tilde{\cal K}^a_r\right)
    \\
    \!\!&&\!\!\qquad\qquad\quad
    + \left(v^s \tensor{\epsilon}{_s_{(i}^{(k}}
    - \zeta \delta_{(i}^{(k}\right) \delta_{j)r} \tensor{\epsilon}{^{l)}^p^q} \left({\cal P}^{-1}\right)_c^r {\cal P}^d_p \partial_d {\cal P}_q^c
    + \left( v^s \tensor{\epsilon}{_s_{(i}^{(k}}
    + \zeta \delta_{(i}^{(k}\right) \delta^{l)}_{|r|} \tensor{\epsilon}{_{j)}^p^q} \left({\cal P}^{-1}\right)_c^r {\cal P}^d_p \partial_d {\cal P}_q^c \Bigg]
    \,.\nonumber
\end{eqnarray}

The second-class constraint matrix is given by
\begin{eqnarray}
    \tensor{C}{_i_j^k^l}(x,y)
    \!\!&=&\!\!
    \left(\begin{matrix}
        \{\mathfrak{K}_{ij}(x),\mathfrak{K}^{kl}(y)\}
        & \{\mathfrak{K}_{ij}(x),{\cal T}^{kl}(y)\}
        \\
        \{{\cal T}_{ij}(x),\mathfrak{K}^{kl}(y)\}
        & \{{\cal T}_{ij}(x),{\cal T}^{kl}(y)\}
    \end{matrix}\right)\bigg|_{\rm SCS}
    \nonumber\\
    \!\!&=&\!\!
    \left(\begin{matrix}
        0
        & - \frac{1}{1+\zeta^2} \tensor{V}{_i_j^k^l} \delta^3(x-y)
        \\
        \frac{1}{1+\zeta^2} \tensor{V}{_i_j^k^l} \delta^3(x-y)
        & \tensor{X}{_i_j^k^l} \delta^3(x-y) + \tensor{Y}{^d_i_j^k^l} \frac{\partial \delta^3(x-y)}{\partial x^d}
    \end{matrix}\right)
    \,,
\end{eqnarray}
and its inverse by
\begin{eqnarray}
    \tensor{(C^{-1})}{_i_j^k^l}(x,y)
    \!\!&=&\!\!
    \left(\begin{matrix}
        \tensor{H}{_i_j^k^l}(x,y)
        & (1+\zeta^2) \tensor{(V^{-1})}{_i_j^k^l} \delta^3(x-y)
        \\
        - (1+\zeta^2) \tensor{(V^{-1})}{_i_j^k^l} \delta^3(x-y)
        & 0
    \end{matrix}\right)
    \,,
\end{eqnarray}
with
\begin{equation}
    H (x,y) = \left(1+\zeta^2\right)^2 \left[(V^{-1}) X (V^{-1})
    - (V^{-1}) \frac{\partial Y^d}{\partial x^d} (V^{-1}) \right] \delta^3(x-y)
    + \left(1+\zeta^2\right)^2 (V^{-1}) Y^d (V^{-1}) \frac{\partial \delta^3(x-y)}{\partial x^d}\,,
\end{equation}
\end{widetext}
where the expressions are ordered according to the contraction of their internal indices, suppressed for brevity, such that
\begin{eqnarray}
    \!\!&&\!\!
    \int{\rm d}^3z\; \tensor{(C^{-1})}{_i_j^p^q}(x,z) \tensor{C}{_p_q^k^l}(z,y)
    \\
    \!\!&&\!\!\qquad\qquad\qquad\qquad\qquad
    = 
    \left(\begin{matrix}
        1
        && 0
        \\
        0
        && 1
    \end{matrix}\right) \delta^{(k}_i\delta^{l)}_j \delta^3(x-y)\,.\nonumber
\end{eqnarray}

The Dirac bracket, for any phase-space functionals ${\cal O}$ and ${\cal U}$ is given by
\begin{equation}
    \{{\cal O},{\cal U}\}_{\rm D} = \{{\cal O},{\cal U}\}
    - \{{\cal O},{\cal U}\}_{\rm C}|_{\rm SCS}
    \,,
\end{equation}
with correction bracket
\begin{widetext}
\begin{eqnarray}
    \{{\cal O},{\cal U}\}_{\rm C} = \int{\rm d}^3z_1{\rm d}^3z_2
    \left(\begin{matrix}
        \{{\cal O},\mathfrak{K}^{ij}(z_1)\}\\
        \{{\cal O},{\cal T}^{ij}(z_1)\}
    \end{matrix}\right)^{\rm T}
    \tensor{(C^{-1})}{_i_j^k^l}(z_1,z_2)
    \left(\begin{matrix}
        \{\mathfrak{K}_{kl}(z_2),{\cal U}\}\\\{{\cal T}_{kl}(z_2),{\cal U}\}
    \end{matrix}\right)\,.
\end{eqnarray}
\end{widetext}

If ${\cal U}$ is a first-class constraint, then the correction bracket $\{{\cal O},{\cal U}\}_{\rm C}$ is nontrivial on shell---on the first-class and second-class constraint surfaces---only if ${\cal O}$ depends on ${\cal B}_{ij}$.
If ${\cal O}$ too is a first-class constraint, then the correction bracket is proportional to first-class and second-class constraints, and hence the set of constraints given by $G_i$, $L_i$, $H_a$, and $H$ remains first class on the second-class constraint surface when using Dirac brackets.
Furthermore, the above implies that the correction brackets cannot contribute to the covariance conditions of the metric (\ref{eq:Spacetime covariance condition - reduced}) and the tetrad (\ref{eq:Tetrad covariance condition - reduced - 1})-(\ref{eq:Tetrad covariance condition - reduced - 4}), and they can contribute to the covariance condition of the connection (\ref{eq:Connection covariance condition - reduced}) only via its dependence on ${\cal B}_{ij}$.
However, the only possible contribution for the latter is proportional to
\begin{eqnarray}\label{eq:Cov cond T - Second class}
    \frac{\partial\{{\cal T}_{ij},H[\epsilon^{\bar{0}}]\}}{\partial(\partial_c\epsilon^{\bar{0}})} \bigg|_{\rm OS}
    \!\!&=&\!\! 0
    \,,
\end{eqnarray}
which vanishes on shell because ${\cal T}_{ij}$ is related to the torsion tensor by (\ref{eq:Torsion-spatial-sym}):
As a function of the connection and the tetrad, the torsion tensor automatically satisfies the covariance equation
\begin{equation}
    \delta_{\epsilon,\theta,\beta} T^I_{\mu\nu} \big|_{\rm OS} = \mathcal{L}_\xi T^I_{\mu\nu}
    + \delta^{\rm SO(1,3)}_{\theta-\xi^\mu \Gamma_\mu,\beta-\xi^\mu K_\mu} T^I_{\mu\nu} \big|_{\rm OS}
\end{equation}
as generated by the first-class constraints with Poisson brackets because the tetrad and the connection satisfy their own covariance conditions in the same context; in particular, the normal part of this equation for the relevant spatial components implies
\begin{equation}\label{eq:Cov cond T - spatial}
    \frac{\partial\{T_{a b}^i,H[\epsilon^{\bar{0}}]\}}{\partial(\partial_d \epsilon^{\bar{0}})} \bigg|_{\rm OS}
    = - 2 T_{\bar{0} [a}^i \delta_{b]}^d \bigg|_{\rm OS} \,,
\end{equation}
while the transformation of the rest of the phase-space variables in (\ref{eq:Torsion-spatial-sym}) result in terms proportional to $T^k_{ab}$; therefore, both terms vanish on shell if the second-class constraints are imposed---having a fully vanishing torsion---and hence all contributions to (\ref{eq:Cov cond T - Second class}) vanish.
Therefore, all the covariance conditions hold using the Dirac bracket too.

Finally, the Dirac observables (\ref{eq:Lorentz observable}) and (\ref{eq:Gauss observable}) still commute with the first-class constraints using Dirac brackets because they are independent of ${\cal B}_{ij}$ on the second-class constraint surface.

\section{Cosmological constant}
\label{sec:Cosmological}

A cosmological constant can be easily incorporated into the previous analyses.
Its contribution to the action is given by the addition of
\begin{eqnarray}\label{eq:Cosmo action}
    S_\Lambda[e,\omega] \!\!&=&\!\! \int {\rm d}^4x\; \frac{|\det e|}{16\pi G} \left[-2\Lambda\right]
\end{eqnarray}
to (\ref{eq:Holst action}).
This implies a contribution to the Hamiltonian constraint (\ref{eq:Hamiltonian constraint}) given by the addition of
\begin{equation}\label{eq:Cosmo Hamiltonian}
    H_\Lambda = \frac{\sqrt{\det {\cal P}}}{\gamma} \sqrt{8\pi G}\Lambda\,.
\end{equation}

As previously shown, $\sqrt{\det {\cal P}}/\gamma$ is invariant to Lorentz transformations and hence
\begin{eqnarray}
    \{H_\Lambda[N],G_i[\theta^i]\}\!\!&=&\!\!0
    \,,\\
    \{H_\Lambda[N],L_i[\beta^i]\}\!\!&=&\!\!0
    \,.
\end{eqnarray}
This, in turn, implies that $\{\sqrt{\det {\cal P}}/\gamma,H_a[\epsilon^a]\}=\mathcal{L}_{\vec{\epsilon}}\left(\sqrt{\det {\cal P}}/\gamma\right)$ and hence
\begin{equation}
    \{H_\Lambda[N],H_a[\epsilon^a]\} = - H_\Lambda[\epsilon^a\partial_aN]\,.
\end{equation}
Finally, using
\begin{equation}
    \frac{\partial \{\sqrt{\det {\cal P}}/\gamma,H^{(1)}[\epsilon^{\bar{0}}]+H^{(2)}[\epsilon^{\bar{0}}]\}}{\partial (\partial_d \epsilon^{\bar{0}})} = 0\,,
\end{equation}
we obtain
\begin{eqnarray}
    \!\!&&\!\!\{H_\Lambda[N],H^{(1)}[\epsilon^{\bar{0}}]+H^{(2)}[\epsilon^{\bar{0}}]\}
    \\
    \!\!&&\!\!\qquad\qquad
    +\{H^{(1)}[N]+H^{(2)}[N],H_\Lambda[\epsilon^{\bar{0}}]\} = 0\,.\nonumber
\end{eqnarray}

All the above, together with
\begin{equation}
    \{H_\Lambda[N],H_\Lambda[\epsilon^{\bar{0}}]\}=0\,,
\end{equation}
implies that the algebra (\ref{eq:HaHa-Poisson})-(\ref{eq:LL-Poisson}) is preserved when $H_\Lambda$ is incorporated into the Hamiltonian constraint $H=H^{(1)}+H^{(2)}+H_\Lambda$.
Furthermore, because $H_\Lambda$ contains no spatial derivatives, it does not contribute to any of the covariance conditions.
It does not contribute to the second-class constraints either because it does not depend on ${\cal B}_{ij}$.
Finally, its addition preserves the commutation of the Dirac observables (\ref{eq:Lorentz observable}) and (\ref{eq:Gauss observable}) with the first-class constraints.

\section{Geometry and canonical quantization}
\label{sec:Geometry}

\subsection{Area and volume functionals}

The tetrads $e^\mu_I$ constitute four orthonormal vectors---one timelike ($I=0$) and the three spacelike ($I=i=1,2,3$)---and represent a valid dynamical frame (subject to transformations generated by the first-class constraints), such that they may be used to compute a geometric area (internal-)2-form of a two-dimensional surface ${\cal S}$ of the spacetime manifold,
\begin{eqnarray}
    {\cal A}_{IJ}[{\cal S}] = - \int_{\cal S}{\rm d}^2s\, \epsilon_{\mu\nu\alpha\beta} e^\mu_I e^\nu_J w_1^\alpha w_2^\beta
    \,,
\end{eqnarray}
where $w_1$ and $w_2$ coordinatize ${\cal S}$, and
\begin{equation}
    w_1^\alpha = \frac{\partial x^\alpha}{\partial w_1}
    \quad,\quad
    w_2^\beta = \frac{\partial x^\beta}{\partial w_2}\,.
\end{equation}

Consider a coordinate system $x^\mu=(t,x,y,z)$ and a spacelike surface ${\cal S}$ such that $w_a=\tensor{\varepsilon}{_a_b_c} w_1^b w_2^c$ is its spatial co-normal vector, where $\tensor{\varepsilon}{_a_b_c}$ is the Levi--Civita symbol.
In this case, the nontrivial components of the area 2-form are given by
\begin{eqnarray}\label{eq:Spacelike area - spatial contribution}
    {\cal A}_{0i}[{\cal S}] \!\!&=&\!\! 8\pi G \int_{{\cal S}} {\rm d}^2w\, w_a {\cal P}^a_i\,,\\
    {\cal A}_i[{\cal S}] \!\!&=&\!\! 8\pi G \int_{{\cal S}} {\rm d}^2w\, w_a\tensor{\epsilon}{_i^j^k}{\cal P}^a_jv_k\\
    \!\!&=&\!\! 8\pi G \int_{{\cal S}} {\rm d}^2w\, w_a{\cal K}^a_i \bigg|_{\rm SCS}\,,\nonumber
\end{eqnarray}
where we defined ${\cal A}_i = \frac{1}{2} \tensor{\epsilon}{_i^j^k} {\cal A}_{jk}$.
The Lorentz-invariant area is given by the magnitude of the area 2-form,
\begin{equation}\label{eq:Lorentz-invariant area}
    {\cal A}[{\cal S}] = \int_{{\cal S}} {\rm d}^2w\; \sqrt{\tensor{{\cal A}}{_0^i}{\cal A}_{0i}-{\cal A}^i{\cal A}_i}
    \,,
\end{equation}
where ${\cal A}_{0i}$ is the spacelike contribution and ${\cal A}_i$ is the timelike contribution.
In fact, the area functional (\ref{eq:Lorentz-invariant area}) is invariant to the full canonical gauge transformations because it is suitably smeared in its spatial indices: Its local version ${\cal A}(x)=\sqrt{\tensor{{\cal A}}{_0^i}(x){\cal A}_{0i}(x)-{\cal A}^i(x){\cal A}_i(x)}$ constitutes a covariant measure for two-dimensional integrations.
For a spacelike surface ${\cal S}$, the invariant area equals the following expression,
\begin{eqnarray}\label{eq:Lorentz-invariant area - spacelike}
    \!\!{\cal A}[{\cal S}] \!\!&=&\!\!
    8\pi G \int_{{\cal S}} {\rm d}^2w \sqrt{w_aw_b\gamma^{-2}\left(\delta^{ij}+\gamma^2 v^iv^j\right){\cal P}^a_i{\cal P}^b_j}
    \qquad\nonumber\\
    \!\!&=&\!\! 8\pi G \int_{{\cal S}} {\rm d}^2w \sqrt{\delta^{ij}w_aw_b\left({\cal P}^a_i{\cal P}^b_j-{\cal K}^a_i{\cal K}^b_j\right)} \bigg|_{\rm SCS}
    \!.
\end{eqnarray}

Similarly, the geometric volume of a three-dimensional region ${\cal R}$ coordinatized by $w_1$, $w_2$, and $w_3$ is given by
\begin{eqnarray}
    {\cal V}_I[{\cal R}] = - \int_{\cal R}{\rm d}^3w\; \epsilon_{\mu\nu\alpha\beta} e^\mu_I w^\nu_1 w_2^\alpha w_3^\beta
    \,,
\end{eqnarray}
where
\begin{equation}
    w_i^\alpha = \frac{\partial x^\alpha}{\partial w_i}
    \,.
\end{equation}
If ${\cal R}$ is spacelike, then we have the components
\begin{eqnarray}
    {\cal V}_0[{\cal R}] \!\!&=&\!\! (8\pi G)^{3/2} \int_{\cal V}{\rm d}^3w \sqrt{\det{\cal P}}
    \,,\\
    {\cal V}_i[{\cal R}] \!\!&=&\!\! (8\pi G)^{3/2} \int_{\cal R}{\rm d}^3w \sqrt{\det{\cal P}} v^i
    \,,
\end{eqnarray}
and the Lorentz-invariant volume is given by the magnitude,
\begin{eqnarray}
    {\cal V}[{\cal R}] \!\!&=&\!\! \int_{\cal R}{\rm d}^3w \sqrt{{\cal V}_0^2-{\cal V}_i{\cal V}^i}
    \nonumber\\
    \!\!&=&\!\! (8\pi G)^{3/2} \int_{\cal R}{\rm d}^3w \frac{\sqrt{\det{\cal P}}}{\gamma}
    \,.
\end{eqnarray}
This volume functional is invariant to the full canonical gauge transformations: Its local version ${\cal V}(x)=\sqrt{{\cal V}_0(x)^2-{\cal V}_i(x){\cal V}^i(x)}$ constitutes a covariant measure for three-dimensional integrations.

We can now derive the expected Lorentz contraction from canonical gauge transformations.
First, consider a spacelike surface ${\cal S}$ and two observer frames ${\cal O}'$ and ${\cal O}$, with the former being at rest with ${\cal S}$ and moving at a speed $v^i$ with respect to ${\cal O}$.
In the frame ${\cal O}$, we have $v^i|_{\cal O}=0$ and hence ${\cal K}^a_i|_{\cal O} = 0$.
Using
\begin{eqnarray}
    \{ {\cal P}^a_i,L_j[\beta^j]\} \!\!&=&\!\! 
    {\cal K}^a_m \tensor{\epsilon}{^m_n_i} \beta^n
    \,,\\
    \{ {\cal K}^a_i,L_j[\beta^j]\} \!\!&=&\!\! 
    - {\cal P}^a_m \tensor{\epsilon}{^m_n_i} \beta^n
    \,,
\end{eqnarray}
we can obtain the relation of these phase-space variables between the two frames: If $v^i=v \delta^i_1$, then we use the rapidity $\beta^i=\eta \delta^i_1$---so that $\gamma=\cosh\eta$---and hence
\begin{eqnarray}
    {\cal P}^a_1 |_{{\cal O}'} \!\!&=&\!\! e^{\{\cdot,L_j[\beta^j]\}} {\cal P}^a_1 |_{{\cal O}} = {\cal P}^a_1 |_{{\cal O}}
    \,,\\
    {\cal P}^a_2 |_{{\cal O}'} \!\!&=&\!\! e^{\{\cdot,L_j[\beta^j]\}} {\cal P}^a_2 |_{{\cal O}} = \cosh(\eta) {\cal P}^a_2 |_{{\cal O}}
    \,,\\
    {\cal P}^a_3 |_{{\cal O}'} \!\!&=&\!\! e^{\{\cdot,L_j[\beta^j]\}} {\cal P}^a_3 |_{{\cal O}} = \cosh(\eta) {\cal P}^a_3 |_{{\cal O}}
\end{eqnarray}
and
\begin{eqnarray}
    {\cal K}^a_1 |_{{\cal O}'} \!\!&=&\!\! e^{\{\cdot,L_j[\beta^j]\}} {\cal K}^a_1 |_{{\cal O}} = 0
    \,,\\
    {\cal K}^a_2 |_{{\cal O}'} \!\!&=&\!\! e^{\{\cdot,L_j[\beta^j]\}} {\cal K}^a_2 |_{{\cal O}} = \sinh(\eta) {\cal P}^a_3 |_{{\cal O}}
    \,,\\
    {\cal K}^a_3 |_{{\cal O}'} \!\!&=&\!\! e^{\{\cdot,L_j[\beta^j]\}} {\cal K}^a_3 |_{{\cal O}} = - \sinh(\eta) {\cal P}^a_2 |_{{\cal O}}\,.
\end{eqnarray}
Therefore,
\begin{eqnarray}
    {\cal A}_{01}[{\cal S}] \!\!&=&\!\! {\cal A}_{01}'[{\cal S}]
    \,,\\
    {\cal A}_{02}[{\cal S}] \!\!&=&\!\! {\cal A}_{02}'[{\cal S}] / \gamma
    \,,\\
    {\cal A}_{03}[{\cal S}] \!\!&=&\!\! {\cal A}_{03}'[{\cal S}] / \gamma\,,
\end{eqnarray}
where the prime denotes the measured value in ${\cal O}'$, indeed reflects the expected Lorentz contraction of the area components perpendicular to the boost axis.
Because ${\cal K}^a_i$ is non-vanishing in the ${\cal O}'$ frame according to ${\cal O}$, there are timelike contributions to the area 2-form.
However, the magnitude remains Lorentz invariant by use of the hyperbolic functions identity $\cosh^2(\eta)-\sinh^2(\eta)=1$.
A similar calculation goes through for the volume functional.

\subsection{Implications for loop quantum gravity}

A cornerstone of canonical loop quantum gravity \cite{rovelli2004quantum,thiemann2008modern} is the introduction of the holonomy-flux algebra, which provides a mathematical basis for loop quantization.
The traditional Ashtekar--Barbero configuration variable ${\cal A}_a^i$ is related to the new variables on the second-class constraint surface by
\begin{equation}
    {\cal A}_a^i = \Gamma_a^i - \zeta^{-1} K_a^i
    = - \zeta^{-1} A_a^i
    = - \zeta^{-1} {\cal D}_a^i\,,
\end{equation}
while its momentum $E_i^a$ is related to the densitized triad and to the new variables by
\begin{equation}
    E_i^a = - 8 \pi G {\cal P}^a_i\,,
\end{equation}
upon fixing the time gauge, which sets a vanishing velocity $v_i=0$ and turns the full configuration variable $B_a^i$ non-dynamical---see App.~\ref{App:Time gauge} for more details on the phase-space reduction due to the imposition of the time gauge.
These variables satisfy the bracket
\begin{equation}
    \{{\cal A}_a^i(x),E^b_j(y)\} = \frac{8\pi G}{\zeta} \delta^b_a\delta^i_j \delta^3(x-y)\,.
\end{equation}
Notice the $8\pi G/\zeta$ factor compared to a basic bracket.

The holonomy $h_e[{\cal A}]$ is the parallel transport by the connection ${\cal A}_a^i$ along a curve $e$ in the spatial manifold. For a curve $e: [0,1] \mapsto \Sigma$, it is defined as
\begin{equation}\label{eq:Holonomy LQG}
    h_e[{\cal A}] = P \exp \left( \alpha \int_e {\cal A}_a^i \tau_i {\rm d} x^a \right) \in {\rm SU}(2)\,,
\end{equation}
where $P$ denotes path-ordering, and $\tau_i$ are the generators of $\mathfrak{su}(2)$ ($\tau_i = -i \sigma_i / 2$ with $\sigma_i$ being Pauli matrices in the fundamental representation).
For later discussions, we have included a constant parameter $\alpha\in\mathbb{R}$ in the argument of the holonomy, which is undetermined in loop quantum gravity and is typically (and implicitly) fixed to unity.

The flux $E_i[{\cal S}]$ is the densitized triad smeared over a two-dimensional surface ${\cal S}$ in the spatial manifold, defined as
\begin{equation}
    E_i[{\cal S}] = \int_{\cal S} {\rm d}^2w\, w_a E_i^a\,,
\end{equation}
where $w_a$ is the normal to the surface ${\cal S}$, and ${\rm d}^2w$ is the coordinate area element.

The holonomy-flux algebra is the result of applying the classical Poisson bracket to the holonomy and flux functionals,
\begin{equation}\label{eq:Holonomy-flux - LQG}
    \{ h_e[{\cal A}], E_i[{\cal S}] \} = \frac{8\pi \alpha G}{\zeta} \int_e {\rm d} \lambda \dot{e}^a w_a \delta(e(\lambda), {\cal S}) \tau_i h_e[{\cal A}]
    \,,
\end{equation}
where $\delta(e(\lambda), {\cal S})$ has support where the curve $e$ intersects the surface ${\cal S}$ at point $e(\lambda)$, and $\dot{e}^a(\lambda)$ is the tangent vector to the edge parametrized by $\lambda\in[0,1]$.

Adapting the above traditional procedure to the new variables of the extended phase space, we have two connections and hence two different holonomy functionals
\begin{eqnarray}\label{eq:Holonomy A}
    h_e[A] \!\!&=&\!\! P \exp \left( \alpha_1\int_e A_a^i \tau_i {\rm d} x^a \right)
    \,,\\
    h_e[B] \!\!&=&\!\! P \exp \left( \alpha_2\int_e B_a^i \bar{\tau}_i {\rm d} x^a \right)
    \,,\label{eq:Holonomy B}
\end{eqnarray}
with rotation and boost generators $\tau_i$ and $\bar{\tau}_i$, respectively, as well as undetermined constants $\alpha_1,\alpha_2\in\mathbb{R}$; alternatively, one could construct a single holonomy functional
\begin{equation}\label{eq:Holonomy single}
    h_e[A,B] = P \exp \left[\alpha \int_e \left(A_a^i \tau_i + B_a^i \bar{\tau}_i\right) {\rm d} x^a \right]
\end{equation}
with $\alpha\in\mathbb{R}$.
Similarly, we have two flux functionals,
\begin{eqnarray}\label{eq:Flux P}
    {\cal P}_i[{\cal S}] \!\!&=&\!\! \int_{\cal S} {\rm d}^2w\, w_a {\cal P}_i^a \,,\\
    {\cal K}_i[{\cal S}] \!\!&=&\!\! \int_{\cal S} {\rm d}^2w\, w_a {\cal K}_i^a \,.\label{eq:Flux K}
\end{eqnarray}
This implies the extended holonomy-flux algebra
\begin{eqnarray}\label{eq:Holonomy-flux algebra - 1}
    \!\!\{ h_e[A], {\cal P}_i[{\cal S}] \} \!\!&=&\!\! \alpha_1 \int_e {\rm d} \lambda\, \dot{e}^a w_a \delta(e(\lambda), {\cal S}) {\tau}_i h_e[A]
    \,,\qquad\\
    \!\!\{ h_e[B], {\cal K}_i[{\cal S}] \} \!\!&=&\!\! \alpha_2 \int_e {\rm d} \lambda\, \dot{e}^a w_a \delta(e(\lambda), {\cal S}) \bar{\tau}_i h_e[B]
    \,,\qquad\label{eq:Holonomy-flux algebra - 2}
\end{eqnarray}
using the holonomies (\ref{eq:Holonomy A}) and (\ref{eq:Holonomy B}); and similarly if the single holonomy (\ref{eq:Holonomy single}) is used.
Because these variables are connected to the $(K,\tilde{\cal P})$, $(\Gamma,\tilde{\cal K})$ variables by a canonical transformation, the holonomy-flux algebra holds when the connections and their momenta are exchanged for the latter pairs, but it is not clear whether they should share the same generators in the holonomy or if they must be replaced by a suitable linear combination---in fact, use of the latter variables seems more natural because $K_a^i$ and $\Gamma_a^i$ are direct components of the connection 1-form, but the former are closer to the traditional Ashtekar--Barbero variables.
Furthermore, while the holonomy-flux algebra (\ref{eq:Holonomy-flux algebra - 1})-(\ref{eq:Holonomy-flux algebra - 2}) closes using the Poisson bracket of the extended phase space at the kinematical level, this is not the case if Dirac brackets are used instead, so it does not hold on shell.
On the other hand, one could instead work on the reduced phase space and define the holonomies in terms of the corresponding variables.
However, the use of the reduced phase-space variables ${\cal D}_a^i$ and ${\cal E}^i$ is not as natural as the use of $A_a^i$ and $B_a^i$ (or $K_a^i$ and $\Gamma_a^i$) in the holonomies (\ref{eq:Holonomy A}) and (\ref{eq:Holonomy B}) because the former have a complicated relation to both the connection $\omega_a^{IJ}$ and the tetrad $e^a_I$---furthermore, ${\cal E}^i$ is a densitized scalar and would instead require the use of a point-holonomy.

Another central result in canonical loop quantum gravity is the derivation of a discrete spectrum of the area operator \cite{rovelli1995discreteness}, though this is done at the kinematical level.
As a first step, a graph $\Gamma$ with $L$ links and $N$ nodes is introduced in the spatial manifold with the links and nodes broadly denoted by $e$ and $u$.
To each link a spin state $\ket{j,m}$ is assigned, and to each node an intertwiner $\iota^{m_1m_2\dotsi m_\ell}_{k_u}$, where $\ell$ is the number of links connecting to the node and $k_u$ is its intertwine number; a graph with these spin assignments is referred to as a spin network.
Based on this spin network, a wavefunction is introduced in the connection representation; for this to be SU(2) invariant, the wavefunction
\begin{eqnarray}\label{eq:Wavefunction loop}
    \!\!\psi_{j_Lk_{u_N}} (h_e[{\cal A}]) \!\!&=&\!\! \iota^{m_1m_2m_3m_4}_{k_{u_1}} \dotsi \iota^{m_{L-3}m_{L-2}m_{L-1}m_L}_{k_{u_N}}
    \quad\\
    \!\!&&\!\!\times
    D^{j_1}_{m_1n_1}(h_{e_1}[{\cal A}])\dotsi D^{j_L}_{m_Ln_L}(h_{e_L}[{\cal A}])\,,\nonumber
\end{eqnarray}
---where $D^{j}_{mn}$ are the Wigner matrices in the $j$ representation---does not depend directly on the connection but on holonomies integrated along the links of $\Gamma$ with its SU(2) indices contracting those of the intertwiners at the nodes that the respective links connect to.

The flux $E_i[{\cal S}]$ is then promoted to an operator $\hat{E}_i[{\cal S}]$, which acts on spin network states according to the quantization of the holonomy-flux algebra (\ref{eq:Holonomy-flux - LQG}): For a link $e$ piercing the surface ${\cal S}$ at a single point, the action of the flux operator is given by
\begin{equation}
    \hat{E}_i[{\cal S}] \ket{j,m} = -\frac{8\pi \alpha \ell_{\rm P}^2}{\zeta} i \tau_i \ket{j,m}
\end{equation}
if the edge $e$ is oriented up (outgoing relative to the normal $w_a$), where $\ell_{\rm P}=\sqrt{\hbar G}$, and $i\hbar$ enters due to promotion of the Poisson bracket to the usual quantum commutator.
For multiple intersections, the total flux is the sum of the above result over all intersection points.

To obtain the spectrum of the area operator, we must compute the spectrum of the quadratic flux operator; for a single intersection, this is
\begin{equation}
    \hat{E}_i[{\cal S}] \hat{E}^i[{\cal S}]\ket{j,m} = - \left(\frac{8\pi\alpha\ell_P^2}{\zeta}\right)^2 \tau^i \tau_i\ket{j,m}\,.
\end{equation}
Since the SU(2) generators satisfy $-\tau^i \tau_i = C(j)$, where $C(j) = j(j + 1)$ is the Casimir operator for the spin-$j$ representation, we obtain
\begin{equation}
    \sqrt{\hat{E}^i [{\cal S}]\hat{E}_i[{\cal S}]}\ket{j,m} = \frac{8\pi |\alpha| \ell_{\rm P}^2}{|\zeta|} \sqrt{j(j + 1)}\ket{j,m}\,.
\end{equation}
Therefore, the eigenvalues of the area operator (\ref{eq:Lorentz-invariant area - spacelike}) are given by
\begin{equation}\label{eq:Standard area spectrum}
    A[{\cal S}] = \frac{8\pi |\alpha| \ell_P^2}{|\zeta|} \sqrt{j(j + 1)}
\end{equation}
for a single intersection, and the corresponding sum if the surface is punctured multiple times.
Notice that $\alpha$ and $\zeta$ remain in the final result because of the definition of the holonomy (\ref{eq:Holonomy LQG}).
The choice $\alpha=\zeta$ eliminates the appearance of the Barbero--Immirzi parameter in the area spectrum, constituting a quantization ambiguity not because of the value of $\zeta$ as it is commonly stated, but because of the freedom to rescale the holonomy's argument by the parameter $\alpha$.

Having reviewed the traditional ingredients of the kinematical loop quantization, we are now ready to extend such results using the new variables of our larger phase space in a Dirac quantization procedure.
Because we have retained the full Lorentz group, which allows boosted frames, the spin network must specify not only spin parameters $(j,m)$ for unitary representations of SU(2) but also boosting parameters for unitary representations of SL$(2,\mathbb{C})$, which are further labeled by a positive real number $\rho$ and non-negative half-integer $k$.
The Hilbert space $V^{(\rho,k)}$ of the $(\rho,k)$ representation can be decomposed as
\begin{equation}
    V^{(\rho,k)} = \oplus_{j=k}^\infty V_j\,,
\end{equation}
where $V_j$ is a $2j+1$ dimensional space associated to the spin $j$ irreducible representation of SU(2).
We can then choose a basis of states $\ket{\rho,k;j,m}$, where $j=k,k+1,\dots$ and $m=-j,\dots,j$.
The two Casimir operators of SL$(2,\mathbb{C})$ are given by
\begin{eqnarray}\label{eq:Casimir SL(2,C) - 1}
    C^{(1)}=\bar{\tau}_i\bar{\tau}^i-\tau_i\tau^i \!\!&=&\!\! \rho^2-k^2+1
    \,,\\
    C^{(2)}=\bar{\tau}_i\tau^i\!\!&=&\!\! \rho k\,.\label{eq:Casimir SL(2,C) - 2}
\end{eqnarray}

The fluxes (\ref{eq:Flux P}) and (\ref{eq:Flux K}) are then promoted to operators, which act on the states as
\begin{eqnarray}
    \hat{{\cal P}}_i[{\cal S}] \ket{\rho,k;j,m} \!\!&=&\!\! - \alpha_1 \hbar i \tau_i \ket{\rho,k;j,m}
    \,,\\
    \hat{{\cal K}}_i[{\cal S}] \ket{\rho,k;j,m} \!\!&=&\!\! - \alpha_2 \hbar i \bar{\tau}_i \ket{\rho,k;j,m}
    \,,
\end{eqnarray}
at a kinematical level where the holonomy-flux algebra (\ref{eq:Holonomy-flux algebra - 1})-(\ref{eq:Holonomy-flux algebra - 2}) holds and if the states were constructed from the holonomies (\ref{eq:Holonomy A}) and (\ref{eq:Holonomy B}).
In this case, the invariant area eigenvalues are given by
\begin{eqnarray}\label{eq:Area eigenvalue - Lorentz invariant}
    A[{\cal S}] 
    \!\!&=&\!\! 8\pi |\alpha| \ell_P^2 \sqrt{C^{(1)}}
    \nonumber\\
    \!\!&=&\!\! 8\pi |\alpha| \ell_P^2 \sqrt{\rho^2-k^2+1}\,,
\end{eqnarray}
if $\alpha_1=\alpha_2=\alpha$, a choice that can be made under the grounds of obtaining an invariant eigenvalue.
In addition, imposing the second-class constraint $\mathfrak{K}_{ij}=0$ implies that $w_aw_b\delta^{ij}{\cal P}^a_i{\cal K}^a_j=0$, which, promoted to a flux-like operator annihilating physical states, implies
\begin{equation}\label{eq:Second-class constraint - quantum}
    C^{(2)}= \rho k = 0\,.
\end{equation}
To preserve real area eigenvalues, we may set $k=0$, resulting in the area spectrum:
\begin{equation}\label{eq:Area eigenvalue - Lorentz invariant - red}
    A[{\cal S}] = 8\pi|\alpha| \ell_P^2 \sqrt{\rho^2+1}\,.
\end{equation}
(This result differs from the one obtained in \cite{Alexandrov2}, which is based on the SO(4,$\mathbb{C}$) gauge-fixed system.)
However, as we noted previously, the holonomy-flux algebra (\ref{eq:Holonomy-flux algebra - 1})-(\ref{eq:Holonomy-flux algebra - 2}) does not hold on shell where Dirac brackets must be used instead, which would imply corrections to the action of the flux operators in dynamical solutions.

The large quantization ambiguity does not allow us to compute explicitly the final spectrum of the area operator.
Nevertheless, we can still deduce several useful properties as follows.
Being Lorentz invariant, its eigenvalues must be given by a suitable combination of the Casimirs (\ref{eq:Casimir SL(2,C) - 1}) and (\ref{eq:Casimir SL(2,C) - 2}) such that it always yields real numbers.
This was the case for the kinematical result (\ref{eq:Area eigenvalue - Lorentz invariant - red}) restricted to the second-class constraint surface by (\ref{eq:Second-class constraint - quantum}).
This implies a nonzero minimum area eigenvalue given by $8\pi|\alpha| \ell_P^2$.
The spectrum is in principle continuous because of its dependence on $\rho$.
A deeper understanding of the full quantization procedure will be needed to see whether $\rho$ is discretized by other fundamental restrictions such as the solution to the constraints.

In the standard formulation of canonical loop quantum gravity, the time gauge is imposed as a starting point, hence $v^i=0$ and ${\cal K}^a_i=0$; the area operator in such case is necessarily defined for purely spatial surfaces at rest with the observer, in which case it corresponds to the Lorentz-invariant area operator associated to (\ref{eq:Lorentz-invariant area - spacelike}) in that specific gauge and hence its spectrum can indeed match that of the Lorentz-invariant area.
Although it is not clear how this matching could occur without prior gauge fixing due to the quantization ambiguities---the main challenge is that the new kinematical wavefunction depends not only on $A_a^i$ but also on $B_a^i$ and hence it does not reduce to the traditional wavefunction (\ref{eq:Wavefunction loop}) in any obvious way upon fixing the time gauge---it does offer a possibility to understand how loop quantization can imply not necessarily a nonzero minimum spatial area for arbitrary spacelike surfaces, but rather a nonzero minimum invariant area of such surface.
Being Lorentz invariant, its measurement is agreed upon by observers in different boosted frames, though the measurement of the spatial contribution to the area operator associated to (\ref{eq:Spacelike area - spatial contribution}) can yield different results for different frames.

This procedure clarifies a common conceptual critique of loop quantum gravity, which argues that a nonzero minimum area eigenvalue is inconsistent with Lorentz covariance because a boost should contract it to a lower value that is not in the spectrum \cite{AreaCritique}.
This argument is ill-posed in light of the fact that the area operator (\ref{eq:Lorentz-invariant area - spacelike}) is the Lorentz-invariant one and can indeed have a minimum finite value.
The spatial part of the area of a spacelike surface would correspond to the operator version of (\ref{eq:Spacelike area - spatial contribution}), which is not Lorentz invariant and does suffer Lorentz contractions on boosted frames as shown in the previous subsection in the classical context.
The argument against the discreteness of area operators, amended by the above insights, can be rephrased as follows: A measurement of the Lorentz-invariant operator yielding a finite, minimum value is inconsistent with the measurement of the operator corresponding to its spatial contribution in a different frame, which should result in values lower than that of the Lorentz-invariant area.
This argument falls apart as a basic logical fallacy that compares the eigenvalues of two different operators, one that is Lorentz-invariant and one that is not.
(The argument is ill-posed even if translated to the classical context of special relativity because the Lorentz contraction is the result of two different measurements---corresponding to the length of two different curves or the area of two different surfaces in the manifold---which need not yield identical results even if measured by a single observer in the same coordinate system. Therefore, this argument does not constitute a conceptual critique, but a misconception of what the Lorentz contraction is.)
A similar resolution was discussed in \cite{AreaCritique} at a formal level that did not include an explicit Lorentz-boost operator, leaving itself open to a technical version of the critique that would correctly point out the lack of gauge freedom to perform boosts in canonical loop quantum gravity; this is resolved by not fixing the time gauge, hence preserving the Lorentz first-class constraint (operator) to complete the generators of the Lorentz group as shown here.

We stress the observation that there is no zero area eigenvalue in the Lorentz-invariant spectrum (\ref{eq:Area eigenvalue - Lorentz invariant - red}), in contrast to the standard area spectrum (\ref{eq:Standard area spectrum}) with zero spin, $j=0$.
Therefore, the construction of an area operator requires some care in order to have cylindrical consistency: A consistent operator defined in the traditional spin network basis treats an edge with zero spin like a non-existing edge on the graph \cite{rovelli1995discreteness}; in the present case, an additional projector is required to remove zero-spin edges.
(Alternatively, if the generators $\bar{\tau}^i$ and $\tau^i$ are swapped between the holonomies (\ref{eq:Holonomy A}) and (\ref{eq:Holonomy B}), the invariant area spectrum changes to $8\pi |\alpha| \ell_P^2 \sqrt{k^2-\rho^2-1}$.
Real eigenvalues are then obtained by instead setting $\rho=0$ and restricting $k\ge1$, resulting in the discrete spectrum
\begin{equation}\label{eq:Area eigenvalue - Lorentz invariant - red - 2}
    A[{\cal S}] = 8\pi|\alpha| \ell_P^2 \sqrt{k^2-1}\,.
\end{equation}
In this case, zero is in the spectrum for $k=1$, similar to $j=0$ in the traditional result (\ref{eq:Standard area spectrum}).
However, both spectra, the continuous (\ref{eq:Area eigenvalue - Lorentz invariant - red}) and the discrete (\ref{eq:Area eigenvalue - Lorentz invariant - red - 2}), differ from the standard expression (\ref{eq:Standard area spectrum}).)

The previous results can also be used to analyze certain aspects of the spinfoam framework, which works directly with the representations of ${\rm SL}(2,\mathbb{C})$ \cite{spinfoam}.
For instance, in the leading spinfoam model of Engle--Pereira--Rovelli--Livine (EPRL), the Barbero--Immirzi parameter plays a central role in what is referred to as the $Y_\gamma$ map, which is used to connect the model to the canonical approach by embedding the latter's Hilbert space, based on SU(2) spin networks, into the Hilbert space of the spinfoam formulation.
The $Y_\gamma$ map is defined as follows.
For a given SU(2) representation with spin $j$, the $Y_\gamma$ map embeds it into an ${\rm SL}(2,\mathbb{C})$ representation with labels $(\rho, j)$, where the parameter $\rho$ is related to $j$ via the Barbero--Immirzi parameter:
\begin{equation}\label{eq:Boost parameter - Y_gamma}
    \rho = \zeta^{-1} j\,.
\end{equation}
(The $\gamma$ in $Y_\gamma$ refers to the Barbero--Immirzi parameter in the common notation of loop quantum gravity, which in this work we replaced for $\zeta^{-1}$ and reserved $\gamma=1/\sqrt{1-v^2}$ for the usual boosting function.)

More precisely, the $Y_\gamma$ map acts on the SU(2) Hilbert space $V_j$ and maps it to a subspace of the ${\rm SL}(2,\mathbb{C})$ representation space $V^{(\zeta^{-1} j, j)}$ where the simplicity constraint is satisfied:
\begin{eqnarray}
    Y_\gamma \!\!&:&\!\! V_j \to V^{(\zeta^{-1} j, j)}
    \\
    \!\!&:&\!\! \ket{j,m}_{\rm SU(2)}\mapsto\ket{\rho=\zeta^{-1} j, k=j;j,m}_{{\rm SL}(2,\mathbb{C})}
    \,.\nonumber
\end{eqnarray}
In the classical system, the simplicity constraint ensures that the bi-vector $B^{IJ}=(\star e\wedge e)^{IJ}+\zeta e^I\wedge e^J$, where $\star$ is the Hodge dual, satisfies
\begin{equation}\label{eq:Simplicity constraint}
    B^{0i} = - \frac{\zeta^{-1}}{2} \tensor{\epsilon}{^i_j_k} B^{jk}\,.
\end{equation}
The $ Y_\gamma $ map enforces this by selecting the appropriate coherent states in the ${\rm SL}(2,\mathbb{C})$ representation that satisfy this condition.
(The simplicity constraint (\ref{eq:Simplicity constraint}) of the EPRL model is different to what is referred to by the same name in typical Palatini formulations with $BF$ actions, where it is used to ensure that $B^{IJ}_{\mu\nu}$ is a simple bi-vector in terms of the co-tetrad.)

However, given the new understanding of the complete gauge content in our canonical approach, it is clear that the simplicity constraint (\ref{eq:Simplicity constraint}) is not related to any of the first-class nor second-class constraints; therefore, the $Y_\gamma$ map is ill-posed as a fundamental ingredient for quantum gravity.
Indeed, in the variables of our work, the simplicity constraint (\ref{eq:Simplicity constraint}) is given by
\begin{equation}\label{eq:Simplicity constraint - new variables}
    \tilde{\cal K}^a_i=-\zeta \tilde{\cal P}^a_i\,,
\end{equation}
which, after imposing the second-class constraint $\mathfrak{K}_{ij}=0$, and using the definitions (\ref{eq:Momenta Gauss}) and (\ref{eq:Momenta Lorentz}), together with the relation (\ref{eq:KP relation}), it follows that it is equivalent to the time-gauge fixing condition $v_i=0$, such that ${\cal K}^a_i=0$, $\tilde{\cal P}^a_i={\cal P}^a_i$, and hence (\ref{eq:Simplicity constraint - new variables}) holds.
We conclude that the $Y_\gamma$ map breaks Lorentz covariance due to its equivalence to the time-gauge fixing in the canonical formulation.

\section{Conclusions}
\label{sec:Conclusions}

The extended phase-space system we have introduced can be seen as a new gravitational theory with six additional degrees of freedom, implying a geometric system with an enlarged dynamical content of the torsion.
Imposing specific second-class constraints---corresponding to the vanishing of the torsion components (\ref{eq:Torsion-spatial-sym})---the Hilbert--Palatini theory (including the Barbero--Immirzi and cosmological constants) is recovered with the same equations of motion generated by the corresponding Dirac brackets or by the standard Poisson brackets if the phase space is reduced by solving the second-class constraints.
We have shown that the system, both in the extended and in the reduced phase space, is covariant: The canonical gauge transformations correspond to linear combinations of spacetime diffeomorphisms and Lorentz transformations when the equations of motion and the first-class constraints hold regardless of the enforcement of the second-class constraints.

Beyond clarifying the canonical formulation of the tetrad-connection system in the classical context, including an explicit analysis of its gauge content, we have discussed important implications for the loop quantization approach, which in its traditional formulation, as well as in the spinfoam formulation, breaks Lorentz covariance by fixing the time gauge.
We expect that the adoption of the right variables preserving Lorentz covariance presented in this work can open the pathway to a fully covariant canonical quantization of gravity.

\section*{Acknowledgements}
The author thanks Martin Bojowald, Idrus Husin Belfaqih, and Manuel D\'iaz for useful discussions and going through a draft version of this paper, and Sergei Alexandrov for correspondence.
This work was supported in part by NSF grant PHY-2206591.

\appendix

\section{Canonical decomposition}
\label{sec:Canonical decomposition}

\subsection{Strength tensor}
In terms of the canonical variables, the components of the strength tensor are given by
\begin{eqnarray}\label{eq:Fta0i decomposition}
    F_{ta}^{0i}
    \!\!&=&\!\! \dot{K}_a^i - \partial_{a} \omega_{t}^{0i} - \omega_{a}^{ik} \omega_{t}^{0l} \delta_{kl} + \omega_{a}^{0k} \omega_{t}^{il}  \delta_{kl} \nonumber\\
    \!\!&=&\!\! \dot{K}_a^i - \partial_{a} K_t^i - \tensor{\epsilon}{^i_j_k} \left(K_t^j \Gamma_a^k + \Gamma_t^j K_a^k\right)
    \,,
\end{eqnarray}
and
\begin{eqnarray}\label{eq:Ftai decomposition}
    {\cal F}_{ta}^i \!\!&=&\!\!
    \dot{\Gamma}_a^i - \frac{1}{2} \tensor{\epsilon}{^i_k_l} \partial_{a} \omega_{t}^{kl} - \tensor{\epsilon}{^i_k_l} \omega_{a}^{0k} \omega_{t}^{0l} + \tensor{\epsilon}{^i_k_l} \omega_{a}^{pk} \omega_{t}^{ml} \delta_{pm}
    \nonumber\\
    \!\!&=&\!\!
    \dot{\Gamma}_a^i - \partial_{a} \Gamma_t^i - \tensor{\epsilon}{^i_j_k} \left(\Gamma_t^j \Gamma_a^k-K_t^j K_a^k\right)
    \,,
\end{eqnarray}
for the time-space components, and
\begin{eqnarray}
    F_{ab}^{0i}
    \!\!&=&\!\! 2 \partial_{[a} K_{b]}^i - 2 \omega_{[a}^{0k} \omega_{b]}^{il} \delta_{kl}
    \nonumber\\
    \!\!&=&\!\! 2 \partial_{[a} K_{b]}^i - 2 \tensor{\epsilon}{^i_j_k} K_{[a}^j \Gamma_{b]}^k
    \,,
\end{eqnarray}
and
\begin{eqnarray}
    {\cal F}^i_{ab}
    \!\!&=&\!\! 2 \partial_{[a} \Gamma_{b]}^i + \tensor{\epsilon}{^i_k_l} \omega_{[a}^{0k} \omega_{b]}^{0l} - \tensor{\epsilon}{^i_k_l} \omega_{[a}^{kj} \omega_{b]}^{lp} \delta_{jp}
    \nonumber\\
    \!\!&=&\!\! 2 \partial_{[a} \Gamma_{b]}^i + \tensor{\epsilon}{^i_j_k} \left(K_a^j K_b^k - \Gamma_a^j \Gamma_b^k\right)
\end{eqnarray}
for the spatial components.

The relevant linear combinations of the curvature components involving the Barbero--Immirzi parameter can be written more compactly in the $A,B$ variables, given by (\ref{eq:F lin comb - 1}) and (\ref{eq:F lin comb - 2}).
In terms of the $K,\Gamma$ variables, we obtain the following,
\begin{eqnarray}
    F_{ta}^{0i}
    - \frac{\zeta}{2} \tensor{\epsilon}{^i_k_l} F_{ta}^{kl}
    \!\!&=&\!\!
    \dot{K}_a^i
    - \zeta \dot{\Gamma}_a^i - \partial_{a} \left(K_t^i
    - \zeta \Gamma_t^i\right)
    \nonumber\\
    \!\!&&\!\!
    + \left(\Gamma^j
    + \zeta K_a^j \right) \tensor{\epsilon}{^i_j_k} K_t^k
    \nonumber\\
    \!\!&&\!\!
    + \left(K_a^j
    - \zeta \Gamma^j \right) \tensor{\epsilon}{^i_j_k} \Gamma_t^k\,,
\end{eqnarray}
\begin{eqnarray}
    \frac{1}{2} \tensor{\epsilon}{^i_k_l} F_{ta}^{kl}+\zeta F_{ta}^{0i}
    \!\!&=&\!\!
    \dot{\Gamma}_a^i
    + \zeta \dot{K}_a^i - \partial_{a} \left(\Gamma_t^i
    + \zeta K_t^i\right)
    \nonumber\\
    \!\!&&\!\!
    + \left(\Gamma^j + \zeta K_a^j\right) \tensor{\epsilon}{^i_j_k} \Gamma_t^k
    \nonumber\\
    \!\!&&\!\!
    - \left( K_a^j - \zeta \Gamma^j\right) \tensor{\epsilon}{^i_j_k} K_t^k\,,
\end{eqnarray}
\begin{eqnarray}\label{Fab0i-Fabi lin comb BI decomposition 1}
    F^{0i}_{ab}
    - \zeta {\cal F}^i_{ab} \!\!&=&\!\!
    2 \partial_{[a} K_{b]}^i
    - 2 \zeta \partial_{[a} \Gamma_{b]}^i
    \\
    \!\!&&\!\! + \tensor{\epsilon}{^i_j_k} \left[ - 2 K_{[a}^j \Gamma_{b]}^{k}
    + \zeta \left( \Gamma_{[a}^j \Gamma_{b]}^k - K_{[a}^j K_{b]}^k\right)\right]\,,\nonumber
\end{eqnarray}
\begin{eqnarray}\label{Fab0i-Fabi lin comb BI decomposition 2}
    {\cal F}^i_{ab} + \zeta F^{0i}_{ab}
    \!\!&=&\!\!
    2 \partial_{[a} \Gamma_{b]}^i
    + 2 \zeta \partial_{[a} K_{b]}^i
    \\
    \!\!&&\!\!
    + \tensor{\epsilon}{^i_j_k} \left[K_{[a}^j K_{b]}^k - \Gamma_{[a}^j \Gamma_{b]}^k - 2 \zeta K_{[a}^j \Gamma_{b]}^{k}\right]\,.\nonumber
\end{eqnarray}

\subsection{Tetrad}

Using the decomposition of the tetrad components given in Section~\ref{sec:Foliation}, we now gather combinations that are quadratic in the tetrad and relevant for the decomposition of the action.
We get
\begin{eqnarray}
    \frac{N \gamma/(8\pi G)}{\det \varepsilon} \left(e^t_0 e^a_i - e^t_i e^a_0\right)
    \!\!&=&\!\! \frac{\gamma^2}{8\pi G \det \varepsilon} \varepsilon^a_k \left(\delta^k_i-v^kv_i\right) \nonumber\\
    \!\!&=&\!\! {\cal P}^a_i
\end{eqnarray}
and
\begin{eqnarray}
    \frac{|\det e|}{8\pi G} e^t_k e^a_l \tensor{\epsilon}{^k^l_i}
    \!\!&=&\!\! \frac{\gamma^2 \det \varepsilon}{8\pi G} \varepsilon^a_k \tensor{\epsilon}{^k^l_i} v_l \nonumber\\
    \!\!&=&\!\! {\cal K}^a_i
\end{eqnarray}
for time-space components---which determine the momenta---and
\begin{widetext}
\begin{eqnarray}\label{eq:e0[aeb]i decomposition}
    \frac{|\det e|}{8\pi G} e^{[a}_0 e^{b]}_i
    \!\!&=&\!\!
    \frac{\gamma^2}{8\pi G \det \varepsilon} e^{[a}_0 N^{b]} v_i
    + \frac{\gamma N}{8\pi G \det \varepsilon} e^{[a}_0 \varepsilon^{b]}_i
    \nonumber\\
    \!\!&=&\!\!
    - \frac{\gamma^2}{8\pi G \det \varepsilon} N^{[a} \varepsilon^{b]}_k \left(\delta^k_i-v^k v_i\right)
    + \frac{N}{\gamma} \frac{\gamma^2 /(8\pi G)}{\det \varepsilon} \varepsilon^{[a}_i \varepsilon^{b]}_j v^j
    \nonumber\\
    \!\!&=&\!\!
    - N^{[a} {\cal P}^{b]}_i
    + N \frac{\sqrt{(\det {\cal P})^{-1}}}{\gamma\sqrt{8\pi G}} {\cal P}^{[a}_p {\cal P}^{b]}_q \left(\delta^p_i+\gamma^2v^pv_i\right) \left(\delta^q_j+\gamma^2v^qv_j\right) v^j
    \nonumber\\
    \!\!&=&\!\!
    - N^{[a} {\cal P}^{b]}_i
    - N \frac{\sqrt{(\det {\cal P})^{-1}}}{\gamma\sqrt{8\pi G}} \gamma^2 v^q {\cal P}^{[a}_q {\cal P}^{b]}_i\,,
\end{eqnarray}
and
\begin{eqnarray}\label{eq:em[aeb]n decomposition}
    \frac{1}{2} \tensor{\epsilon}{^m^n_i} \frac{|\det e|}{8\pi G} e^{[a}_m e^{b]}_n
    \!\!&=&\!\! \frac{\gamma^2/(8\pi G)}{\det \varepsilon} \left( N^{[a} \varepsilon^{b]}_n \tensor{\epsilon}{^m^n_i} v_m
    + \frac{N}{\gamma} \frac{1}{2} \tensor{\epsilon}{^m^n_i} \varepsilon^{[a}_m \varepsilon^{b]}_n\right)
    \nonumber\\
    \!\!&=&\!\! - N^{[a} {\cal K}^{b]}_i
    + \frac{N}{\gamma} \frac{\gamma^2/(8\pi G)}{\det \varepsilon} \frac{1}{2} \tensor{\epsilon}{^m^n_i} \varepsilon^{[a}_m \varepsilon^{b]}_n
    \nonumber\\
    \!\!&=&\!\! - N^{[a} {\cal K}^{b]}_i
    + N \frac{\sqrt{(\det {\cal P})^{-1}}}{\gamma\sqrt{8\pi G}} \frac{1}{2} \tensor{\epsilon}{^m^n_i} {\cal P}^{[a}_p {\cal P}^{b]}_q \left(\delta^p_m\delta^q_n+\gamma^2\left(v^pv_m\delta^q_n + \delta^p_m v^qv_n\right)\right)
    \nonumber\\
    \!\!&=&\!\! - N^{[a} {\cal K}^{b]}_i
    + N \frac{\sqrt{(\det {\cal P})^{-1}}}{\gamma\sqrt{8\pi G}} \left( \frac{1}{2} {\cal P}^{[a}_p {\cal P}^{b]}_q \tensor{\epsilon}{^p^q_i} - \gamma^2 v^q {\cal P}^{[a}_q {\cal K}^{b]}_i\right)
\end{eqnarray}
\end{widetext}
for spatial components.

The relevant linear combinations of these terms involving the Barbero--Immirzi parameter are given by
\begin{eqnarray}
    \!\!&&\!\!\frac{|\det e|}{8\pi G} \left(e^{[a}_0 e^{b]}_i + \frac{\zeta}{2} \tensor{\epsilon}{^m^n_i} e^{[a}_m e^{b]}_n \right)
    =
    - N^{[a} \tilde{\cal P}^{b]}_i
    \\
    \!\!&&\!\!\qquad
    - N \frac{\gamma\sqrt{(\det {\cal P})^{-1}}}{\sqrt{8\pi G}} \left( v^q {\cal P}^{[a}_q \tilde{\cal P}^{b]}_i
    - \frac{\zeta}{2\gamma^2} {\cal P}^{[a}_p {\cal P}^{b]}_q \tensor{\epsilon}{^p^q_i} \right)
    \nonumber
\end{eqnarray}
and
\begin{eqnarray}
    \!\!&&\!\!\frac{|\det e|}{8\pi G} \left(\frac{1}{2} \tensor{\epsilon}{^m^n_i} e^a_m e^b_n - \zeta e^a_0 e^b_i \right) =
    - N^{[a} \tilde{\cal K}^{b]}_i
    \\
    \!\!&&\!\!\qquad
    - N \frac{\gamma\sqrt{(\det {\cal P})^{-1}}}{\sqrt{8\pi G}}
    \left(v^q {\cal P}^{[a}_q \tilde{\cal K}^{b]}_i
    - \frac{1}{2\gamma^2} {\cal P}^{[a}_p {\cal P}^{b]}_q \tensor{\epsilon}{^p^q_i}\right)
    \nonumber\,.
\end{eqnarray}

\subsection{Action}

Using the decompositions of the previous subsections, we can now decompose the action as
\begin{eqnarray}\label{eq:Holst action canonical 1 - app}
    S[e,\omega]
    \!\!&=&\!\! \int {\rm d}^4x \Bigg[ \tilde{\cal P}^a_i F_{ta}^{0i}
    + \tilde{\cal K}^a_i {\cal F}_{ta}^i
    \\
    \!\!&&\!\!\qquad\quad
    + \frac{|\det e|}{8\pi G} e^a_0 e^b_i \left(F^{0i}_{ab}-\zeta{\cal F}^i_{ab}\right)
    \nonumber\\
    \!\!&&\!\!\qquad\quad
    + \frac{|\det e|}{8\pi G} \frac{1}{2} \tensor{\epsilon}{^m^n_i} e^a_m e^b_n \left({\cal F}^i_{ab}+\zeta F^{0i}_{ab}\right)
    \Bigg]
    \,.\nonumber
\end{eqnarray}

The first line of (\ref{eq:Holst action canonical 1 - app}) can be rewritten as
\begin{equation}
    \int {\rm d}^4x \left[ \tilde{\cal P}^a_i \dot{K}_a^i
    + \tilde{\cal K}^a_i \dot{\Gamma}_a^i\right]
    - L_i \left[ K_t^i \right]
    - G_i \left[ \Gamma_t^i \right]
    \nonumber\,,
\end{equation}
where, upon use of (\ref{eq:Fta0i decomposition}) and (\ref{eq:Ftai decomposition}), we obtain the Lorentz--Gauss constraints (\ref{eq:Lorentz constraint}) and (\ref{eq:Gauss constraint}).

The second and third lines of (\ref{eq:Holst action canonical 1 - app}) can be written as
\begin{equation}
    - H[N] - H_a[N^a]\,,
\end{equation}
where, upon use of (\ref{Fab0i-Fabi lin comb BI decomposition 1}), (\ref{Fab0i-Fabi lin comb BI decomposition 2}), (\ref{eq:e0[aeb]i decomposition}), and (\ref{eq:em[aeb]n decomposition}), we obtain the vector and Hamiltonian constraints (\ref{eq:Vector constraint}) and (\ref{eq:Hamiltonian constraint}).

\subsection{Torsion}

The components of the torsion tensor, upon foliating the manifold, are given by
\begin{eqnarray}
    \tensor{T}{^0_a_b} \!\!&=&\!\! \partial_{[a} (\varepsilon^j_{b]}v_j)
    + K^i_{[a} \varepsilon^j_{b]}
    \,,\\
    \tensor{T}{^i_a_b} \!\!&=&\!\! \partial_{[a} \varepsilon_{b]}^i
    + \left(v_j K^i_{[a}
    + \tensor{\epsilon}{^i^j_k} \Gamma^k_{[a}\right) \varepsilon^j_{b]}
    \,.
\end{eqnarray}
for the purely spatial components.
These are used to derive the expression (\ref{eq:Torsion-spatial-sym}) related to the second-class constraint.

\subsection{Bianchi identity}

In computing the constraint algebra---in particular, the bracket $\{H_a[N^a],H_b[\epsilon^b]\}$---it is useful to have an expression of the Bianchi identity of the strength tensor
\begin{equation}
    D_\alpha F^{IJ}_{\beta\gamma}
    + D_\beta F^{IJ}_{\gamma\alpha}
    + D_\gamma F^{IJ}_{\alpha\beta} = 0
\end{equation}
or, explicitly,
\begin{eqnarray}\label{eq:Bianchi - 1}
    3 \partial_{[\alpha} F^{IJ}_{\beta\gamma]} = - 3 \left( \omega_{[\alpha}^{IK} F^{LJ}_{\beta\gamma]}
    - \omega^{JK}_{[\alpha} F^{LI}_{\beta\gamma]} \right) \eta_{KL}
\end{eqnarray}
in terms of the phase-space variables for the spatial components:
\begin{eqnarray}\label{eq:Bianchi - 2}
    3 \partial_{[a} F^{0i}_{bc]}
    \!\!&=&\!\! 3 \tensor{\epsilon}{^i_j_k} \left(\Gamma^j_{[a} F^{0k}_{bc]} + K_{[a}^j {\cal F}^k_{bc]}\right)
    \,,\\
    3 \partial_{[a} {\cal F}^i_{bc]}
    \!\!&=&\!\! 3 \tensor{\epsilon}{^i_j_k} \left(- K_{[a}^j F^{0k}_{bc]} + \Gamma_{[a}^j {\cal F}^k_{bc]}\right)
    \,.\label{eq:Bianchi - 3}
\end{eqnarray}

\subsection{Lorentz transformations}
\label{app:Lorentz transformations}

The connection 1-form transforms inhomogeneously under proper Lorentz transformations:
\begin{equation}
    \tensor{\omega}{_\mu^I_J} \to
    \tensor{\Lambda}{^I_K} \partial_\mu \tensor{(\Lambda^{-1})}{^K_J}
    + \tensor{\Lambda}{^I_K} \tensor{\omega}{_\mu^K_L} \tensor{(\Lambda^{-1})}{^L_J}\,.
\end{equation}
For an infinitesimal Lorentz transformation $\tensor{\Lambda}{^I_J}\approx\tensor{\delta}{^I_J}+\tensor{\Omega}{^I_J}$ generated by $\tensor{\Omega}{^I_J}$, this is given by
\begin{equation}
    \tensor{\omega}{_\mu^I_J} \to \tensor{\omega}{_\mu^I_J} -
    \partial_\mu \tensor{\Omega}{^I_J}
    + \tensor{\Omega}{^I_K} \tensor{\omega}{_\mu^K_J}
    -\tensor{\omega}{_\mu^I_K} \tensor{\Omega}{^K_J}
    \,.
\end{equation}
Defining
\begin{equation}
    \beta^i = - \Omega^{0i}\quad,\quad \theta^i = - \frac{1}{2} \tensor{\epsilon}{^i_j_k} \Omega^{jk}\,,
\end{equation}
we obtain the transformations
\begin{eqnarray}
    K_t^i &\to& K_t^i +
    \dot{\beta}^i
    + \tensor{\epsilon}{^i_j_k} \left(\theta^j K_t^k + \beta^j \Gamma_t^k\right)
    \,,\\
    \Gamma_t^i &\to& \Gamma_t^i +
    \dot{\theta}^i
    + \tensor{\epsilon}{^i_j_k} \left(\theta^j \Gamma_t^k - \beta^j K_t^k\right)\,,
\end{eqnarray}
for the time components and
\begin{eqnarray}
    K_a^i &\to& K_a^i +
    \partial_a\beta^i
    + \tensor{\epsilon}{^i_j_k} \left( \theta^j K_a^k
    + \beta^j \Gamma_a^k\right)
    \,,\\
    \Gamma_a^i &\to& \Gamma_a^i +
    \partial_a \theta^i
    + \tensor{\epsilon}{^i_j_k} \left(\theta^j \Gamma_a^k - \beta^j K_a^k\right)\,,
\end{eqnarray}
for the spatial components.
Therefore, we obtain
\begin{eqnarray}
    \delta^{\rm SO(1,3)}_{\theta,\beta} K_t^i \!\!&=&\!\! \dot{\beta}^i
    + \tensor{\epsilon}{^i_j_k} \left( \beta^j \Gamma_t^k
    + \theta^j K_t^k\right)
    \,,\\
    \delta^{\rm SO(1,3)}_{\theta,\beta} \Gamma_t^i \!\!&=&\!\!
    \dot{\theta}^i
    + \tensor{\epsilon}{^i_j_k} \left(\theta^j \Gamma_t^k - \beta^j K_t^k\right)\,,
\end{eqnarray}
and
\begin{eqnarray}
    \delta^{\rm SO(1,3)}_{\theta,\beta} K_a^i \!\!&=&\!\! \partial_a\beta^i
    + \tensor{\epsilon}{^i_j_k} \left( \theta^j K_a^k
    + \beta^j \Gamma_a^k\right)
    \,,\\
    \delta^{\rm SO(1,3)}_{\theta,\beta} \Gamma_a^i \!\!&=&\!\!
    \partial_a \theta^i
    + \tensor{\epsilon}{^i_j_k} \left(\theta^j \Gamma_a^k - \beta^j K_a^k\right)\,.
\end{eqnarray}

On the other hand, the tetrad has the simpler transformation
\begin{eqnarray}
    \delta^{\rm SO(1,3)}_{\theta,\beta} e^\mu_I \!\!&=&\!\! -\tensor{\Omega}{^J_I}e^\mu_J
    \\
    \!\!&=&\!\! \left(\delta^0_I e^\mu_i - \delta_{Ii} e^\mu_0\right) \beta^i
    - \delta_{Ii} e^\mu_j \tensor{\epsilon}{^i^j_k} \theta^k
    \,.\nonumber
\end{eqnarray}

\subsection{Lie derivatives}
\label{app:Lie derivatives}

The Lie derivative of the connection 1-form
\begin{eqnarray}
    \mathcal{L}_\xi \omega_\mu^{IJ} = \xi^\nu \partial_\nu \omega_\mu^{IJ} + \omega_\nu^{IJ} \partial_\mu \xi^\nu\,,
\end{eqnarray}
generated by the four-vector $\xi^\mu$, can be written in terms of the gauge functions $(\epsilon^{\bar{0}},\epsilon^a)$ using (\ref{eq:Diffeomorphism generator projection}),
\begin{eqnarray}
    \mathcal{L}_\xi \omega_t^{IJ} \!\!&=&\!\! \left(\xi^t \omega_t^{IJ}\right)^\bullet
    + \xi^a \partial_a \omega_t^{IJ} + \omega_a^{IJ} \dot{\xi}^a
    \\
    \!\!&=&\!\! \left(\xi^\mu \omega_\mu^{IJ}\right)^\bullet
    + \left(\epsilon^a-\frac{\epsilon^{\bar{0}}}{N}N^a\right) \left(\partial_a \omega_t^{IJ}
    - \dot{\omega}_a^{IJ}\right)
    \,,\nonumber\\
    \mathcal{L}_\xi \omega_a^{IJ} \!\!&=&\!\! \frac{\epsilon^{\bar{0}}}{N} \left(\dot{\omega}_a^{IJ}
    - \mathcal{L}_{\vec N} \omega_a^{IJ}\right)
    + \left(\omega_t^{IJ} - N^b \omega_b^{IJ}\right) \partial_a \xi^t
    \nonumber\\
    \!\!&&\!\!
    + \mathcal{L}_{\vec \epsilon} \omega_a^{IJ}\,.
\end{eqnarray}

From this we derive the Lie derivatives in terms of relevant canonical variables and Lorentz transformations:
\begin{eqnarray}
    \mathcal{L}_\xi K_t^i \!\!&=&\!\! \left(\xi^t K_t^i\right)^\bullet
    + \xi^a \partial_a K_t^i + K^i_a \dot{\xi}^a
    \\
    \!\!&=&\!\! \left(\xi^tK_t^i+\xi^a K^i_a\right)^\bullet
    + \xi^a \left(\partial_a K_t^i
    - \dot{K}_a^i\right)
    \nonumber\\
    \!\!&=&\!\! \left(\xi^\mu K_\mu^i\right)^\bullet
    - \xi^a \left(F_{ta}^{0i}-\tensor{\epsilon}{^i_j_k} \left(\Gamma_a^j K_t^k + K_a^j \Gamma_t^k\right)\right)
    \nonumber\\
    \!\!&=&\!\! \left(\xi^\mu K_\mu^i\right)^\bullet
    + \tensor{\epsilon}{^i_j_k} \left(\xi^\mu K_\mu^j \Gamma_t^k+\xi^\mu\Gamma_\mu^j K_t^k\right)
    \nonumber\\
    \!\!&&\!\!
    + \epsilon^a N^b F_{ab}^{0i}
    + \left(\epsilon^{\bar{0}}N^a-N\epsilon^a\right) F_{\bar{0} a}^{0i}
    \nonumber\\
    \!\!&=&\!\! \delta^{\rm SO(1,3)}_{\xi^\mu \Gamma_\mu,\xi^\mu K_\mu} K_t^i
    + \epsilon^a N^b F_{ab}^{0i}
    + \left(\epsilon^{\bar{0}}N^a-N\epsilon^a\right) F_{\bar{0} a}^{0i}
    \,,\nonumber
\end{eqnarray}
\begin{eqnarray}
    \mathcal{L}_\xi \Gamma_t^i \!\!&=&\!\! \left(\xi^t \Gamma_t^i\right)^\bullet
    + \xi^a \partial_a \Gamma_t^i + \Gamma_a^i \dot{\xi}^a
    \\
    \!\!&=&\!\! \left(\xi^\mu\Gamma_\mu^i\right)^\bullet
    + \xi^a \left(\partial_a \Gamma_t^i
    - \dot{\Gamma}_a^i\right)
    \nonumber\\
    \!\!&=&\!\! \left(\xi^\mu\Gamma_\mu^i\right)^\bullet
    + \tensor{\epsilon}{^i_j_k} \left(\xi^\mu\Gamma_\mu^j\Gamma_t^k-\xi^\mu K_\mu^jK_t^k\right)
    - \xi^a {\cal F}_{ta}^i
    \nonumber\\
    \!\!&=&\!\! \delta^{\rm SO(1,3)}_{\xi^\mu\Gamma_\mu,\xi^\mu K_\mu} \Gamma_t^i
    + \epsilon^a N^b{\cal F}_{ab}^i
    + \left(\epsilon^{\bar{0}}N^a-N\epsilon^a\right) {\cal F}_{\bar{0}a}^i
    \nonumber
\end{eqnarray}
for the Lagrange multipliers and
\begin{eqnarray}
    \mathcal{L}_\xi K^i_a \!\!&=&\!\! \frac{\epsilon^{\bar{0}}}{N} \left(\dot{K}_a^i
    - \mathcal{L}_{\vec N} K^i_a\right)
    + \left(K_t^i - N^b K^i_b\right) \partial_a \xi^t
    \nonumber\\
    \!\!&&\!\!
    + \mathcal{L}_{\vec \epsilon} K^i_a
    \,,\\
    \mathcal{L}_\xi \Gamma^i_a \!\!&=&\!\! \frac{\epsilon^{\bar{0}}}{N} \left(\dot{\Gamma}_a^i
    - \mathcal{L}_{\vec N} \Gamma^i_a\right)
    + \left(\Gamma_t^i - N^b \Gamma^i_b\right) \partial_a \xi^t
    \nonumber\\
    \!\!&&\!\!
    + \mathcal{L}_{\vec \epsilon} \Gamma^i_a
\end{eqnarray}
for the configuration variables.

Similarly, the Lie derivative of the tetrad is given by
\begin{eqnarray}
    \mathcal{L}_\xi e^t_I \!\!&=&\!\! \frac{\epsilon^{\bar{0}}}{N} \dot{e}^t_I
    + \left(\epsilon^b-\frac{\epsilon^{\bar{0}}}{N}N^b\right) \partial_b e^t_I
    \\
    \!\!&&\!\!
    - \frac{\epsilon^{\bar{0}}}{N} e^t_I \left(\frac{\dot{\epsilon}^0}{\epsilon^{\bar{0}}} - \frac{\dot{N}}{N}\right)
    - \frac{\epsilon^{\bar{0}}}{N} e^b_I \left(\frac{\partial_b \epsilon^{\bar{0}}}{\epsilon^{\bar{0}}} - \frac{\partial_b N}{N}\right)
    \,,\nonumber
\end{eqnarray}
and
\begin{eqnarray}
    \mathcal{L}_\xi \left(e^a_I + N^a e^t_I\right)
    \!\!&=&\!\! \frac{\epsilon^{\bar{0}}}{N} \left(e^a_I + N^a e^t_I\right)^\bullet
    \nonumber\\
    \!\!&&\!\!
    + e^t_I q^{a b} \left(\epsilon^{\bar{0}} \partial_b N - N \partial_b \epsilon^{\bar{0}} \right)
    \nonumber\\
    \!\!&&\!\!
    - \frac{\epsilon^{\bar{0}}}{N} \mathcal{L}_{\vec{N}} \left(e^a_I+N^a e^t_I\right)
    \nonumber\\
    \!\!&&\!\!
    + \mathcal{L}_{\vec{\epsilon}} \left(e^a_I+N^a e^t_I\right)
    \,.
\end{eqnarray}
And the Lie derivative of the co-tetrad is given by
\begin{eqnarray}
    \mathcal{L}_\xi e_t^0 \!\!&=&\!\! \left(\gamma \epsilon^{\bar{0}}\right)^\bullet
    + v_i \varepsilon_a^i \dot{\epsilon}^a
    + \frac{\epsilon^{\bar{0}}}{N} N^a \left(v_i \varepsilon_a^i\right)^\bullet
    \nonumber\\
    \!\!&&\!\!
    + \left(\epsilon^a-\frac{\epsilon^{\bar{0}}}{N} N^a\right) \partial_a \left(\gamma N
    + v_j N^b \varepsilon_b^j\right)\,,
    \nonumber\\
    \mathcal{L}_\xi e_t^i \!\!&=&\!\! \left(\gamma v^i \epsilon^{\bar{0}}\right)^\bullet
    + \varepsilon^i_a \dot{\epsilon}^a
    + \frac{\epsilon^{\bar{0}}}{N} N^a \dot{\varepsilon}^i_a
    \nonumber\\
    \!\!&&\!\!
    + \left(\epsilon^a-\frac{\epsilon^{\bar{0}}}{N} N^a\right) \partial_a \left(\gamma N v^i
    + N^b \varepsilon_b^i\right)
    \nonumber
\end{eqnarray}
for the time components and
\begin{eqnarray}
    \mathcal{L}_\xi e_a^0 \!\!&=&\!\! \frac{\epsilon^{\bar{0}}}{N} \left(v_i \varepsilon_a^i\right)^\bullet
    + \gamma \epsilon^{\bar{0}} \left(\frac{\partial_a \epsilon^{\bar{0}}}{\epsilon^{\bar{0}}} - \frac{\partial_a N}{N}\right)
    \nonumber\\
    \!\!&&\!\!
    - \frac{\epsilon^{\bar{0}}}{N} \mathcal{L}_{\vec N} \left(v_i \varepsilon_a^i\right)
    + \mathcal{L}_{\vec \epsilon}\, \left(v_i \varepsilon_a^i\right)\,,
    \nonumber\\
    \mathcal{L}_\xi e_a^i \!\!&=&\!\! \frac{\epsilon^{\bar{0}}}{N} \dot{\varepsilon}_a^i
    + \gamma v^i \epsilon^{\bar{0}} \left(\frac{\partial_a \epsilon^{\bar{0}}}{\epsilon^{\bar{0}}} - \frac{\partial_a N}{N}\right)
    \nonumber\\
    \!\!&&\!\!
    - \frac{\epsilon^{\bar{0}}}{N} \mathcal{L}_{\vec N} \varepsilon_a^i
    + \mathcal{L}_{\vec \epsilon}\, \varepsilon_a^i
    \nonumber
    \nonumber
\end{eqnarray}
for the spatial components.

\subsection{Covariance conditions}
\label{sec:Covariance condition}

Using 
\begin{equation}
    \frac{\partial (\gamma^2 v^{(m}v^{n)})}{\partial v^k} =
    2 \gamma^2 v^{(m} \left( \delta^{n)}_k+\gamma^2 v^{n)} v_k\right)
    \,,
\end{equation}
we obtain the relevant derivatives of the inverse spatial metric with respect to the phase-space variables:
\begin{eqnarray}
    \frac{\partial q^{ab}}{\partial v^k} \!\!&=&\!\! \frac{(\det {\cal P})^{-1}}{8\pi G} {\cal P}^a_m {\cal P}^b_n \frac{\partial (\gamma^2 v^{(m}v^{n)})}{\partial v^k}
    \\
    \!\!&=&\!\! 
    2 \gamma^2 \frac{(\det {\cal P})^{-1}}{8\pi G} \left( v^m {\cal P}^{(a}_m {\cal P}^{b)}_k+{\cal P}^{a}_m {\cal P}^{b}_n \gamma^2 v^m v^n v_k\right)
    \nonumber\\
    \!\!&=&\!\! 
    2 \gamma^2 q^{ab}
    \nonumber\\
    \!\!&&\!\!
    + 2 \gamma^2 \frac{(\det {\cal P})^{-1}}{8\pi G} \left( v^m {\cal P}^{(a}_m {\cal P}^{b)}_k-{\cal P}^{a}_m {\cal P}^{b}_n \delta^{mn} v_k\right)
    \nonumber
\end{eqnarray}
as well as
\begin{eqnarray}
    \frac{{\rm d} q^{ab}}{{\rm d} {\cal P}^c_i} &=&
    \frac{\partial q^{ab}}{\partial {\cal P}^c_i}
    + \frac{\partial q^{ab}}{\partial v^k} \frac{\partial v^k}{\partial {\cal P}^c_i} 
    \,,\\
    \frac{\partial q^{ab}}{\partial {\cal K}^c_i} &=& \frac{\partial q^{ab}}{\partial v^k} \frac{\partial v^k}{\partial {\cal K}^c_i}
    \,.
\end{eqnarray}

Because the Hamiltonian constraint contains only first-order spatial derivatives, the bracket $\{q^{ab},H[\epsilon^{\bar{0}}]\}$ can contain at most first-order spatial derivatives of the gauge function $\epsilon^{\bar{0}}$.
Using
\begin{eqnarray}\label{eq:d{P,H}/de0}
    \frac{\partial \{{\cal P}^b_j,H[\epsilon^{\bar{0}}]\}}{\partial (\partial_d \epsilon^{\bar{0}})} \!\!&=&\!\! 2 \frac{\gamma \sqrt{(\det {\cal P})^{-1}}}{\sqrt{8\pi G}} v^q {\cal P}^{[d}_q {\cal P}^{b]}_j
    \,,\\
    \frac{\partial \{{\cal K}^b_j,H[\epsilon^{\bar{0}}]\}}{\partial (\partial_d \epsilon^{\bar{0}})} \!\!&=&\!\!
    2 \frac{\gamma \sqrt{(\det {\cal P})^{-1}}}{\sqrt{8\pi G}} v^q {\cal P}^{[d}_q {\cal K}^{b]}_j
    \\
    \!\!&&\!\!
    + \frac{\gamma\sqrt{(\det {\cal P})^{-1}}}{\sqrt{8\pi G}} \frac{1}{\gamma^2} {\cal P}^b_p {\cal P}^d_q \tensor{\epsilon}{^p^q_j}
    \,,\nonumber
\end{eqnarray}
we get
\begin{eqnarray}\label{eq:d{v,H}/de0}
    \frac{\partial \{v_q,H[\epsilon^{\bar{0}}]\}}{\partial (\partial_d \epsilon^{\bar{0}})} \!\!&=&\!\! \frac{1}{\gamma^2} \frac{\gamma\sqrt{(\det {\cal P})^{-1}}}{\sqrt{8\pi G}} {\cal P}^d_q
    \,,\\
    \frac{\partial \{\det {\cal P},H[\epsilon^{\bar{0}}]\}}{\partial (\partial_d \epsilon^{\bar{0}})} \!\!&=&\!\! 2 \frac{\gamma \sqrt{\det {\cal P}}}{\sqrt{8\pi G}} v^q {\cal P}^d_q
    \,,
\end{eqnarray}
and hence
\begin{equation}
    \frac{\partial \{\gamma,H[\epsilon^{\bar{0}}]\}}{\partial (\partial_d \epsilon^{\bar{0}})} =
    \frac{\gamma^2\sqrt{(\det {\cal P})^{-1}}}{\sqrt{8\pi G}} v^q {\cal P}^d_q
    \,,
\end{equation}
\begin{equation}
    \frac{\partial \{\gamma/\sqrt{\det {\cal P}},H[\epsilon^{\bar{0}}]\}}{\partial (\partial_d \epsilon^{\bar{0}})} =
    0\,.
\end{equation}

Using all the above, it follows that the relevant terms cancel out in the transformation of the structure function such that
\begin{eqnarray}
    \frac{\partial \{q^{ab},H[\epsilon^{\bar{0}}]\}}{\partial (\partial_d \epsilon^{\bar{0}})}
    \!\!&=&\!\! 0\,,
\end{eqnarray}
on and off the second-class and first-class constraint surfaces.
Therefore, the metric's covariance condition (\ref{eq:Spacetime covariance condition}) indeed holds.

Similarly, the relevant part of the tetrad's covariance condition (\ref{eq:Tetrad covariance condition}) is given by the normal transformations.
The time components of this covariance condition reduce to the single equation
\begin{eqnarray}
    \!\!\!\!&&\!\!\frac{\{v_i,H[\epsilon^{\bar{0}}]\}}{\epsilon^{\bar{0}}}
    - \frac{\sqrt{(\det {\cal P})^{-1}}}{\gamma\sqrt{8\pi G}} {\cal P}^b_i \frac{\partial_b \epsilon^{\bar{0}}}{\epsilon^{\bar{0}}} \bigg|_{\rm OS}
    \\
    \!\!\!\!&&\!\!\qquad\qquad\qquad
    = \frac{\{v_i,H[N]\}}{N}
    - \frac{\sqrt{(\det {\cal P})^{-1}}}{\gamma\sqrt{8\pi G}} {\cal P}^b_i \frac{\partial_b N}{N} \bigg|_{\rm OS}
    ,\nonumber
\end{eqnarray}
and the spatial components reduce to
\begin{eqnarray}
    \!\!&&\!\!
    \left(\delta^a_c\delta^k_i
    - \frac{1}{2} {\cal P}^a_i ({\cal P}^{-1})^k_c \right) \frac{\{{\cal P}^c_k,H[\epsilon^{\bar{0}}]\}}{\epsilon^{\bar{0}}}
    \\
    \!\!&&\!\!
    + \frac{\gamma \sqrt{(\det {\cal P})^{-1}}}{\sqrt{8\pi G}} v^j {\cal P}^a_j {\cal P}^b_i \frac{\partial_b \epsilon^{\bar{0}}}{\epsilon^{\bar{0}}} \bigg|_{\rm OS}
    \nonumber\\
    \!\!&&\!\!\qquad\qquad\qquad
    = \left(\delta^a_c\delta^k_i
    - \frac{1}{2} {\cal P}^a_i ({\cal P}^{-1})^k_c \right) \frac{\{{\cal P}^c_k,H[N]\}}{N}
    \nonumber\\
    \!\!&&\!\!\qquad\qquad\qquad\quad
    + \frac{\gamma \sqrt{(\det {\cal P})^{-1}}}{\sqrt{8\pi G}} v^j {\cal P}^a_j {\cal P}^b_i \frac{\partial_b N}{N} \bigg|_{\rm OS}
    \,.\nonumber
\end{eqnarray}
Because these conditions must hold for arbitrary $\epsilon^{\bar{0}}$ and $N$, they imply the equations
\begin{eqnarray}
    \frac{\partial \{v_i,H[\epsilon^{\bar{0}}]\}}{\partial(\partial_c\epsilon^{\bar{0}})} \bigg|_{\rm OS} 
    \!\!&=&\!\! \frac{\sqrt{(\det {\cal P})^{-1}}}{\gamma\sqrt{8\pi G}} {\cal P}^c_i \bigg|_{\rm OS} 
    \,,\\
    \frac{\partial \{v_i,H[\epsilon^{\bar{0}}]\}}{\partial(\partial_{c_1}\partial_{c_2}\epsilon^{\bar{0}})} \bigg|_{\rm OS} 
    \!\!&=&\!\! 0
    \,,
\end{eqnarray}
and
\begin{eqnarray}
    \!\!&&\!\!
    \left(\delta^a_c\delta^k_i
    - \frac{1}{2} {\cal P}^a_i ({\cal P}^{-1})^k_c \right) \frac{\partial\{{\cal P}^c_k,H[\epsilon^{\bar{0}}]\}}{\partial(\partial_d \epsilon^{\bar{0}})} \bigg|_{\rm OS}
    \nonumber\\
    \!\!&&\!\!\qquad=
    - \frac{\gamma \sqrt{(\det {\cal P})^{-1}}}{\sqrt{8\pi G}} v^j {\cal P}^a_j {\cal P}^d_i \bigg|_{\rm OS}
    \,,\\
    \!\!&&\!\!
    \left(\delta^a_c\delta^k_i
    - \frac{1}{2} {\cal P}^a_i ({\cal P}^{-1})^k_c \right) \frac{\partial\{{\cal P}^c_k,H[\epsilon^{\bar{0}}]\}}{\partial(\partial_{d_1} \partial_{d_2} \epsilon^{\bar{0}})} \bigg|_{\rm OS} = 0
    \,.\qquad
\end{eqnarray}
Using (\ref{eq:d{v,H}/de0}) and (\ref{eq:d{P,H}/de0}) we conclude that the covariance condition of the tetrad (\ref{eq:Tetrad covariance condition}) is indeed satisfied.

\section{Constraints}

\subsection{Gauss and Lorentz constraints}

We gather the transformations generated by the Lorentz--Gauss constraints of several phase-space variables:
\begin{eqnarray}
    \{\tilde{\cal P}^a_i,G_j[\theta^j]\} \!\!&=&\!\!
    - \tilde{\cal P}^a_m \tensor{\epsilon}{^m_n_i} \theta^n
    \,,\label{eq:[P~,J]}\\
    \{\tilde{\cal K}^a_i,G_j[\theta^j]\} \!\!&=&\!\! 
    - \tilde{\cal K}^a_m \tensor{\epsilon}{^m_n_i} \theta^n
    \,,
\end{eqnarray}
\begin{eqnarray}
    \{ \tilde{\cal P}^a_i,L_j[\beta^j]\} \!\!&=&\!\! 
    \tilde{\cal K}^a_m \tensor{\epsilon}{^m_n_i} \beta^n
    \,,\label{eq:[P~,L]}\\
    \{ \tilde{\cal K}^a_i,L_j[\beta^j]\} \!\!&=&\!\! 
    - \tilde{\cal P}^a_m \tensor{\epsilon}{^m_n_i} \beta^n
    \,,
\end{eqnarray}
\begin{eqnarray}
    \{K_a^i,G_j[\theta^j]\} \!\!&=&\!\!
    - K^m_a \tensor{\epsilon}{^i_m_n} \theta^n
    \,,\\
    \{\Gamma_a^i,G_j[\theta^j]\} \!\!&=&\!\!
    \partial_a \theta^i
    - \Gamma^m_a \tensor{\epsilon}{^i_m_n} \theta^n
    \,,\label{eq:[G~,J]}
\end{eqnarray}
\begin{eqnarray}
    \{K_a^i,L_j[\beta^j]\} \!\!&=&\!\!
    \partial_a \beta^i
    - \Gamma^m_a \tensor{\epsilon}{^i_m_n} \beta^n
    \,,\\
    \{\Gamma_a^i,L_j[\beta^j]\} \!\!&=&\!\!
    K^m_a \tensor{\epsilon}{^i_m_n} \beta^n
    \,,
\end{eqnarray}
\begin{eqnarray}
    \{ {\cal P}^a_i,L_j[\beta^j]\} \!\!&=&\!\! 
    {\cal K}^a_m \tensor{\epsilon}{^m_n_i} \beta^n
    \,,\\
    \{ {\cal K}^a_i,L_j[\beta^j]\} \!\!&=&\!\! 
    - {\cal P}^a_m \tensor{\epsilon}{^m_n_i} \beta^n
    \,.
\end{eqnarray}

Using the above, we now compute the brackets of relevant combinations:
\begin{eqnarray}
    \{ \sqrt{\det {\cal P}} ,L_j[\beta^j]\}
    \!\!&=&\!\! - \sqrt{\det {\cal P}}\; v_j \beta^j
    \,,\\
    \{v_i , L_j[\beta^j]\} \!\!&=&\!\! - \left( \delta_{ij} - v_i v_j \right) \beta^j
    \,,\\
    \{\sqrt{\det {\cal P}}/\gamma , L_j[\beta^j]\} \!\!&=&\!\! 0
    \,,
\end{eqnarray}
\begin{eqnarray}
    \{ v^q {\cal P}_q^a , L_j[\beta^j]\} \!\!&=&\!\!
    - {\cal P}_q^a \left( \delta^q_j - v^q v_j \right) \beta^j
    + {\cal K}^a_m v^q \tensor{\epsilon}{^m_n_q} \beta^n
    \nonumber\\
    \!\!&=&\!\!
    - \frac{{\cal P}_n^a \beta^n}{\gamma^2}
    \,,
\end{eqnarray}
\begin{equation}
    \{ v^q {\cal K}_q^a , L_j[\beta^j]\} =
    0\,.
\end{equation}

Of particular importance is the transformation of the spatial components of the strength tensor:
\begin{widetext}
\begin{eqnarray}\label{eq:[F0i,J]}
    \{F^{0i}_{ab},G_j[\theta^j]\} \!\!&=&\!\!  
    - 2 \left( \partial_{[a} K^m_{b]} \tensor{\epsilon}{^i_m_n}
    + K^i_{[a} \Gamma_{b]}^k \delta_{kn}
    - K_{[a}^k \Gamma^i_{b]} \delta_{kn}\right) \theta^n
    = - F^{0m}_{ab} \tensor{\epsilon}{^i_m_n} \theta^n
    \,,\\
    \{F^{0i}_{ab},L_j[\beta^j]\} \!\!&=&\!\!
    - 2 \left( \partial_{[a} \Gamma^m_{b]} \tensor{\epsilon}{^i_m_n}
    + K_{[a}^k K^i_{b]} \delta_{kn}
    - \Gamma^k_{[a} \Gamma_{b]}^i \delta_{nk} \right) \beta^n
    = - {\cal F}^m_{ab} \tensor{\epsilon}{^i_m_n} \beta^n
    \,,\\
    \{{\cal F}^i_{ab},G_j[\theta^j]\} \!\!&=&\!\! 
    - 2 \left( \partial_{[a} \Gamma^m_{b]} \tensor{\epsilon}{^i_m_n}
    + K_{[a}^k K^i_{b]} \delta_{kn}
    - \Gamma_{[a}^k \Gamma^i_{b]} \delta_{kn}
    \right) \theta^n
    = - {\cal F}^m_{ab} \tensor{\epsilon}{^i_m_n} \theta^n
    \,,\\
    \{{\cal F}^i_{ab},L_j[\beta^j]\} \!\!&=&\!\! 
    2 \left(\partial_{[a} K^m_{b]} \tensor{\epsilon}{^i_m_n}
    + K^i_{[a} \Gamma_{b]}^k \delta_{kn}
    - K_{[a}^k \Gamma^i_{b]} \delta_{kn}
    \right) \beta^n
    = F^{0m}_{ab} \tensor{\epsilon}{^i_m_n} \beta^n
    \,,\label{eq:[Fi,L]}
\end{eqnarray}
\end{widetext}
where we used
\begin{eqnarray*}
    F^{0m}_{ab} \tensor{\epsilon}{^i_m_n} \!\!&=&\!\! 2 \left(\partial_{[a} K_{b]}^m \tensor{\epsilon}{^i_m_n}
    + K_{[a}^i \Gamma_{b]}^k \delta_{kn}
    - K_{[a}^k \Gamma_{b]}^i \delta_{kn}\right)
    \,,\\
    {\cal F}^m_{ab} \tensor{\epsilon}{^i_m_n} \!\!&=&\!\! 2 \left(\partial_{[a} \Gamma_{b]}^m \tensor{\epsilon}{^i_m_n}
    + K_{[a}^k K_{b]}^i \delta_{kn}
    - \Gamma_{[a}^k \Gamma_{b]}^i \delta_{kn}\right)\,.
\end{eqnarray*}

\subsection{Vector constraint}

Similarly, the transformation of basic phase-space variables generated by the vector constraint are given by
\begin{eqnarray}
    \{ K_a^i, H_c[N^c]\} \!\!&=&\!\! N^c F^{0i}_{ca}
    \,,\\
    \{\Gamma_a^i, H_c[N^c]\} \!\!&=&\!\! N^c {\cal F}^i_{ca}
    \,,
\end{eqnarray}
and
\begin{widetext}
\begin{eqnarray}
    \{\tilde{\cal P}^a_i , H_c[N^c]\} \!\!&=&\!\! 
    \mathcal{L}_{\vec{N}} \tilde{\cal P}^a_i
    - N^a \partial_c \tilde{\cal P}^c_i
    - 2 N^{[a} \left( 
    \tilde{\cal P}^{c]}_m \Gamma_c^n
    - \tilde{\cal K}^{c]}_m K_c^n \right) \tensor{\epsilon}{^m_n_i}
    \nonumber\\
    \!\!&=&\!\! 
    \mathcal{L}_{\vec{N}} \tilde{\cal P}^a_i
    + N^a L_i
    + N^c \left( 
    \tilde{\cal P}^a_m \Gamma_c^n
    - \tilde{\cal K}^a_m K_c^n \right) \tensor{\epsilon}{^m_n_i}
    \,,\\
    \{\tilde{\cal K}^a_i , H_c[N^c]\} \!\!&=&\!\!
    \mathcal{L}_{\vec{N}} \tilde{\cal K}^a_i
    - N^a \partial_c \tilde{\cal K}^c_i
    - 2 N^{[a} \left( 
    \tilde{\cal P}^{c]}_m K_c^n
    + \tilde{\cal K}^{c]}_m \Gamma_c^n\right) \tensor{\epsilon}{^m_n_i}
    \nonumber\\
    \!\!&=&\!\!
    \mathcal{L}_{\vec{N}} \tilde{\cal K}^a_i
    + N^a G_i
    + N^c \left( 
    \tilde{\cal P}^a_m K_c^n
    + \tilde{\cal K}^a_m \Gamma_c^n\right) \tensor{\epsilon}{^m_n_i}\,.
\end{eqnarray}
Using this, we obtain the corresponding transformation of the strength tensor components
\begin{eqnarray}
    \{F^{0i}_{ab},H_c[N^c]\}
    \!\!&=&\!\!
    \mathcal{L}_{\vec N} F^{0i}_{ab}
    - N^c \left(\partial_a F^{0i}_{bc} + \partial_c F^{0i}_{ab} + \partial_b F^{0i}_{ca}\right)
    + 2 N^c \left(\Gamma_{[a}^j F^{0k}_{b]c} + K_{[a}^j {\cal F}^k_{b]c} \right) \tensor{\epsilon}{^i_j_k}
    \nonumber\\
    \!\!&=&\!\!
    \mathcal{L}_{\vec N} F^{0i}_{ab}
    - 2 N^c \tensor{\epsilon}{^i_j_k} \left( \Gamma_{[b}^{j} F^{0k}_{c]a}
    + \frac{1}{2} \Gamma_b^{j} F^{0k}_{ac}
    + K_{[b}^j {\cal F}^k_{c]a}
    + \frac{1}{2} K_b^j {\cal F}^k_{ac}
    \right)
    \,,\\
    \{{\cal F}^i_{ab},H_c[N^c]\}
    \!\!&=&\!\!
    \mathcal{L}_{\vec N} {\cal F}^i_{ab}
    - N^c \left(\partial_a {\cal F}^i_{bc} + \partial_c {\cal F}^i_{ab} + \partial_b {\cal F}^i_{ca}\right)
    - 2 N^c \left(K_{[a}^j F^{0k}_{b]c} - \Gamma_{[a}^j {\cal F}^k_{b]c}\right) \tensor{\epsilon}{^i_j_k}
    \nonumber\\
    \!\!&=&\!\!
    \mathcal{L}_{\vec N} {\cal F}^i_{ab}
    - 2 N^c \tensor{\epsilon}{^i_j_k} \left(
    - K_{[b}^{j} F^{0k}_{c]a}
    - \frac{1}{2} K_b^{j} F^{0k}_{ac}
    + \Gamma_{[b}^j {\cal F}^k_{c]a}
    + \frac{1}{2} \Gamma_b^j {\cal F}^k_{ac} \right)\,,
\end{eqnarray}
where we used the Bianchi identities (\ref{eq:Bianchi - 2}) and (\ref{eq:Bianchi - 3}).

It follows that
\begin{eqnarray}\label{eq:Comb Ha - 1}
    \{N^a\tilde{\cal P}^b_i F^{0i}_{ab},H_c[\epsilon^c]\}
    \!\!&=&\!\!
    N^a \mathcal{L}_{\vec \epsilon} \left(\tilde{P}^b_iF^{0i}_{ab}\right)
    - N^a \epsilon^b \partial_c \tilde{\cal P}^c_i F^{0i}_{ab}
    - 2 N^a \epsilon^c \tilde{P}^b_i \tensor{\epsilon}{^i_j_k} \left( \frac{1}{2} \Gamma_b^{j} F^{0k}_{ac}
    + K_{[b}^j {\cal F}^k_{c]a}
    + \frac{1}{2} K_b^j {\cal F}^k_{ac}
    \right)
    \nonumber\\
    \!\!&&\!\!
    - 2 N^a \epsilon^c \tilde{\cal K}^b_i K_{[b}^j F^{0k}_{c]a}\tensor{\epsilon}{^i_j_k}
    \,,\\
    \{N^a \tilde{\cal K}^b_i {\cal F}^i_{ab},H_c[\epsilon^c]\}
    \!\!&=&\!\!
    N^a \mathcal{L}_{\vec{\epsilon}} \left(\tilde{\cal K}^b_i {\cal F}^i_{ab}\right)
    - N^a \epsilon^b \partial_c \tilde{\cal K}^c_i {\cal F}^i_{ab}
    - 2 N^a \epsilon^c \tilde{\cal K}^b_i \tensor{\epsilon}{^i_j_k} \left(
    - K_{[b}^{j} F^{0k}_{c]a}
    - \frac{1}{2} K_b^{j} F^{0k}_{ac}
    + \frac{1}{2} \Gamma_b^j {\cal F}^k_{ac} \right)
    \nonumber\\
    \!\!&&\!\!
    + 2 N^a \epsilon^c \tilde{\cal P}^b_i K_{[b}^j {\cal F}^k_{c]a} \tensor{\epsilon}{^i_j_k}\,.\label{eq:Comb Ha - 2}
\end{eqnarray}
\end{widetext}

\subsection{Hamiltonian constraint}
\label{sec:Hamiltonian constraint - transf}

Relevant transformations of basic phase-space variables generated by the second component of the Hamiltonian constraint are given by
\begin{eqnarray}
    \{A_a^i, H^{(2)}[N]\} \!\!&=&\!\! - N \frac{\sqrt{\det {\cal P}}}{\gamma\sqrt{8\pi G}} {\cal P}^b_q \tensor{\epsilon}{^i^q_k} \left({\cal F}^k_{cd} + \zeta F^{0k}_{cd}\right)
    \nonumber\\
    \!\!&&\!\!
    + \frac{1}{\sqrt{\det {\cal P}}} \frac{\partial \sqrt{\det {\cal P}}}{\partial {\cal P}^a_i} H^{(2)} N
    \\
    \!\!&&\!\!
    + \frac{\gamma^2 v^2}{2} \left(\delta^i_m-\frac{v_mv^i}{v^2}\right) \left({\cal P}^{-1}\right)^m_a H^{(2)} N
    \,,\nonumber
\end{eqnarray}
\begin{eqnarray}
    \{B_a^i, H^{(2)}[N]\} \!\!&=&\!\! N \frac{1}{2} \gamma^2 v^k \tensor{\epsilon}{_k_m^i} \left({\cal P}^{-1}\right)^m_a H^{(2)}
    \,,
\end{eqnarray}
\begin{eqnarray}
    \{{\cal K}^a_i, H^{(2)}[N]\}
    \!\!&=&\!\! - \partial_c \left(N \frac{\sqrt{(\det {\cal P})^{-1}}}{\gamma\sqrt{8\pi G}} {\cal P}^c_p {\cal P}^a_q \tensor{\epsilon}{^p^q_i} \right)
    \nonumber\\
    \!\!&&\!\!
    - 2 N \frac{\sqrt{(\det {\cal P})^{-1}}}{\gamma\sqrt{8\pi G}} {\cal P}^d_{[p} {\cal P}^a_{i]} \Gamma_d^p
    \,,
\end{eqnarray}
\begin{eqnarray}
    \{{\cal P}^a_i,H^{(2)}[N]\}
    \!\!&=&\!\!
    2 N \frac{\sqrt{(\det {\cal P})^{-1}}}{\gamma\sqrt{8\pi G}} {\cal P}^d_{[p} {\cal P}^a_{i]} K_d^p\,.
\end{eqnarray}

Therefore,
\begin{widetext}
\begin{eqnarray}
    \{{\cal P}^a_i,H[N]\} \!\!&=&\!\!
    2 \partial_c \left(N \frac{\sqrt{(\det {\cal P})^{-1}}}{\gamma\sqrt{8\pi G}} \gamma^2 v^q {\cal P}^{[c}_q {\cal P}^{a]}_i \right)
    \\
    \!\!&&\!\!
    - 2 N \frac{\sqrt{(\det {\cal P})^{-1}}}{\gamma\sqrt{8\pi G}} \frac{1}{1+\zeta^2} \tensor{\epsilon}{^m_l_i} \Bigg[ \gamma^2 v^q {\cal P}^{[a}_q {\cal P}^{b]}_m \left(B_b^l-\zeta A_b^l\right)
    - \left(\gamma^2 v^q {\cal P}^{[a}_q {\cal K}^{b]}_m
    - \frac{1}{2} {\cal P}^a_p {\cal P}^b_q \tensor{\epsilon}{^p^q_m}\right) \left( A_b^l
    + \zeta B_b^l \right)
    \Bigg]
    \,,\nonumber
\end{eqnarray}
\begin{eqnarray}
    \{{\cal K}^a_i,H[N]\} \!\!&=&\!\!
    2 \partial_c \left(N \frac{\sqrt{(\det {\cal P})^{-1}}}{\gamma\sqrt{8\pi G}} \left(\gamma^2 v^q {\cal P}^{[c}_q {\cal K}^{a]}_i
    - \frac{1}{2} {\cal P}^c_p {\cal P}^a_q \tensor{\epsilon}{^p^q_i}\right) \right)
    \\
    \!\!&&\!\!
    - 2 N \frac{\sqrt{(\det {\cal P})^{-1}}}{\gamma\sqrt{8\pi G}} \frac{1}{1+\zeta^2} \tensor{\epsilon}{^m_l_i} \Bigg[ \gamma^2 v^q {\cal P}^{[a}_q {\cal P}^{b]}_m \left(A_b^l + \zeta B_b^l\right)
    + \left(\gamma^2 v^q {\cal P}^{[a}_q {\cal K}^{b]}_m
    - \frac{1}{2} {\cal P}^{[a}_p {\cal P}^{b]}_q \tensor{\epsilon}{^p^q_m}\right) \left(B_b^l - \zeta A_b^l \right)
    \Bigg]\,.\nonumber
\end{eqnarray}
\end{widetext}

\subsection{Constraint brackets}
\label{app:Constraint brackets}

Using (\ref{eq:[P~,J]})-(\ref{eq:[G~,J]}), the brackets (\ref{eq:JJ})-(\ref{eq:LL}) are straightforward to show.

Using (\ref{eq:[P~,J]})-(\ref{eq:[P~,L]}) and (\ref{eq:[F0i,J]})-(\ref{eq:[Fi,L]}), it follows that the Lorentz--Gauss constraints commute with the vector constraint, hence showing (\ref{eq:HaJ}) and (\ref{eq:HaL}).

Using (\ref{eq:Comb Ha - 1}) and (\ref{eq:Comb Ha - 2}), together with the Bianchi identity components (\ref{eq:Bianchi - 1}) and (\ref{eq:Bianchi - 2}), the bracket of the vector constraint with itself yields
\begin{eqnarray}
    \!\!&&\!\!
    \{H_a[N^a],H_c[M^c]\}
    \nonumber\\
    \!\!&&\!\!\qquad
    = - H_a \left[\mathcal{L}_{\vec M} N^a\right]
    - \int{\rm d}^3x N^a M^b \bigg[F^{0i}_{ab} \partial_c \tilde{\cal P}^c_i
    \nonumber\\
    \!\!&&\!\!\qquad\qquad\quad
    + \tilde{P}^c_i \tensor{\epsilon}{^i_j_k} \left( \Gamma_c^{j} F^{0k}_{ab}
    + K_c^j {\cal F}^k_{ab}
    \right)
    \nonumber\\
    \!\!&&\!\!\qquad\qquad\quad
    + {\cal F}^i_{ab} \partial_c \tilde{\cal K}^c_i
    + \tilde{\cal K}^c_i \tensor{\epsilon}{^i_j_k} \left(\Gamma_c^j {\cal F}^k_{ab} - K_c^{j} F^{0k}_{ab}\right)\bigg]
    \nonumber\\
    \!\!&&\!\!\qquad
    = - H_a \left[\mathcal{L}_{\vec M} N^a\right]
    + L_i[N^a M^b F^{0i}_{ab}]
    \nonumber\\
    \!\!&&\!\!\qquad\quad
    + G_i[N^a M^b {\cal F}^i_{ab}]
    \,,
\end{eqnarray}
which shows (\ref{eq:HaHa}).

To compute $\{H[N],H[\epsilon^{\bar{0}}]\}$, we proceed in steps.
First, using the transformations generated by the vector constraint, we obtain
\begin{eqnarray}
    \!\!\!\!&&\!\!
    \{H^{(1)}[N],H^{(1)}[\epsilon^{\bar{0}}]\}
    \nonumber\\
    \!\!\!\!&&\!\!\qquad
    =
    \int{\rm d}^3x N  \frac{\gamma \sqrt{(\det {\cal P})^{-1}}}{\sqrt{8\pi G}} v^q {\cal P}^a_q \{H_a,H^{(1)}[\epsilon^{\bar{0}}]\}
    \nonumber\\
    \!\!\!\!&&\!\!\qquad\quad
    + H_a\left[N\{ \frac{\gamma \sqrt{(\det {\cal P})^{-1}}}{\sqrt{8\pi G}} v^q {\cal P}^a_q,H^{(1)}[\epsilon^{\bar{0}}]\}\right]
    \nonumber\\
    \!\!\!\!&&\!\!\qquad
    = - H_a\left[ \frac{(\det {\cal P})^{-1}}{8\pi G} \gamma^2 v^pv^q {\cal P}^a_p {\cal P}^b_q \left(\epsilon^{\bar{0}}\partial_bN-N\partial_b\epsilon^{\bar{0}}\right) \right]
    \nonumber\\
    \!\!\!\!&&\!\!\qquad\quad
    - H_a\left[\epsilon^{\bar{0}}\{\frac{\gamma \sqrt{(\det {\cal P})^{-1}}}{\sqrt{8\pi G}} v^q {\cal P}^a_q,H^{(1)}[N]\}\right]
    \nonumber\\
    \!\!\!\!&&\!\!\qquad\quad
    + H_a\left[N\{ \frac{\gamma \sqrt{(\det {\cal P})^{-1}}}{\sqrt{8\pi G}} v^q {\cal P}^a_q,H^{(1)}[\epsilon^{\bar{0}}]\}\right]
    \nonumber\\
    \!\!\!\!&&\!\!\qquad\qquad = 0
    \,.
\end{eqnarray}
and
\begin{eqnarray}
    \!\!&&\!\!\{H^{(2)}[N],H^{(1)}[\epsilon^{\bar{0}}]\}
    \\
    \!\!&&\!\!\qquad
    = H_a \left[\{H^{(2)}[N],\epsilon^{\bar{0}} \frac{\gamma \sqrt{(\det {\cal P})^{-1}}}{\sqrt{8\pi G}} v^q {\cal P}^a_q\}\right]
    \nonumber\\
    \!\!&&\!\!\qquad\quad
    + \int{\rm d}^3y\;\{H^{(2)}[N],H_a\} \epsilon^{\bar{0}} \frac{\gamma \sqrt{(\det {\cal P})^{-1}}}{\sqrt{8\pi G}} v^q {\cal P}^a_q
    \nonumber\\
    \!\!&&\!\!\qquad
    = - H^{(2)}\left[\mathcal{L}_{\epsilon^{\bar{0}} \frac{\gamma \sqrt{(\det {\cal P})^{-1}}}{\sqrt{8\pi G}} v^q {\cal P}^a_q} N\right]
    \nonumber\\
    \!\!&&\!\!\qquad\quad
    - H_a \left[\epsilon^{\bar{0}}\{\frac{\gamma \sqrt{(\det {\cal P})^{-1}}}{\sqrt{8\pi G}} v^q {\cal P}^a_q,H^{(2)}[N]\}\right]
    \nonumber\\
    \!\!&&\!\!\qquad\quad
    + G_k\left[\frac{\gamma \sqrt{(\det {\cal P})^{-1}}}{\sqrt{8\pi G}} v^q {\cal P}^a_q \epsilon^{\bar{0}}\{\Gamma^k_a,H^{(2)}[N]\}\right]
    \nonumber\\
    \!\!&&\!\!\qquad\quad
    + L_k\left[\frac{\gamma \sqrt{(\det {\cal P})^{-1}}}{\sqrt{8\pi G}} v^q {\cal P}^a_q \epsilon^{\bar{0}}\{K^k_a,H^{(2)}[N]\}\right]\,.\nonumber
\end{eqnarray}

It is useful to define the operation
\begin{equation}
    \delta^{(2)}_{\epsilon^{\bar{0}},N} {\cal O} = \epsilon^{\bar{0}} \{{\cal O},H^{(2)}[N]\} - N \{{\cal O},H^{(2)}[\epsilon^{\bar{0}}]\}\,.
\end{equation}
Using the transformations of App.~\ref{sec:Hamiltonian constraint - transf}, we obtain
\begin{equation}
    \delta^{(2)}_{\epsilon^{\bar{0}},N} {\cal P}^a_i=0\,,
\end{equation}
\begin{equation}
    \delta^{(2)}_{\epsilon^{\bar{0}},N} {\cal K}^b_n
    = - \frac{\sqrt{(\det {\cal P})^{-1}}}{\sqrt{8\pi G}} \frac{{\cal P}^a_p {\cal P}^b_q \tensor{\epsilon}{^p^q_n}}{\gamma} \left(\epsilon^{\bar{0}}\partial_aN-N\partial_a\epsilon^{\bar{0}}\right)
    \,,
\end{equation}
\begin{equation}
    \delta^{(2)}_{\epsilon^{\bar{0}},N} v_q = \frac{\sqrt{(\det {\cal P})^{-1}}}{\sqrt{8\pi G}} \frac{1}{\gamma} {\cal P}^b_q \left(\epsilon^{\bar{0}}\partial_bN-N\partial_b\epsilon^{\bar{0}}\right)\,,
\end{equation}
and hence
\begin{eqnarray}
    \delta^{(2)}_{\epsilon^{\bar{0}},N} \gamma \!\!&=&\!\! 
    \delta^{(2)}_{\epsilon^{\bar{0}},N} \gamma
    = \gamma^3 v^q\delta^{(2)}_{\epsilon^{\bar{0}},N} v_q
    \\
    \!\!&=&\!\! \frac{\sqrt{(\det {\cal P})^{-1}}}{\sqrt{8\pi G}} \gamma^2 v^q {\cal P}^b_q \left(\epsilon^{\bar{0}}\partial_bN-N\partial_b\epsilon^{\bar{0}}\right)\,.\nonumber
\end{eqnarray}

Using the above results we can now compute the following brackets,
\begin{eqnarray}
    \!\!&&\!\!\{H^{(2)}[N],H^{(1)}[\epsilon^{\bar{0}}]\}+\{H^{(1)}[N],H^{(2)}[\epsilon^{\bar{0}}]\}
    \\
    \!\!&&\!\!
    \qquad= - H^{(2)} \left[\frac{\sqrt{(\det {\cal P})^{-1}}}{\sqrt{8\pi G}} \gamma v^q {\cal P}^a_q (\epsilon^{\bar{0}}\partial_a N-N\partial_a \epsilon^{\bar{0}})\right]
    \nonumber\\
    \!\!&&\!\!\qquad\quad
    - H_a\left[\frac{\sqrt{(\det {\cal P})^{-1}}}{\sqrt{8\pi G}} {\cal P}^a_q\delta^{(2)}_{\epsilon^{\bar{0}},N} (\gamma v^q)\right]
    \nonumber\\
    \!\!&&\!\!
    \qquad= - H^{(2)} \left[\frac{\sqrt{(\det {\cal P})^{-1}}}{\sqrt{8\pi G}} \gamma v^q {\cal P}^a_q (\epsilon^{\bar{0}}\partial_a N-N\partial_a \epsilon^{\bar{0}})\right]
    \nonumber\\
    \!\!&&\!\!
    \qquad\quad
    - H_a\left[q^{ab} \left(\epsilon^{\bar{0}}\partial_bN-N\partial_b\epsilon^{\bar{0}}\right)\right]
    \,,\nonumber
\end{eqnarray}
and
\begin{eqnarray}
    \!\!\!\!\!\!&&\!\!\!\!\!\!
    \{H^{(2)}[N],H^{(2)}[\epsilon^{\bar{0}}]\}
    \\
    \!\!\!\!\!\!&&\!\!\!\!\!\!\qquad
    = H^{(2)} \left[-\frac{N}{\gamma} \{\gamma,H^{(2)}[\epsilon^{\bar{0}}]\}\right]
    \nonumber\\
    \!\!\!\!\!\!&&\!\!\!\!\!\!\qquad\quad
    - \int {\rm d}^4x\; \frac{\sqrt{(\det {\cal P})^{-1}}}{\sqrt{8\pi G}} \frac{{\cal P}^a_p {\cal P}^b_q \tensor{\epsilon}{^p^q_i}}{\gamma} N \partial_a \{B^i_b,H^{(2)}[\epsilon^{\bar{0}}]\}
    \nonumber\\
    \!\!\!\!\!\!&&\!\!\!\!\!\!\qquad
    = H^{(2)} \left[\frac{1}{2\gamma} \delta^{(2)}_{\epsilon^{\bar{0}},N} \gamma\right]
    \nonumber\\
    \!\!\!\!\!\!&&\!\!\!\!\!\!\qquad\quad
    + H^{(2)} \left[\frac{\sqrt{(\det {\cal P})^{-1}}}{\sqrt{8\pi G}} \frac{\gamma}{2} v^q {\cal P}^a_q \left(\epsilon^{\bar{0}} \partial_a N-N \partial_a \epsilon^{\bar{0}}\right)\right]
    \nonumber\\
    \!\!\!\!\!\!&&\!\!\!\!\!\!\qquad
    = H^{(2)}\left[\frac{\sqrt{(\det {\cal P})^{-1}}}{\sqrt{8\pi G}} \gamma v_q {\cal P}^c_q (\epsilon^{\bar{0}}\partial_c N-N\partial_c\epsilon^{\bar{0}}) \right]
    \,.\nonumber
\end{eqnarray}
Combining all the brackets above results in the full bracket (\ref{eq:HH}).

\subsection{Second-class constraint}

Relevant functional derivatives of the second-class constraint (\ref{eq:Torsion-spatial-sym}) are given by
\begin{eqnarray}
    \frac{\delta {\cal T}^{ij}(x)}{\delta {\cal P}^a_r(z)} \!\!&=&\!\! \bigg[\left(\delta^{ij} \delta^r_q - \delta^{r(i} \delta_q^{j)} \right) \Gamma_a^q
    + \delta_m^{(i} \tensor{\epsilon}{^{j)}^r^q} \left({\cal P}^{-1}\right)_c^m \partial_a {\cal P}^c_q
    \nonumber\\
    \!\!&&\!\!\quad
    - \left({\cal P}^{-1}\right)_a^m \left({\cal P}^{-1}\right)_c^r \delta_m^{(i} \tensor{\epsilon}{^{j)}^p^q} {\cal P}^d_p \partial_d {\cal P}_q^c\bigg] \delta^3(x-z)
    \nonumber\\
    \!\!&&\!\!
    + \left({\cal P}^{-1}\right)_a^m \delta_m^{(i} \tensor{\epsilon}{^{j)}^p^r} {\cal P}^d_p \frac{\partial \delta^3(x-z)}{\partial x^d}\,,
\end{eqnarray}
\begin{equation}
    \frac{\delta {\cal T}^{ij}(x)}{\delta {\cal K}^a_r(z)} = - \delta^{r(i} \delta_q^{j)} K_a^q \delta^3(x-z)\,,
\end{equation}
\begin{equation}
    \frac{\delta {\cal T}_{ij}(x)}{\delta B^r_a(z)} = \frac{1}{1+\zeta^2} \left[ \delta_{ij} {\cal P}^a_r - \delta_{r(i} \tilde{\cal P}^a_{j)} \right] \delta^3(x-z)
    \,,
\end{equation}
\begin{equation}
    \frac{\delta {\cal T}_{ij}(x)}{\delta A^r_a(z)} = \frac{1}{1+\zeta^2} \left[
    - \delta_{r(i} \tilde{\cal K}^a_{j)}
    - \zeta \delta_{ij} {\cal P}^a_r \right] \delta^3(x-z)\,,
\end{equation}

\begin{equation}
    \frac{\delta {\cal T}_{ij}(x)}{\delta K^r_a(z)} = \frac{1}{1+\zeta^2} \left[
    - \delta_{r(i} \tilde{\cal K}^a_{j)}
    - \zeta \delta_{r(i} \tilde{\cal P}^a_{j)}\right] \delta^3(x-z)
    \,,
\end{equation}
and hence
\begin{eqnarray}
    \frac{\partial {\cal T}_{ij}}{\partial {\cal B}_{kl}}
    \!\!&=&\!\!
    \left({\cal P}^{-1}\right)^n_b \delta_{n(k} \frac{\partial {\cal T}_{ij}}{\partial B_b^{l)}}
    - \delta_{n(k} {\cal K}^b_{l)} \left({\cal P}^{-1}\right)^r_b \left({\cal P}^{-1}\right)^n_c \frac{\partial {\cal T}_{ij}}{\partial A_c^r}
    \nonumber\\
    \!\!&=&\!\! \frac{1}{1+\zeta^2} \Bigg[ \delta_{ij} \delta^{kl}
    - \delta_{(i}^k \delta^{l}_{j)}
    + \tensor{\epsilon}{_{(i}^p^{(k}} \tensor{\epsilon}{^{l)}^q_{j)}} v_p v_q \Bigg]
    \,.
\end{eqnarray}
Using the above, we obtain
\begin{widetext}
\begin{eqnarray}
    \int{\rm d}^3 z \frac{\delta {\cal T}_{ij}(x)}{\delta A^r_a(z)} \frac{\delta {\cal T}^{kl}(y)}{\delta {\cal P}^a_r(z)} \!\!&=&\!\! - \frac{1}{1+\zeta^2} \Bigg[ \delta_{r(i} \tilde{\cal K}^a_{j)}
    \left(\delta^{kl} \delta^r_p - \delta_p^{(k} \delta^{l)r} \right) \Gamma_a^p
    + \zeta \delta_{ij} \delta_p^{(k} \delta^{l)r} {\cal K}^a_r K_a^p
    \nonumber\\
    \!\!&&\!\!\qquad\qquad
    + \left(v^s \tensor{\epsilon}{_s_{(i}^{(k}}
    - \zeta \delta_{(i}^{(k}\right) \delta_{j)r} \tensor{\epsilon}{^{l)}^p^q} \left({\cal P}^{-1}\right)_c^r {\cal P}^d_p \partial_d {\cal P}_q^c \Bigg] \delta^3(x-y)
    \nonumber\\
    \!\!&&\!\!
    + \frac{1}{1+\zeta^2} \left[
    v^s \tensor{\epsilon}{_{s}_{(i}^{(k}}
    - \zeta \delta_{(i}^{(k} \right] \tensor{\epsilon}{^{l)}_{j)}^p} {\cal P}^d_p \frac{\partial \delta^3(x-y)}{\partial x^d}\,,
\end{eqnarray}
\end{widetext}
and
\begin{eqnarray}
    \!\!\!\!\!\!&&\!\!\!\!\!\!
    \int{\rm d}^3 z \frac{\delta {\cal T}_{ij}(x)}{\delta B^r_a(z)} \frac{\delta {\cal T}^{kl}(y)}{\delta {\cal K}^a_r(z)} 
    \\
    \!\!\!\!\!\!&&\!\!\!\!\!\!\qquad\quad
    = - \frac{1}{1+\zeta^2} \left( \delta_{ij} {\cal P}^a_r - \delta_{r(i} \tilde{\cal P}^a_{j)} \right)
    \delta^{r(k} \delta_q^{l)} K_a^q \delta^3(x-y)\,.\nonumber
\end{eqnarray}
With this, the bracket (\ref{eq:TT bracket}) follows.

\section{Time gauge}
\label{App:Time gauge}

The time gauge fixes a vanishing velocity variable
\begin{equation}
    v^i=0\,.
\end{equation}
On the second-class constraint surface, where $\mathfrak{K}_{ij}=0$, the Lorentz constraint (\ref{eq:Lorentz constraint}) can be solved for the conjugate of the velocity variable,
\begin{equation}
    {\cal E}_i = - \partial_a {\cal P}^a_i\,,
\end{equation}
and the Gauss constraint (\ref{eq:Gauss constraint}) reduces to
\begin{equation}
    G_i = \zeta \partial_a {\cal P}^a_i
    + \tensor{\epsilon}{_i_j^k} {\cal D}_a^j {\cal P}^a_k\,.
\end{equation}
Therefore, the time gauge reduces the phase space by determining $v_i$ and ${\cal E}^i$, and the Ashtekar--Barbero connection ${\cal A}_a^i$ is identified as a multiple of ${\cal D}_a^i$.


\begin{thebibliography}{10}

\bibitem{ADM}
R. Arnowitt, S. Deser, and C.~W. Misner,  in {\em Gravitation: An Introduction to Current Research}, edited by L. Witten (Wiley, New York, 1962), reprinted in \cite{arnowitt2008republication}.

\bibitem{arnowitt2008republication}
R. Arnowitt, S. Deser, and C.~W. Misner, Gen. Rel. Grav. {\bf 40}, 1997 (2008).

\bibitem{hojman1976geometrodynamics}
S.~A. Hojman, K. Kucha\v{r}, and C. Teitelboim, Ann. Phys. (New York) {\bf 96},  88  (1976).

\bibitem{kuchar1974geometrodynamics}
K.~V. Kucha\v{r}, J. Math. Phys. {\bf 15}, 708 (1974).

\bibitem{EMGCov}
M. Bojowald, and E.~I. Duque, Phys. Rev. D {\bf 108}, 084066 (2023), arXiv:2310.06798.

\bibitem{HypDef}
M. Bojowald, E.~I. Duque, and A. Shah, Phys. Rev. D {\bf 111}, 124048 (2025), arXiv:2410.18807.

\bibitem{Ashtekar1}
A. Ashtekar, Phys. Rev. Lett. \textbf{57}, 2244 (1986).

\bibitem{Ashtekar2}
A. Ashtekar, Phys. Rev. D \textbf{36}, 1587 (1987).

\bibitem{JacobsonSmolin1}
T. Jacobson and L. Smolin, Phys. Lett. B \textbf{196}, 39-42 (1987).

\bibitem{JacobsonSmolin2}
T. Jacobson and L. Smolin, Class. Quantum Grav. \textbf{5}, 583 (1988).

\bibitem{Samuel}
J. Samuel, Pramana \textbf{28}, L429-L432 (1987).

\bibitem{Barbero}
J.~F. Barbero, Phys. Rev. D \textbf{51}, 5507 (1995), arXiv:gr-qc/9410014.

\bibitem{Holst}
S. Holst, Phys. Rev. D \textbf{53}, 5966 (1996), arXiv:gr-qc/9511026.

\bibitem{rovelli2004quantum}
C. Rovelli, \textit{Quantum Gravity}, (Cambridge University Press, 2004).

\bibitem{thiemann2008modern}
T. Thiemann, \textit{Introduction to Modern Canonical Quantum General Relativity}, (Cambridge University Press, 2008).

\bibitem{Alexandrov1}
S. Alexandrov, Class. Quantum Grav. \textbf{17}, 4255 (2000), arXiv:gr-qc/0005085.

\bibitem{Alexandrov3}
S.~Yu. Alexandrov and D.~V. Vassilevich, Phys. Rev. D \textbf{58}, 124029 (1998), 	arXiv:gr-qc/9806001.

\bibitem{Ashtekar3}
A. Ashtekar, A.~P. Balachandran, and S. Jo, Int. J. Mod. Phys. A \textbf{4}, 1493-1514 (1989).

\bibitem{BarberoEuclidean}
J.~F. Barbero, M. Basquens, and E.~J.~S. Villaseñor, Phys. Rev. D \textbf{109}, 064047 (2024), arXiv:2312.12947.

\bibitem{Lagraa}
M.~H. Lagraa, M. Lagraa, and N. Touhami, Class. Quantum Grav. \textbf{34} 115010 (2017), arXiv:1606.06918.

\bibitem{Montesinos}
M. Montesinos, R. Escobedo, J. Romero, and M. Celada, Phys. Rev. D \textbf{101}, 024042 (2020), arXiv:1912.01019.

\bibitem{pons1997gauge}
J.~M. Pons, D.~C. Salisbury, and L.~C. Shepley, Phys. Rev. D \textbf{658} (1997), arXiv:gr-qc/9612037.

\bibitem{salisbury1983realization}
D.~C. Salisbury, and K. Sundermeyer, Phys. Rev. D \textbf{27}, 740 (1983).

\bibitem{rovelli1995discreteness}
C. Rovelli and L. Smolin, Nuclear Physics B \textbf{442}, 593 (1995), arXiv:gr-qc/9411005.

\bibitem{Alexandrov2}
S. Alexandrov and D. Vassilevich, Phys. Rev. D \textbf{64}, 044023 (2001), arXiv:gr-qc/0103105.

\bibitem{AreaCritique}
C. Rovelli and S. Speziale, Phys. Rev. D \textbf{67}, 064019 (2003), arXiv:gr-qc/0205108.

\bibitem{spinfoam}
C. Rovelli and F. Vidotto, \textit{Covariant loop quantum gravity: an elementary introduction to quantum gravity and spinfoam theory}, (Cambridge university press, 2015).

\end{thebibliography}
\end{document}